\documentclass[10pt,conference,letterpaper]{IEEEtran}
\usepackage[letterpaper,left=0.635in,right=0.64in,top=0.75in,bottom=1in]{geometry}
\usepackage[USenglish]{babel}
\usepackage{hyperref}
\usepackage{cite}
\usepackage{amsmath,amssymb,amsfonts}
\usepackage{algorithmic}
\usepackage{graphicx,color}
   \graphicspath{{./png/}}
   \DeclareGraphicsExtensions{.png}
\usepackage{textcomp}

\usepackage{xcolor}
\usepackage{soul}
\usepackage{mathtools}
\usepackage[libertine]{newtxmath}
\usepackage{dblfloatfix} 
\usepackage[numbers,sort&compress,square]{natbib}

\definecolor{DarkGreen}{rgb}{0,0.5,0}
\AtBeginDocument{\definecolor{ojcolor}{cmyk}{0.93,0.59,0.15,0.02}}

\begin{document}
\pagestyle{headings}
\def\cal{\fam2}
\renewcommand{\d}{{\mathrm d}}
\newcommand{\e}{\mathrm{e}}
\newcommand{\rd}{{\rm d}}
\newcommand{\sgn}{{\mathrm{sgn}}}
\newcommand{\const}{{\rm const}}
\newcommand{\half}{{\frac{1}{2}}}
\newcommand{\Ibl}{{\llbracket}}
\newcommand{\Ibr}{{\rrbracket}}
\renewcommand{\mod}{{\rm mod}}
\newcommand{\nint}{{\rm nint}}
\newcommand{\appropto}{\mathrel{\vcenter{
  \offinterlineskip\halign{\hfil$##$\cr
    \propto\cr\noalign{\kern2pt}\sim\cr\noalign{\kern-2pt}}}}}
\newcommand{\Seps}{{{\mathcal{S}}_{\scriptstyle \varepsilon}}}
\newcommand{\CalphaPM}{{{\mathcal{C}}^{\boldmath \alpha^+}_{\boldmath \alpha^-}}}
\newcommand{\CalphaP}{{\prescript{\boldmath \bar{p}}{\boldmath \check{\alpha}}{\mathcal{C}}}}
\newcommand{\beginlabel}[2]{\begin{#1}\label{#2}}

\title{Collision-resistant multi-channel M-ASPM configurations with shared single detection channel}
\author{\IEEEauthorblockN{Alexei V. Nikitin}
\IEEEauthorblockA{
Nonlinear LLC\\
Ashland City, TN, USA\\
E-mail: avn@nonlinearcorp.com}
\and
\IEEEauthorblockN{Ruslan L. Davidchack}
\IEEEauthorblockA{Sch. of of Computing and Mathematical Sci.,\\
U. of Leicester, Leicester, UK\\
E-mail: rld8@leicester.ac.uk}}

\maketitle
\thispagestyle{plain}
\pagestyle{plain}
\begin{abstract}
M-ary Aggregate Spread Pulse Modulation (M-ASPM) is a physical layer (PHY) modulation technique that offers several advantages for low-power wide-area networks (LPWANs). For instance, in conventional LPWAN modulations increasing receiver sensitivity by extending symbol duration—thereby proportionally increasing the time-on-air (ToA)—exacerbates collision exposure. In contrast, M-ASPM payload processing gain can vary over a wide range without impacting the effective packet collision rate. In particular, in this work we demonstrate how short front portions of M-ASPM packets can serve as a separate collision-resistant detection channel that, in addition to performing asynchronous packet detection and synchronization, obtains the carrier frequency offset (CFO) for each packet within a desired range and with the required precision. Then, while raising processing gain, the subsequent payload information can be extracted without expanding the sample window per symbol. Consequently, the receiver sensitivity can be significantly increased without exacerbating packet collisions and thus without reducing network throughput under collision-limited operation. We further establish a multi-channel configuration in which numerous quasi-orthogonal payload channels share a single detection channel that additionally performs payload channel identification and selection. Such sharing is especially useful for scaling and economizing LPWAN deployments under diverse technical requirements and constraints. The presented analysis is validated via extensive simulations under high packet collision rates in wide ranges of payload sizes and processing gains, and for varying noise and interference power levels. The results signify that M-ASPM provides a structurally distinct scaling behavior compared to conventional LPWAN modulations, decoupling range extension from collision-induced throughput degradation.
\end{abstract}


\maketitle

\section{INTRODUCTION AND MOTIVATION} \label{sec:introduction}
Aggregate Spread Pulse Modulation (ASPM)~{\cite{Nikitin20pulsed}} is a spread-spectrum technique in which the transmitted low-rate information is represented by a sparse sequence of pulses embedded within a wideband waveform of a given bandwidth. This waveform is then used to modulate a single-frequency carrier. The demodulated signal in the receiver remains wideband, but it is sparsely sampled to extract the encoded information.

M-ary ASPM (M-ASPM)~{\cite{Nikitin21MaryGlobe, Nikitin2022M-ASPM, Nikitin23_25ASPMpatents}} encodes $\log_2 M$~bits per pulse by providing $M$~distinct ``states" for a single pulse through a combination of its amplitude and arrival time. Various such combinations enable coherent and/or noncoherent modulation schemes with different spectral and energy-per-bit efficiencies. The maximum pulse rate of a given encoding scheme cannot exceed some fraction of the bandwidth, and decreasing this rate below its maximum value proportionally increases the M-ASPM processing gain. Favorably, this increase, and thus the increase in the receiver sensitivity, can be performed in an effectively continuous manner, and without changes in the bandwidth or other physical parameters of the transmitter (Tx) and the receiver (Rx).

Further, by using different pulse shaping filters (PSFs), the statistical properties and the time-domain appearance of the transmitted M-ASPM signal can be varied without changing its spectral composition. For example, the waveform can be made to be statistically indistinguishable from Gaussian noise (e.g., for covert communications)~{\cite{Nikitin20pulsed, Nikitin23_25ASPMpatents}}, or to consist of constant-envelope pulses (e.g., for energy efficiency of transmissions)~{\cite{Nikitin21MaryGlobe, Nikitin2022M-ASPM, Nikitin23_25ASPMpatents, Nikitin2023MASPMpower}}. In addition, properly chosen distinct PSFs enable multiple quasi-orthogonal PSF channels that can be simultaneously used within the same spectral band without excessive mutual interference. Thus, we can address various technical requirements of communication schemes under diverse constraints.

In particular, the noncoherent single-sideband M-ASPM with constant-envelope pulses of the same magnitude~{\cite{Nikitin2023MASPMpower, Nikitin2024implementation, Nikitin2025detection}} is highly appropriate for low-power wide-area networks (LPWANs), and this version of M-ASPM is assumed through the rest of this paper. As illustrated in Fig.~\ref{fig:SER EbN0 nc}, for a given number of bits per pulse, such M-ASPM has the same energy-per-bit efficiency as LoRa (short for ``Long Range," a popular modulation technique for LPWANs~{\cite{Vangelista2017frequency, Baruffa20error}}) with the same number of bits per waveform (i.e., when the LoRa spreading factor (SF) is $\mathrm{SF}=\log_2 M$). Further, the transmission efficiency of such M-ASPM is the same as the efficiency of transmitting a continuous constant-envelope signal (that is, the same as LoRa's).

\begin{figure}[!t]
\centering{\includegraphics[width=8.6cm]{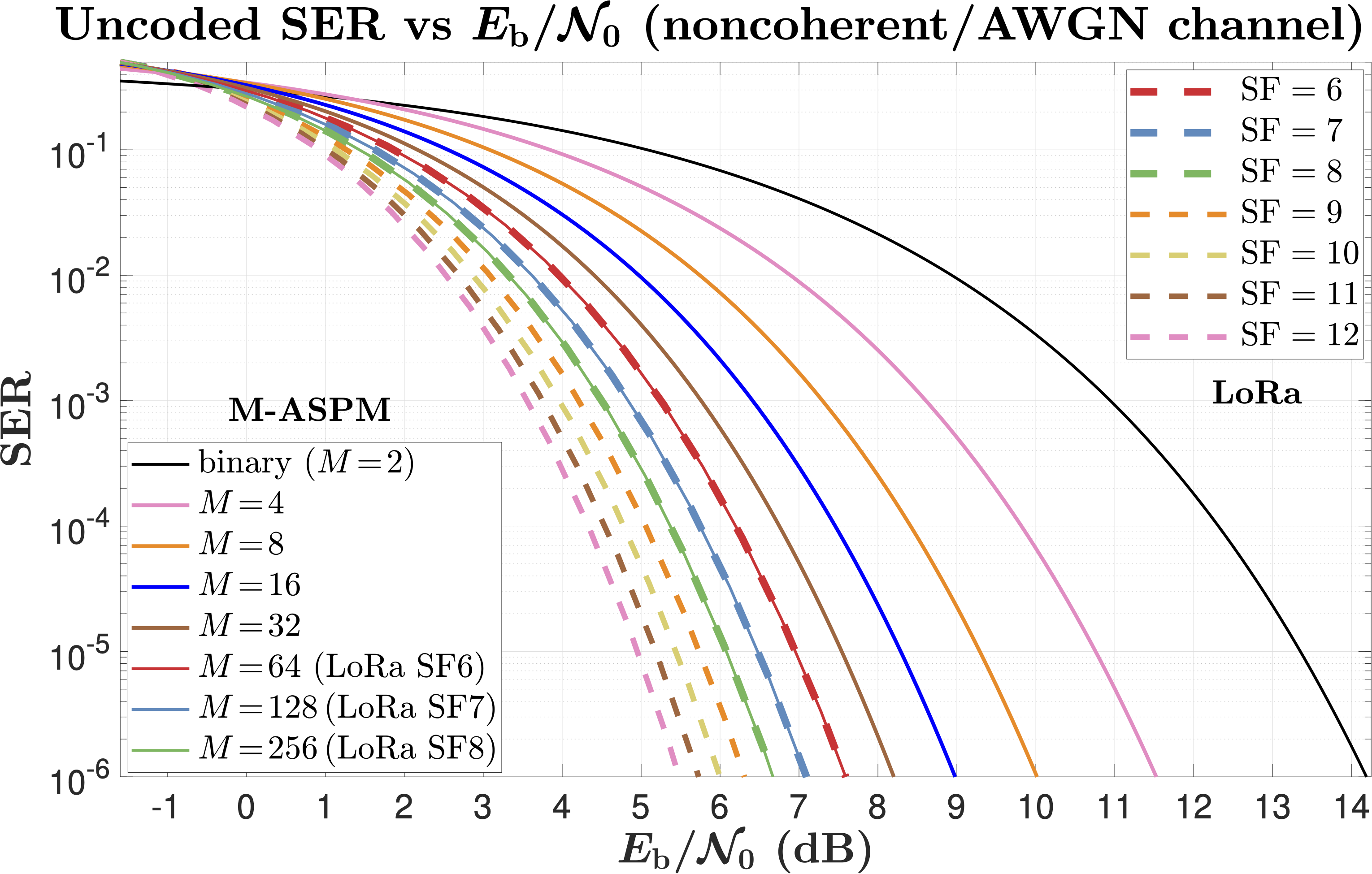}}
\caption{\boldmath Uncoded symbol error rate (SER) vs {\boldmath $E_\mathrm{b}/{\mathcal{N}}_0$} performances of LoRa (dashed lines) and single-sideband M-ASPM (solid lines) for noncoherent detection in AWGN channel.
\label{fig:SER EbN0 nc}}
\end{figure}

However, in LoRa-based long-range wide-area network (LoRaWAN) deployments, improved receiver sensitivity for larger~SF values comes at the expense of greater collision vulnerability due to longer time-on-air (ToA), which increases proportionally to~$2^\mathrm{SF}$. In contrast, in M-ASPM higher receiver sensitivity can be achieved without exacerbating collision exposure, decoupling range extension from collision-induced throughput degradation. The theoretical and practical exploration of this inherent PHY-level M-ASPM property is the main motivation for the present work.

Thus, with a broader goal in mind, in Section~\ref{sec:ASPM collision scaling} we describe M-ASPM encoding and processing that enable such decoupling, explain their PHY origins, and quantify the M-ASPM collision exposure.

Further, as outlined in~\cite{Nikitin2024implementation, Nikitin2025detection}, relatively short low-gain front portions of the transmitted packets can be used for robust asynchronous detection of the packets in the Rx at low computational cost. This significantly simplifies the Rx implementation, since unless the packet is detected, only low-order matched filtering is continuously performed. In addition, packet detection can be combined with measuring the packet-specific carrier frequency offset (CFO) that is used for the subsequent CFO and sampling time offset (STO) corrections, and for packet synchronization together with identification of the payload parameters. Favorably, a short duration of such front portions of the packets makes their contribution to the overall collision impact relatively small, and these portions can employ the same PSFs even for the packets with different PSF channels. This single-channel detection approach greatly economizes M-ASPM networks, especially their high-gain, high-throughput configurations.

Therefore, in Section~\ref{sec:ASPM} we provide a detailed description of such multi-channel M-ASPM configurations with shared single detection channel. Together, Sections~\ref{sec:ASPM collision scaling} and~\ref{sec:ASPM} describe the complete system architecture, combined with explicit algorithms for asynchronous detection, CFO measurement, synchronization, and payload channel identification and sampling.

In Section~\ref{sec:physical parameters} we specify the values and ranges of the physical parameters that are used in the subsequent quantitative examples and simulations. 

In Section~\ref{sec:collisions} we examine various factors contributing to the impact of packet collisions in multi-channel M-ASPM configurations, and provide their overall analytical assessment. Then, in~Section~\ref{sec:simulation settings}, we describe the common setting for simulations presented in the ensuing sections that illustrate and verify this analysis.

The simulations presented in Section~\ref{sec:simulation results} address the main properties of collisions among packets with single- and multi-channel payloads. Further, the simulations in Section~\ref{sec:main tradeoffs} characterize the impact on the M-ASPM collision performance caused by various tradeoffs among the transmit packet rates, payload sizes, the~$M$ and~IpI values, and the degree of orthogonality among the payload channels. In this section we also assess the impact of high-rate packet collisions on the detection channel.

We provide additional discussion of the presented results and their implications in Sections~\ref{sec:secondary tradeoffs} and~\ref{sec:implications}, and conclude the paper in Section~\ref{sec:conclusion}.

Further, all acronyms used in the paper are listed in Appendix~\ref{app:acronyms}, and the mathematical notations are discussed in Appendix~\ref{app:notations}.

\section{M-ASPM AND ITS COLLISION SCALING} \label{sec:ASPM collision scaling}
In this paper, we adopt the noncoherent single-sideband M-ASPM with constant-envelope pulses of the same magnitude~{\cite{Nikitin2023MASPMpower, Nikitin2024implementation, Nikitin2025detection}}. The information is encoded in the ``arrival times" (indices) $k_j$ of the pulses in a digital ``pulse train"~$\hat{x}[k]$, where only a relatively small fraction of samples (``pulses") has non-zero values. Such a ``designed" pulse train can be expressed as
\beginlabel{equation}{eq:ptrain}
  \hat{x}[k] = \sum_{j=1}^{j_\mathrm{max}} \Ibl k\!=\!k_j\Ibr\, (-1)^j\,,
\end{equation}
where $k$~is the sample index (``time"), $k_j$~is the sample index of the \mbox{$j$-th}~pulse, and $\Ibl\dots\Ibr$ is the {\it Iverson bracket\/}~\cite{Knuth92two} which is equal to~1 if the expression inside is true and~0 if it is false.
The positive integer~$j$ represents the sequential pulse counter in the train~$\hat{x}[k]$, and the value of $j_\mathrm{max}$ is determined by the size of the message encoded in the designed train. For example, with M-ary encoding $j_\mathrm{max}=\lceil N_\mathrm{b}/\log_2M\rceil$, where $N_\mathrm{b}$ is the size of the message in bits and $\lceil x\rceil$~is the ceiling function.

Note that in~(\ref{eq:ptrain}) only the positions of the pulses, and not their amplitudes (i.e., the magnitudes and signs), encode information.
The alternating signs of the pulses in~(\ref{eq:ptrain}) simply ensure that~$\hat{x}[k]$ is a zero-mean signal, helping to eliminate a direct current (DC) bias in the modulating signal. This is convenient but not strictly necessary.

To encode $\log_2M$ bits per pulse, for the arrival times in~(\ref{eq:ptrain}) we can use
\beginlabel{equation}{eq:kj}
  k_j = k_0 + (j-1)N_\mathrm{p} + m_j n_\mathrm{off}\,,
\end{equation}
where $k_0$ is some initial (``starting") index, $N_\mathrm{p}$ is the average (``nominal") interpulse interval (IpI), ${n_\mathrm{off}<N_\mathrm{p}/M}$ is the increment in pulse-position offsets, and ${m_j\in \{0,1,2,\dots,M\!-\!1\}}$.
Therefore, $m_j n_\mathrm{off}$ is simply the remainder after division of $k_j\!-\!k_0$ by the IpI~$N_\mathrm{p}$.
The average ``pulse rate"~$f_\mathrm{p}$ in such a train is $f_\mathrm{p}=F_\mathrm{s}/N_\mathrm{p}$, where $F_\mathrm{s}$ is the sample rate, and thus the raw bit rate $f_\mathrm{b}$ is ${f_\mathrm{b}=f_\mathrm{p}\log_2M}$.

Note that, while the time duration of~$\hat{x}[k]$ is approximately ${(j_\mathrm{max}-1)N_\mathrm{p}/F_\mathrm{s}}$ and increases with~$N_\mathrm{p}$, its non-zero values are confined to the intervals $k_0 + (j\!-\!1)N_\mathrm{p} + [0,n_\mathrm{off}(M\!-\!1)]$, $j\in \{1,2,\dots,j_\mathrm{max}\}$, with the total time duration~$j_\mathrm{max}{n_\mathrm{off}(M\!-\!1)/F_\mathrm{s}}$ that does not depend on~$N_\mathrm{p}$.
As shown below in Section~\ref{subsec:Rx processing}, such encoding allows us to preserve, while raising the M-ASPM processing gain, the time duration of the sampling window per symbol for extracting the payload information in the Rx.

\subsection{PULSE-SHAPING FILTERS AND TRANSMITTED SIGNAL} \label{subsec:PSFs}
To enable creation of a band-limited analog signal suitable for modulating a carrier, the train~$\hat{x}[k]$ given by~(\ref{eq:ptrain}) can be ``reshaped" by linear filtering. In particular, the impulse response~$\hat{\zeta}_i[k]$ of such a ``pulse shaping" filter (PSF) can be a nonlinear chirp with the desired autocorrelation function (ACF), e.g.
\beginlabel{equation}{eq:chirp Ni}
  \hat{\zeta}_i[k] = \hat{g}_i[k]+\mathrm{i}\,\hat{h}_i[k] = \frac{1}{\sqrt{L_i}}\, \Ibl 0\!\le\! k\!<\!L_i\Ibr\, \exp \left( \mathrm{i}\,\Phi_i[k]\right)\,,
\end{equation}
where~$\Phi_i[k]$ is the phase and~$L_i$ is the ``duration" (length) of the chirp in samples. To generate such a waveform, one can use, for example, the approach described in~\cite{Doerry2007generating}. In~(\ref{eq:chirp Ni}), the imaginary part of~$\hat{\zeta}_i[k]$ is the discrete Hilbert transform of its real part, i.e., ${\hat{h}_i[k]=H\left\{ \hat{g}_i[k] \right\}}$~\cite{Bracewell2000FourierFULL, Todoran2008discrete}.

Filtering the designed train~$\hat{x}[k]$ with the PSF~$\hat{\zeta}_i[k]$ creates the digital modulating signal~$z_i[k]$ (the ``reshaped train")
\beginlabel{equation}{eq:ptrain filtered}
   z_i[k] = \sqrt{L_i}\, (\hat{x}\ast \hat{\zeta}_i)[k] =  \sqrt{L_i}\, \sum_j \hat{\zeta}_i[k\!-\!k_j]\, (-1)^j\,,
\end{equation}
where the asterisk denotes convolution. After digital-to-analog (D/A) conversion, the real and imaginary parts of the analog signal~$z_i(t)$ can be used for quadrature amplitude modulation of a carrier with frequency~$f_\mathrm{c}$, providing the transmitted waveform ${\mathrm{Re} (z_i(t))\sin(2\pi f_\mathrm{c}t)} + {\mathrm{Im} (z_i(t))\cos(2\pi f_\mathrm{c}t)}$. Since $\hat{h}_i[k]$ is the Hilbert transform of~$\hat{g}_i[k]$, this waveform will occupy only a single sideband with the physical bandwidth~$B$ equal to the baseband bandwidth of~$\hat{\zeta}_i[k]$~{\cite{Bracewell2000FourierFULL}}. In addition, if we require that the chirps in~(\ref{eq:ptrain filtered}) do not overlap (i.e., $L_i\le N_\mathrm{p}-(M\!-\!1)n_\mathrm{off}$), then
the magnitude of the reshaped train is
\beginlabel{equation}{eq:z abs}
   \left| z_i[k] \right| = \sum_j \Ibl 0 \!\le\! k\!-\!k_j \!<\! L_i\Ibr\,,
\end{equation}
and the transmitted signal will consist of constant-envelope pulses of the same magnitude. This is beneficial for the Tx power efficiency, and for simplicity of its implementation. The variance of such a reshaped train is equal to its {\it pulse duty cycle\/}~$D_i=L_i/N_\mathrm{p}<1$, and thus, for a given IpI~$N_\mathrm{p}$, the average power of~$z_i[k]$ is proportional to~$D_i$.

Note that, for a given ACF, we can construct a great variety of distinct PSFs $\hat{\zeta}_i[k]$, $i\in \{1,2,3,\dots\}$ with different lengths and/or time-frequency profiles. Thus, we refer to forming the reshaped train~$z_i[k]$ with a particular PSF~$\hat{\zeta}_i[k]$ as ``using the $i$-th PSF channel."

\subsubsection{Choice of ACF and sampling rate} \label{subsubsec:ACF}
As discussed in~{\cite{Nikitin21MaryGlobe, Nikitin2022M-ASPM}}, a good choice for the ACF of a PSF would be a pulse that combines a small time-bandwidth product (TBP)~\cite{Gabor45theory, Vetterli95wavelets} (e.g., close to that of a Gaussian pulse) with compact frequency support. An example of such ACF would be a raised-cosine (RC) pulse~\cite{Proakis06digital} with a sufficiently large roll-off factor~$0\le \beta\le 1$ (e.g., $\beta\gtrsim 1/5$). Then the sample rate~$F_\mathrm{s}$ in the digital waveforms can be chosen as ${F_\mathrm{s}=2\mathcal{N}_\mathrm{s}B}$, where~$1\le \mathcal{N}_\mathrm{s} = 2/(1\!+\!\beta)\le 2$ is the oversampling factor.

Throughout this paper we use $\beta=1/4$, and thus $\mathcal{N}_\mathrm{s}=8/5$. The main reason for this particular choice is to ensure insensitivity of the received signal to fractional STOs and to simplify compensation for the cumulative STO due to the CFO between the Tx and Rx~{\cite{Nikitin2024implementation}}. In addition, albeit relatively small, $60$\% oversampling above the Nyquist rate significantly simplifies design of anti-aliasing filters in the Rx, so that these filters do not attenuate the signal even at large CFO.

The above choice of the ACF and the oversampling factor limits the minimum value of the pulse-position increment in the encoding~(\ref{eq:kj}) as $n_\mathrm{off}\geq 4$. Further, with $\mathcal{N}_\mathrm{s}=8/5$ the value of the M-ASPM spectral efficiency is
\beginlabel{equation}{eq:eta ASPM}
  \eta =  \frac{f_\mathrm{b}}{B} = \frac{2\mathcal{N}_\mathrm{s}}{N_\mathrm{p}}\,\log_2 M = \frac{16}{5N_\mathrm{p}}\,\log_2 M\,.
\end{equation}

\subsection{PROCESSING OF RECEIVED SIGNAL} \label{subsec:Rx processing}
For noncoherent ('nc') detection, in the receiver's quadrature demodulator, the noisy passband signal is multiplied by the orthogonal sinusoidal signals from a local oscillator (LO), lowpassed (i.e., with anti-aliasing filters), and converted to the in-phase and quadrature digital signals $I[k]$ and $Q[k]$. Then the ``raw" incoming digital signal is $x[k]=I[k]+\mathrm{i}\,Q[k]$.

For known values of the CFO~$\Delta{f}_\mathrm{c}$ and the initial index~$k_0$, for each $j$-th pulse in the payload with the PSF~$\hat{\zeta}_i[k]$ we obtain~$M$ values
\beginlabel{equation}{eq:payloadsampling}
  y_j[m] = \sum_{n=0}^{L_i-1} x[j,m,n]\, \zeta_i[n]\,,
\end{equation}
where $m\in \{0,1,2,\dots,M\!-\!1\}$,
\beginlabel{equation}{eq:zetai}
  \zeta_i[n] = \hat{\zeta}^\ast_i[n] \exp \left( \mathrm{i}\, \frac{ 2\pi\Delta{f}_\mathrm{c}}{F_\mathrm{s}} n \right)\,,
\end{equation}
and
\beginlabel{align}{eq:xjmk}
  &x[j,m,n] =\\ &x[k_0 + (j\!-\!1)N_\mathrm{p} - \nint\left((j\!-\!1)N_\mathrm{p}\Delta{f}_\mathrm{c}/f_\mathrm{c}\right)
  + mn_\mathrm{off} + n]\,.\nonumber
\end{align}
In~(\ref{eq:zetai}) the superscript asterisk denotes the complex conjugate, and in~(\ref{eq:xjmk}) the {\it nearest integer function\/}~$\nint(x)=\lfloor 1/2+x\rfloor$, where $\lfloor x\rfloor$~is the floor function.

Note that~(\ref{eq:payloadsampling}) represents acquiring $M$~samples per pulse in the signal $y_\mathrm{nc}[k]$ obtained as a convolution of~$x[k]$ with~$\zeta_i[L_i\!-\!k]$, with the sampling incorporating the STO compensation.

This process may be referred to as {\em decimated sampling\/} of~$x[k]$ with the PSF~$\hat{\zeta}_i[k]$, and obtaining the initial index~$k_0$ in equation~(\ref{eq:xjmk}) for decimated sampling may be referred to as {\em synchronization\/}. To explicitly emphasize, when needed, that decimated sampling requires synchronization and CFO/STO corrections, it may be referred to as {\em synchronized corrected decimated sampling\/} (SCDS).

Then, for each $j$-th pulse in the payload sequence, the value of~$m_j$ can be determined from the condition
\begin{equation} \label{eq:pl sampling}
  y_j^2[m_j] = \max \{y_j^2[0],y_j^2[1],\dots,y_j^2[M\!-\!1]\}\,.
\end{equation}

Note that, as follows from~(\ref{eq:payloadsampling})--(\ref{eq:pl sampling}), in SCDS the time duration of the sampling window per payload pulse remains~$n_\mathrm{off}M/F_\mathrm{s}$ for any values of the IpI~$N_\mathrm{p}$ and the PSF length~$L_i$.

\subsection{UNCODED SER PERFORMANCE OF {M-ASPM} IN AWGN CHANNEL AND ITS CONTROL BY IPI} \label{subsec:Mary SER}
As shown in~{\cite{Nikitin21MaryGlobe,Nikitin2022M-ASPM}}, in the ideal case of zero CFO and STO the symbol error probability~$P_\mathrm{s}$ of noncoherent M-ASPM in AWGN channel can be expressed as
\begin{equation} \label{eq:ASPM SER binom EbN0}
P_\mathrm{s} \!=\! P_\mathrm{s} \left(\frac{\Gamma}{\eta}\right) \!=\! \frac{1}{M} \sum_{k=2}^M (-1)^k\binom{M}{k} \exp \left( -\frac{k\!-\!1}{k}\, \frac{\Gamma}{\eta}\, \log_2 M \right),
\end{equation}
where~${\binom{n}{m} = \frac{n!}{(n-m)!\,m!}}$ is the binomial coefficient, $\Gamma$~is the signal-to-noise ratio (SNR), and  $\eta = f_\mathrm{b}/B$ is the spectral efficiency (i.e., $\Gamma/\eta=E_\mathrm{b}/{\mathcal{N}}_0$, where $E_\mathrm{b}$~is the energy per bit and ${\mathcal{N}}_0$~is the one-sided power spectral density of the noise). Notably, as illustrated in Fig.~\ref{fig:SER EbN0 nc}, this AWGN symbol error probability for ideal M-ASPM is the same as for ideal noncoherent LoRa when $M=2^\mathrm{SF}$, where $\mathrm{SF}$~is the LoRa {\it spreading factor}~\cite{Baruffa20error}.

Note that~$P_\mathrm{s}$ is a decreasing function of~$\Gamma\eta^{-1}$,
while both the SNR~$\Gamma$ and the spectral efficiency~$\eta$ are inversely proportional to the IpI~$N_\mathrm{p}$. Consequently, for given bandwidth~$B$ and the value of~$M$, the M-ASPM's receiver sensitivity is proportional to the IpI, and can be raised to a desired level by increasing~$N_\mathrm{p}$.
Therefore, unlike LoRa, which operates at maximum spectral efficiency for a given $M=2^\mathrm{SF}$, M-ASPM is a ``true" spread spectrum technique with adjustable {\it processing gain\/}~$G_\mathrm{p}$. This gain can be expressed as the ratio of the bandwidth and the pulse rate, namely\\[-2ex]
\beginlabel{equation}{eq:Gp}
  G_\mathrm{p} =  \frac{B}{f_\mathrm{p}} = \frac{N_\mathrm{p}}{2\mathcal{N}_\mathrm{s}} = \frac{\log_2 M}{\eta} = \frac{5}{16}\,N_\mathrm{p}\,.
\end{equation}
For example, for~$N_\mathrm{p}=4800$ the processing gain is $G_\mathrm{p}=1500$, or $31.8\,$dB.

Also note that, for a given PSF~$\hat{\zeta}_i[k]$, a change in the processing gain does not affect the computational cost of SCDS in the Rx.

\subsection{M-ASPM collision exposure} \label{subsec:collision exposure}
As was just discussed, for a given bandwidth~$B$ the M-ASPM Rx sensitivity is proportional to the processing gain~$G_\mathrm{p}$, while the time duration of the sampling window per payload pulse is not affected by~$G_\mathrm{p}$. On the other hand, for a given pulse rate~$f_\mathrm{p}$ the value of $\Gamma\eta^{-1}$ does not change with the bandwidth. Therefore, as follows from~(\ref{eq:ASPM SER binom EbN0}), neither does the Rx sensitivity. At the same time, the time duration of the sampling window per payload pulse is inversely proportional to the bandwidth. The physical origin of this distinct M-ASPM property is illustrated in Fig.~\ref{fig:collision scaling}.

\begin{figure}[!t]
\centering{\includegraphics[width=8.6cm]{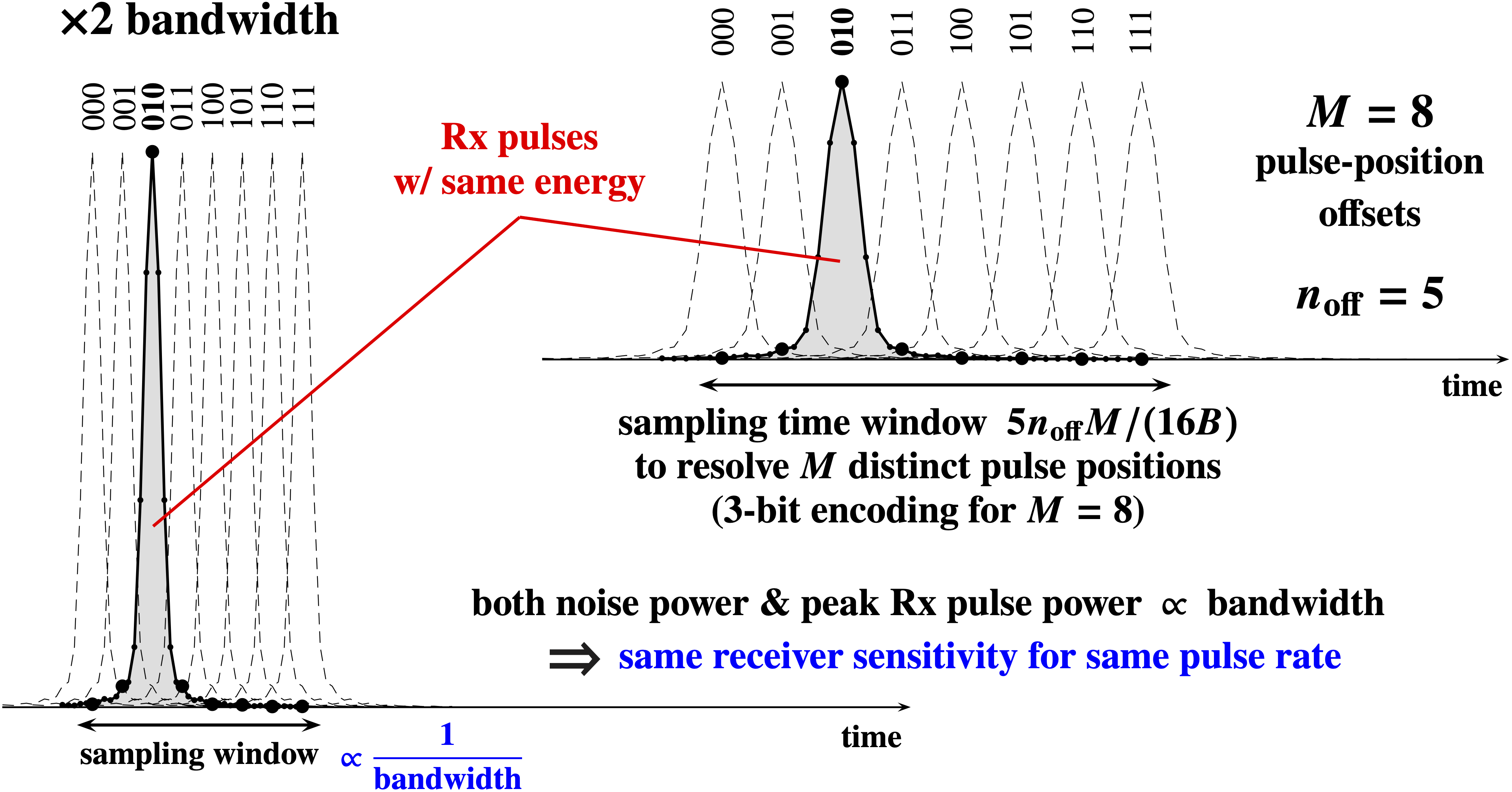}}
\caption{\boldmath While for given pulse rate M-ASPM Rx sensitivity does not change with bandwidth, time duration of sampling window needed to resolve $M$~distinct pulse positions is inversely proportional to bandwidth.
\label{fig:collision scaling}}
\end{figure}

\begin{figure}[!b]
\centering{\includegraphics[width=8.6cm]{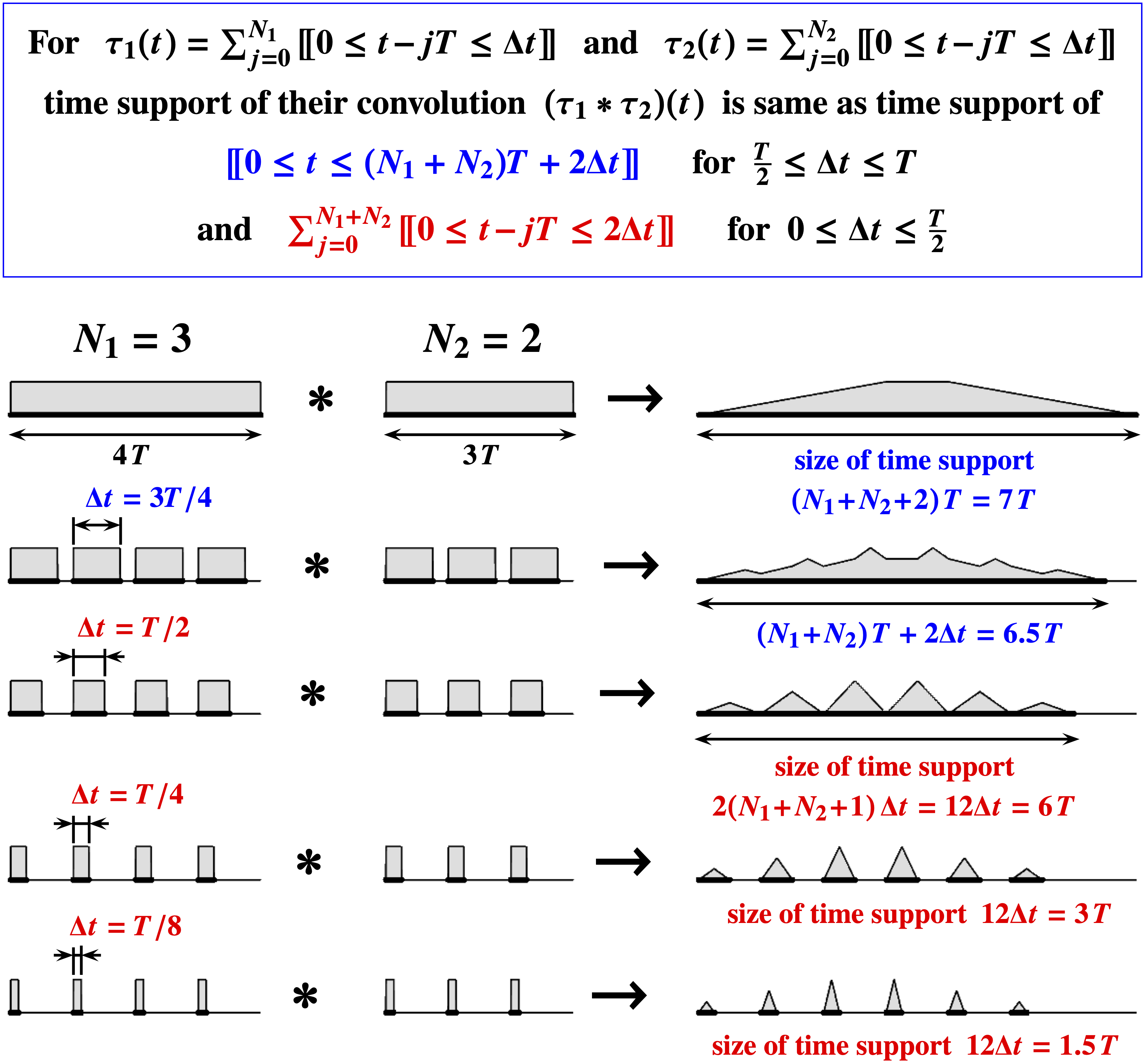}}
\caption{\boldmath Time-domain overlap of colliding signals $\tau_1(t)$ and $\tau_2(t)$ can be quantified by size of time support of their convolution ${(\tau_1\ast \tau_2)(t)}$.
\label{fig:time support}}
\end{figure}

\begin{figure}[!t]
\centering{\includegraphics[width=8.6cm]{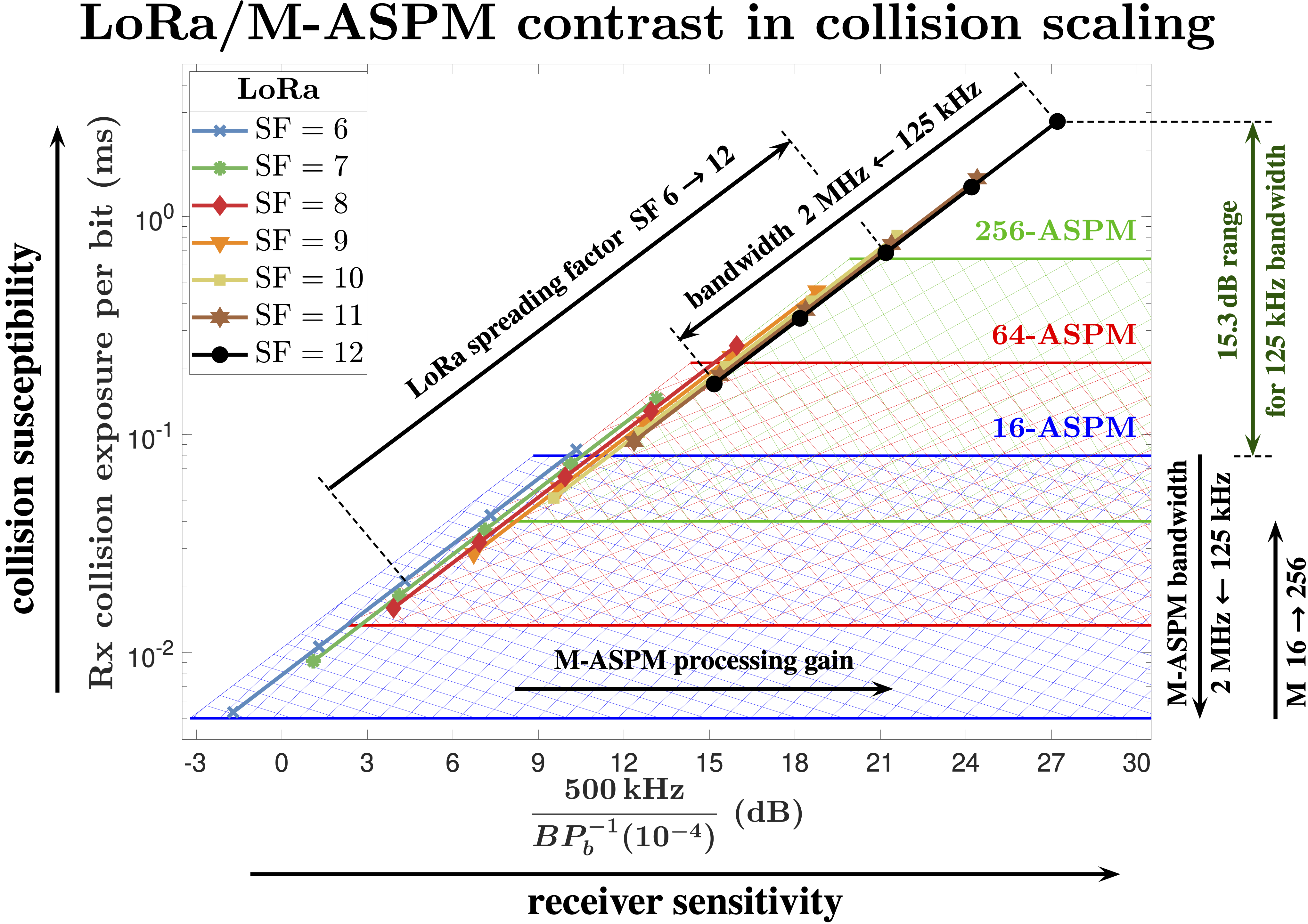}}
\caption{\boldmath At given bandwidth, collision exposure increases with receiver sensitivity for LoRa, and remains unchanged for M-ASPM.
\label{fig:contrast}}
\end{figure}

Note that, with the CFO correction, the TBP of the pulses in the Rx pulse train does not change with bandwidth (see, e.g., \cite{Nikitin2024implementation, Nikitin2025detection}). Therefore, while for a given pulse rate~$f_\mathrm{p}$ the average AWGN SNR~$\Gamma$ is inversely proportional to the bandwidth, the magnitude of the Rx pulses is proportional to the bandwidth and the Rx sensitivity remains unchanged. At the same time, as illustrated in Fig.~\ref{fig:collision scaling}, the width of the Rx pulses is inversely proportional to the bandwidth. As a consequence, the time duration of the sampling window that is needed to resolve $M$~distinct pulse positions (i.e., $n_\mathrm{off}M/F_\mathrm{s}$) is also inversely proportional to the bandwidth.

Further, consider the Rx pulse train of interest containing $N_1$~pulses, that collides with an interfering pulse train composed of $N_2$~pulses. For a sufficiently small CFO, the STO correction term~$\delta{k}_j={\nint\left((j\!-\!1)N_\mathrm{p}\Delta{f}_\mathrm{c}/f_\mathrm{c}\right)}$ in~(\ref{eq:xjmk}) remains much smaller than~$n_\mathrm{off}M$~\cite{Nikitin2024implementation}. Then the time support of the Rx pulse train with $N_1$~pulses is effectively confined to the time support of the function~$\tau_1[k]={\sum_{j=1}^{N_1} \Ibl 0 \!\le\! k\!-\!k_j\!+\!n_\mathrm{off}/2 \!\le\! n_\mathrm{off}M\Ibr}$. For a sufficiently small CFO difference, the time support intervals of the interfering pulse train would be confined to that of~$\tau_2[k]={\sum_{j=1}^{N_2} \Ibl 0 \!\le\! k\!-\!k_j\!+\!n_\mathrm{off}/2\!+\!\Delta{k} \!\le\! n_\mathrm{off}M\Ibr}$, where~$\Delta{k}$ is  due to the arrival time difference. Since, with SCDS, a collision of any pulse of interest with zero-valued portions of an interfering pulse train in the Rx is harmless, a collision can be impactful only when the respective time support intervals of the colliding Rx pulse trains overlap.

As illustrated in Fig.~\ref{fig:time support}, such overlap can be characterized by the time support of the convolution ${(\tau_1\ast \tau_2)[k]}$ of the two non-negative signals $\tau_1[k]$ and $\tau_2[k]$ confining time supports of the colliding Rx pulse trains. One can deduce from this figure that, for two Rx pulse trains containing total $N$~pulses, the combined length (``size") of the time support of ${(\tau_1\ast \tau_2)[k]}$ is $2(N\!-\!1)n_\mathrm{off}M/F_\mathrm{s}$ when $n_\mathrm{off}M\le N_\mathrm{p}/2$. Since the two colliding trains together encode $N\log_2M$ bits, for ${N\gg 1}$ we can define the M-ASPM {\em collision exposure per bit\/} as ${2n_\mathrm{off}M/(F_\mathrm{s}\log_2M)} = {5n_\mathrm{off}M/(8B\log_2M)}$. For a given~$M$, this exposure depends only on the bandwidth, and is independent of the M-ASPM receiver sensitivity.

In contrast, for example, the Tx ToA and the Rx time support size of a LoRa frame are both $2^\mathrm{SF}/B$. Then the LoRa collision exposure per bit is ${2^\mathrm{SF}/(B\times\mathrm{SF})}$ and, for a given bandwidth, it increases with the LoRa receiver sensitivity. This difference in LoRa and M-ASPM collision scaling is quantified in Fig.~\ref{fig:contrast}. Here the receiver sensitivities are calculated for the uncoded AWGN bit error rate $\mathrm{BER} = 10^{-4}$, i.e., for $P_\mathrm{b}=MP_\mathrm{s}/(2(M\!-\!1))=10^{-4}$, where for LoRa $M=2^\mathrm{SF}$.

Note that both LoRa and M-ASPM have the same ranges of energy-per-bit efficiencies, and the ranges of M-ASPM data rates and spectral efficiencies include those of LoRa~\cite{Nikitin2022M-ASPM}. However, as can be seen in Fig.~\ref{fig:contrast}, M-ASPM can offer the collision exposure that is orders of magnitude smaller than LoRa's. For example, for a given bandwidth, 16-ASPM with a single PSF channel can provide collision-limited throughput capacity that is {\em  on par\/} with the capacity of LoRa with $\mathrm{SF}=6$. At the same time, without reducing this capacity, the 16-ASPM receiver sensitivity (and thus the range) can be extended to the range of LoRa with any larger spreading factor, including $\mathrm{SF}=12$ and beyond.

\section{MULTI-CHANNEL M-ASPM WITH SHARED SINGLE DETECTION CHANNEL} \label{sec:ASPM}
Note that sparse SCDS requires accurate synchronization in time (i.e., obtaining the initial index~$k_0$ in~(\ref{eq:xjmk})), and acquiring the value of the CFO~$\Delta{f}_\mathrm{c}$. It further requires knowledge of the PSF~$\hat{\zeta}_i[k]$ and other parameters of the pulse train, i.e., identifying the channel used for payload. Therefore, to benefit from the M-ASPM collision scaling, the practical means for obtaining this information in the Rx must themselves have small collision exposure.

Favorably, in addition to the data-carrying pulse train (``payload") described in Section~\ref{sec:ASPM collision scaling}, an M-ASPM packet can comprise other pulse trains, having relatively short time duration and specifically dedicated to conveying such additional information for SCDS. While occupying the same frequency band and consisting of constant-envelope pulses, these other portions of the packet can employ substantially different IpIs, and use PSFs that are quasi-orthogonal among distinct packet segments. In such an aggregate packet, its separate portions can be tailored to perform different functions in a holistic, coordinated manner.

\subsection{ASYNCHRONOUS DETECTION OF M-ASPM PACKETS COMBINED WITH MEASURING CARRIER FREQUENCY OFFSET} \label{subsec:detection}
For example, as outlined in~{\cite{Nikitin2024implementation, Nikitin2025detection}}, the front segment of a packet (therein referred to as ``leading sequence") can be used for robust and sensitive asynchronous detection of the packet, at low computational cost that is independent of the detection sensitivity. Favorably, as follows from~(\ref{eq:payloadsampling})--(\ref{eq:xjmk}), for a leading sequence with a sufficiently small IpI ${N^\prime_\mathrm{p}\ll N_\mathrm{p}}$, and the respectively short PSF (i.e., with length ${L^\prime\ll L_i}$), the sensitivity of such detection remains largely unaffected by a relatively large CFO~$\Delta{f}_\mathrm{c}$ arising from mismatch in the frequencies of the LOs together with Doppler shifts. Further, this detection can be combined with measuring the CFO within the desired range and with the desired precision. When the CFO value is known, the subsequent segments of the packet, that employ larger IpIs and/or longer PSFs (e.g., the ``timing sequence" for synchronization, and the payload), can be processed without undue deterioration in the quality of the received signal.

However, since the contrast in the processing gains between different portions of the packet can exceed two orders of magnitude, matching the detection sensitivity with that of the payload, while maintaining the portion of the packet dedicated to the detection relatively small, poses a significant challenge. A detection algorithm that overcomes this challenge is presented in~{\cite{Nikitin2025detection}}, together with detailed explanation of the computationally inexpensive nonlinear procedures and tools employed in implementation of its steps. In the current paper we mainly adopt this algorithm, with an important modification (described in Section~\ref{subsubsec:QTF}) that significantly improves its robustness under heavy collisions.

As further emphasized in~{\cite{Nikitin2025detection}}, asynchronous detection of a small-IpI leading sequence requires significantly shorter matched filters in the Rx compared to those used for the synchronization and decoding of the payload (e.g., 30~taps instead of hundreds or thousands). This greatly reduces the Rx implementation cost since, unless the packet is detected, only low-order matched filtering is continuously performed.

\subsection{SYNCHRONIZATION AND PAYLOAD CHANNEL IDENTIFICATION} \label{subsec:synchronization}
The main purpose of synchronization is obtaining the initial index~$k_0$ in equation~(\ref{eq:xjmk}) for decimated sampling of the payload. In addition, obtaining~$k_0$ can be followed by identifying the parameters of the payload sequence such as, for example, the length and the temporal direction of the payload PSF, the payload processing gain, the length of the payload (i.e., the number of pulses in the payload), and/or the value of~$M$. For synchronization and payload channel identification we use a ``timing"~{\cite{Nikitin2024implementation, Nikitin2025detection}} pulse sequence that follows the leading sequence and precedes the payload.

For synchronization we use two pulses with a relatively long PSF~$\hat{\zeta}[k]$ of length $L\gg L^\prime$. The positions of these pulses are separated by~$L$, and the position of the second pulse has a known offset~$-\Delta{k}$ from the initial index~$k_0$ that is used for decimated sampling of the payload.

To compensate for the CFO, after a packet is detected and the measured CFO value~$\Delta{f}_\mathrm{c}$ is obtained, the matched filter~$\zeta[k]$ for the PSF~$\hat{\zeta}[k]$ is modified as
\begin{equation} \label{eq:zeta delta}
  \zeta_\Delta[k] = \zeta[k]\, \exp\left(-\mathrm{i}\, 2\pi \frac{\Delta{f}_\mathrm{c}}{F_\mathrm{s}} k\right)\,,
\end{equation}
and~$\zeta_\Delta[k]$ is then applied to the incoming baseband Rx signal to start producing the filtered signal~$y_\mathrm{nc}[k]$ as
\begin{equation} \label{eq:Rx sync}
  y_\mathrm{nc} = x\ast \zeta_\Delta = \left(I+\mathrm{i}\,Q\right)\ast \zeta_\Delta\,.
\end{equation}

Initially, $y_\mathrm{nc}[k]$ is obtained at full sampling rate (i.e., for all integer values of~$k$ after the beginning of filtering). We continue the filtering until we observe that the two largest peaks of~$y^2_\mathrm{nc}[k]$ on the interval~$[k\!-1\!,k\!+\!L]$ are above some threshold (fence)~$\alpha[k]$ and are $L\pm 1$~samples apart. (Note that, while the CFO compensation has already been performed by modifying the matched filter according to~(\ref{eq:zeta delta}), the fractional STO still causes the $\pm~1$ samples uncertainty in the distance between the synchronization peaks.)

 For example, we obtain the peak power values that are above~$\alpha[k]$ as
\begin{equation} \label{eq:peaks above}
  z^\prime[k] = z_k\, \Ibl  z_k\!>\!\alpha_k\Ibr \Ibl  z_k\!>\! z_{k\!-\!1}\Ibr \Ibl  z_k\!\ge\!z_{k\!+\!1}\Ibr \,,
\end{equation}
where $z_k={y^\prime}^2_\mathrm{nc}[k]$ and the threshold~$\alpha_k=\alpha[k]$ is obtained during the detection as described in~{\cite{Nikitin2024implementation, Nikitin2025detection}}. Then we check if
\begin{equation} \label{eq:two largest peaks}
  \prod_{i=k+2}^{k+L-1} \Ibl \bar{z}^\prime_k>z^\prime_i\Ibr \Ibl z^\prime_{k+L}>z^\prime_i\Ibr = 1\,,
\end{equation}
where
\begin{equation} \label{eq:first peak}
  \bar{z}^\prime_k = \max \{z^\prime_{k-1},z^\prime_k,z^\prime_{k+1}\}\,.
\end{equation}
When the condition~(\ref{eq:two largest peaks}) is met, the value of~$k_0$ can be obtained as
\begin{equation} \label{eq:Dk}
  k_0 = k + \Delta{k}\,.
\end{equation}

For payload channel identification we add a third pulse with the PSF~$\hat{\zeta}[k]$ that has a position offset~$L+i\,n^\prime_\mathrm{off}$ from the second pulse, where $n^\prime_\mathrm{off}\ge 4$, $i\in \{1,2,\dots,N_\mathrm{ch}\}$ is the channel number, and~$N_\mathrm{ch}$ is the total number of payload channels. Then, after synchronization, we use decimated sampling with the filter matched to~$\zeta_\Delta[k]$ given by~(\ref{eq:zeta delta}) to measure this offset and thus identify the $i$-th payload channel that is characterized by its respective set of parameters.

Note that during synchronization we determine the power of the synchronization peaks, i.e., the values of $\bar{z}^\prime_k$ and $z^\prime_{k+L}$ in~(\ref{eq:two largest peaks}). Therefore, as discussed in~\cite{Nikitin2024implementation}, we concurrently obtain a measure of the Rx signal strength. Then, if feedback communication between the uplink nodes and the gateway is available, this measure can be used for implementing the Tx power control discussed in Section~\ref{subsec:power control}.

\subsection{SYNCHRONIZATION CORRECTION FOR PAYLOADS WITH ``FLIP" PSFS} \label{subsec:flip correction}
The PSF~$\hat{\zeta}_i[k]$ can be characterized by its {\em temporal direction\/} as an {\em up-chirp\/} (when the frequency increases with time) or a {\em down-chirp\/} (when the frequency decreases with time). When~$\hat{\zeta}_i[k]$ is an up-chirp, then its matched filter~$\zeta_i[k]=\hat{\zeta}^\ast_i[L_i-k]$ is a down-chirp, and {\em vice versa\/}. Also, when a 2nd~PSF is the matched filter for the 1st~PSF, these two PSFs are ``flip" PSFs. ``Flip" PSFs are characterized by the same frequency passband and the ACF, yet opposite temporal directions.

For non-zero~CFO, the time shifts in the peaks of the filtered Rx signal will be opposite in sign for “flip" PSFs, relative to the positions of these peaks for zero~CFO. Therefore, when the payload PSFs have the opposite temporal direction from those used for synchronization, finite precision of the CFO measurement obtained during the detection (see~{\cite{Nikitin2024implementation, Nikitin2025detection}}, and Section~\ref{subsec:leading}) leads to an additional small (e.g., within $\pm 1$~sample) uncertainty in obtaining~$k_0$.

To correct this, for the payload channels that use PSFs of the opposite temporal direction from those used for synchronization, in the timing sequence we add a pulse with the PSF~$\zeta[k]$ (i.e., the ``flip" PSF of $\hat{\zeta}[k]$) that has a position offset~$L$ from the third pulse. Then, after the payload channel has been identified, we use decimated sampling with~$\zeta_\Delta[k]$ given by~(\ref{eq:zeta delta}) in the $\pm 1$~sample range around the expected position of this pulse (i.e.,  around ${k_0-\Delta{k}+2L+in^\prime_\mathrm{off}}$) to obtain the required correction for~$k_0$ in the payload sampling.

\subsection{SHARING SINGLE DETECTION CHANNEL AMONG M-ASPM PACKETS WITH DIFFERENT PSF CHANNELS USED FOR PAYLOADS} \label{subsec:sharing}
An example of transmitted M-ASPM packet suitable for sharing a single detection channel among multiple payload channels is given in Fig.~{\ref{fig:Tx packet}}, and the respective signal processing in the Rx is illustrated by a diagram shown in~Fig.~{\ref{fig:Rx single}}. The numerical values used in Fig.~{\ref{fig:Tx packet}} are discussed in the next section, as they relate to the analysis and simulations subsequently presented in this paper.

As one can see in the example of Fig.~{\ref{fig:Tx packet}}, the entire transmitted M-ASPM packet is composed of constant-envelope pulses with the same magnitude. This is beneficial for the Tx power efficiency, as it is the same as the efficiency of transmitting a continuous constant-envelope signal. It also simplifies Tx implementation.

The leading sequence (shown in cyan color) is a ``regular" (periodic) sequence of short PSFs with $L^\prime=29$ and 100\%~duty cycle. In the Rx, the in-phase and quadrature baseband digital signals $I[k]$ and $Q[k]$ are filtered with a matched filter~$\zeta^\prime[k]$ of the leading sequence's PSF~$\hat{\zeta}^\prime[k]$ to obtain~$y^\prime_\mathrm{nc}[k]=\left(I+\mathrm{i}\,Q\right)\ast \zeta^\prime$. For asynchronous detection, $y^\prime_\mathrm{nc}[k]$~is obtained at full sampling rate (that is, for all integer values of~$k$). However, filtering the leading sequence requires significantly shorter matched filters in the Rx compared to those used for the synchronization and decoding of the payload (e.g., 30~taps for $L^\prime=29$, instead of hundreds or thousands).

\begin{figure}[!b]
\centering{\includegraphics[width=8.6cm]{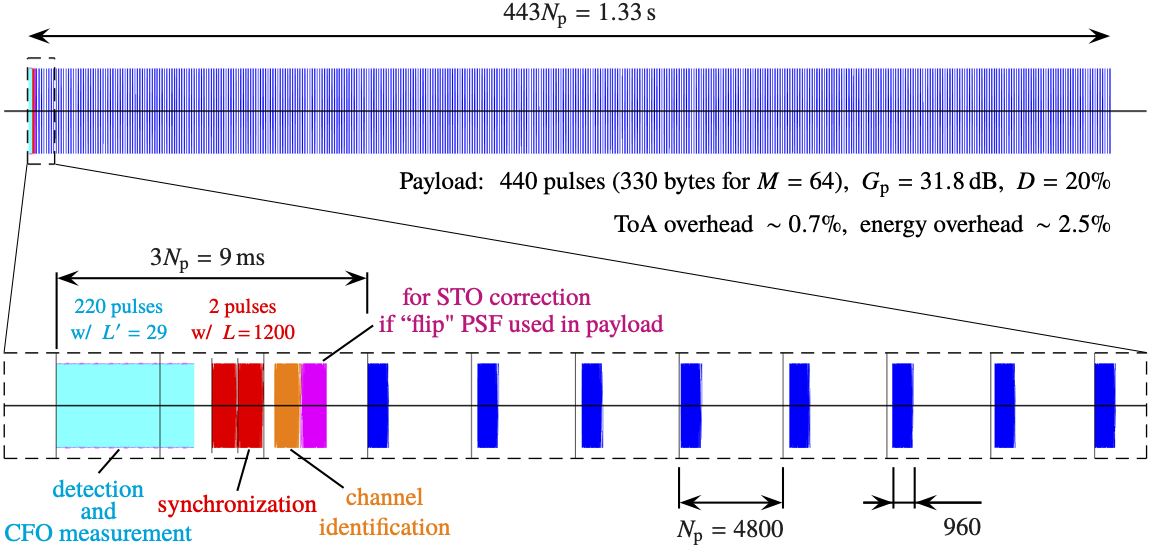}}
\caption{\boldmath Quantitative example of transmitted M-ASPM packet with leading and timing sequences preceding payload.
\label{fig:Tx packet}}
\end{figure}
\begin{figure}[!b]
\centering{\includegraphics[width=8.6cm]{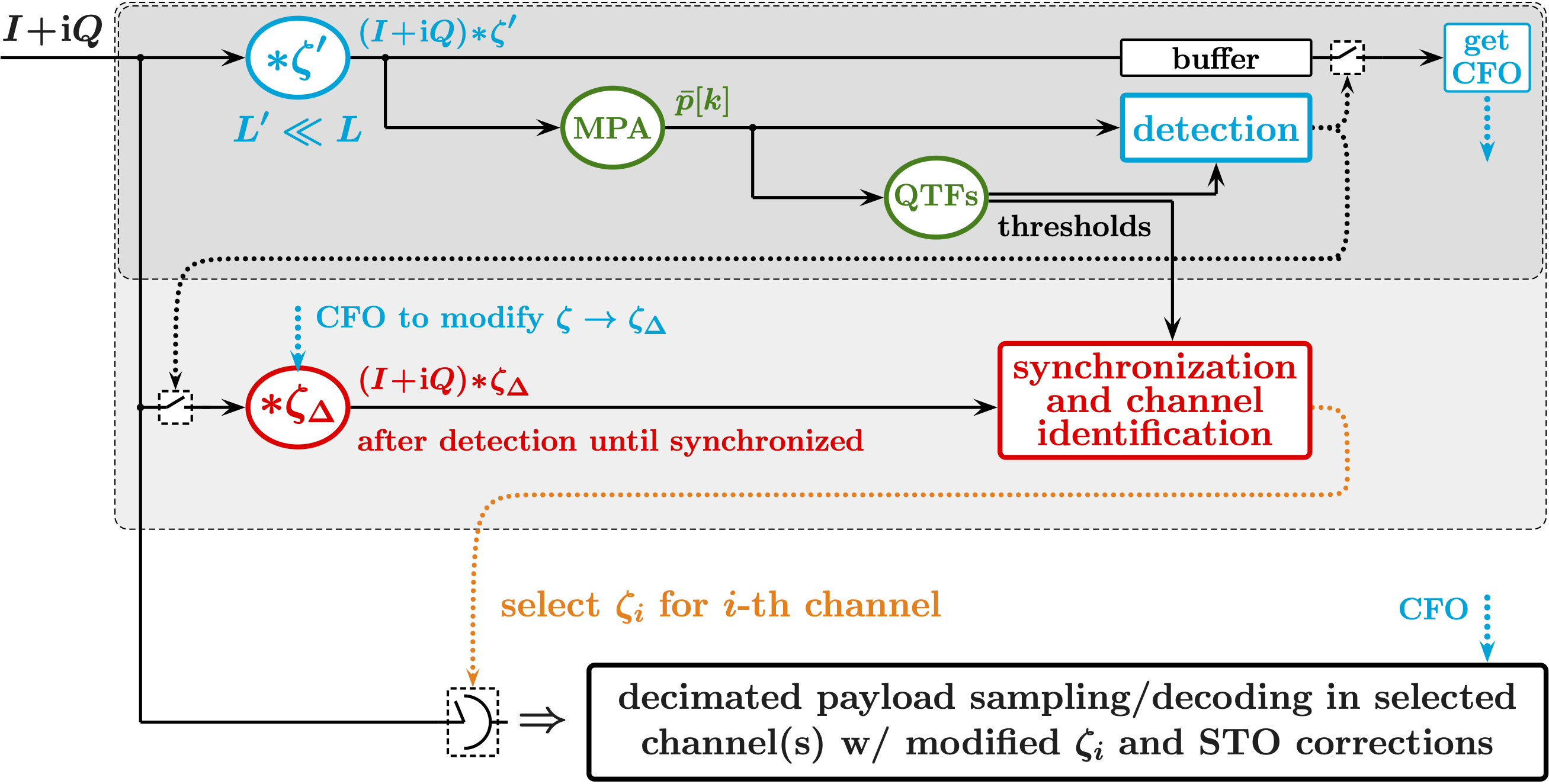}}
\caption{\boldmath Rx signal processing when single detection channel is shared among multiple payload channels.
\label{fig:Rx single}}
\end{figure}

In parallel with obtaining~$y^\prime_\mathrm{nc}[k]$, the modulo power average (MPA) filtering of~$y^\prime_\mathrm{nc}[k]$ is performed, along with applying  quantile tracking filters (QTFs) to the MPA output~$\bar{p}[k]$ for constructing the thresholds for detection and subsequent synchronization~{\cite{Nikitin2024implementation, Nikitin2025detection}. Both the MPA filtering and QTFs are computationally inexpensive ($\mathcal{O}(1)$ per output value), with insignificant contribution to the overall Rx processing cost. In the detection algorithm, calculations of the pulse counting function and its modulo exponential averaging (MEA) are also low-complexity $\mathcal{O}(1)$~operations.

A portion of~$y^\prime_\mathrm{nc}[k]$ (of time duration relatively small compared to duration of the leading sequence) is stored in a data buffer. Once the arrival of the packet is detected, the CFO value~$\Delta{f}_\mathrm{c}$ is calculated using the~$y^\prime_\mathrm{nc}[k]$ values stored in the buffer.

After the packet is detected and~$\Delta{f}_\mathrm{c}$ is obtained, the matched filter~$\zeta[k]$ for the PSF~$\hat{\zeta}[k]$ of the timing sequence is modified (see~(\ref{eq:zeta delta})), and the modified matched filter~$\zeta_\Delta[k]$ is then applied to the incoming baseband Rx signal $x=I+\mathrm{i}\,Q$ to start producing the filtered signal~$y_\mathrm{nc}[k]=(x\ast \zeta_\Delta)[k]$. Using the first two pulses in the timing sequence (shown in red color in~Fig.~{\ref{fig:Tx packet}}) with known position offsets relative to the payload sequence, and the threshold derived from QTF filtering of the MPA output~$\bar{p}[k]$, synchronization is obtained. Subsequently, by measuring the position offset of the third pulse in the timing sequence (shown in orange color), the payload channel number (and thus the parameters of the payload sequence) is identified. For this, we obtain~$N_\mathrm{ch}$ samples, where~$N_\mathrm{ch}$ is the total number of payload channels.

For the payload channels that use PSFs with the opposite temporal direction from those used for synchronization, a ``flip" PSF of $\hat{\zeta}[k]$ (adjacent to the third pulse, and shown in magenta color) is added and used to correct the shift in~$k_0$ due to the finite precision of the CFO measurement. We only need to obtain three samples for the “flip” pulse for this correction.

Then, for the identified payload parameters (i.e., for the obtained channel number~$i$), the SCDS with~$\hat{\zeta}_i[k]$ is performed to extract the payload data.

\section{PHYSICAL PARAMETERS FOR QUANTITATIVE EXAMPLES AND SIMULATIONS} \label{sec:physical parameters}
As was discussed earlier, both noncoherent LoRa and the version of M-ASPM used in this paper are constant-envelope modulations, and both have the same energy-per-bit efficiency for a given number of bits per waveform. Thus, when operating under effectively the same physical conditions (e.g., the same physical frequency band, transmit power, antenna gains, and various system attenuations such as insertion, path, and matching losses, etc.), LoRa may represent a suitable benchmark for M-ASPM.

However, LoRa and M-ASPM are substantially different modulations, and their direct comparison cannot be performed without considering a variety of constraints. Therefore, choosing the same physical domain of operation is useful when such comparison is desirable and appropriate. Consequently, in the quantitative examples and simulations presented in this paper we assume the nominal carrier frequency $f_\mathrm{c}=915\,$MHz, corresponding to the LoRa band in North America. Further, since for any~$M$ the M-ASPM spectral efficiency can be reduced by increasing the IpI, the value of $\Gamma\eta^{-1}$ in~(\ref{eq:ASPM SER binom EbN0}) can be increased to a desired level without reducing the signal bandwidth. Thus, we choose the maximum bandwidth $B=500\,$kHz among those used by LoRa. For this bandwidth, as follows from the discussion in Section~\ref{subsubsec:ACF}, the sampling rate is~$F_\mathrm{s}=1.6\,$MHz.

Throughout the paper, we use the $\pm 30$~parts per million (ppm) CFO range (i.e., $\pm 27.5\,$kHz). In simulations the CFO values are uniformly distributed in this interval (except when considering, as in Fig.~\ref{fig:CFO impact}, the impact of the CFO range on collisions, where this range varies.) This represents a moderate practical LO~mismatch combined with Doppler shift. In the Rx (with the LO frequency~$f_\mathrm{c}$), we assume that the bandwidths of the bandpass and the lowpass/anti-aliasing filters are sufficiently large to accommodate the CFO $\Delta{f}_\mathrm{c}=f^\prime_\mathrm{c}-f_\mathrm{c}$ in the desired range without attenuation.

\begin{figure}[!t]
\centering{\includegraphics[width=8.6cm]{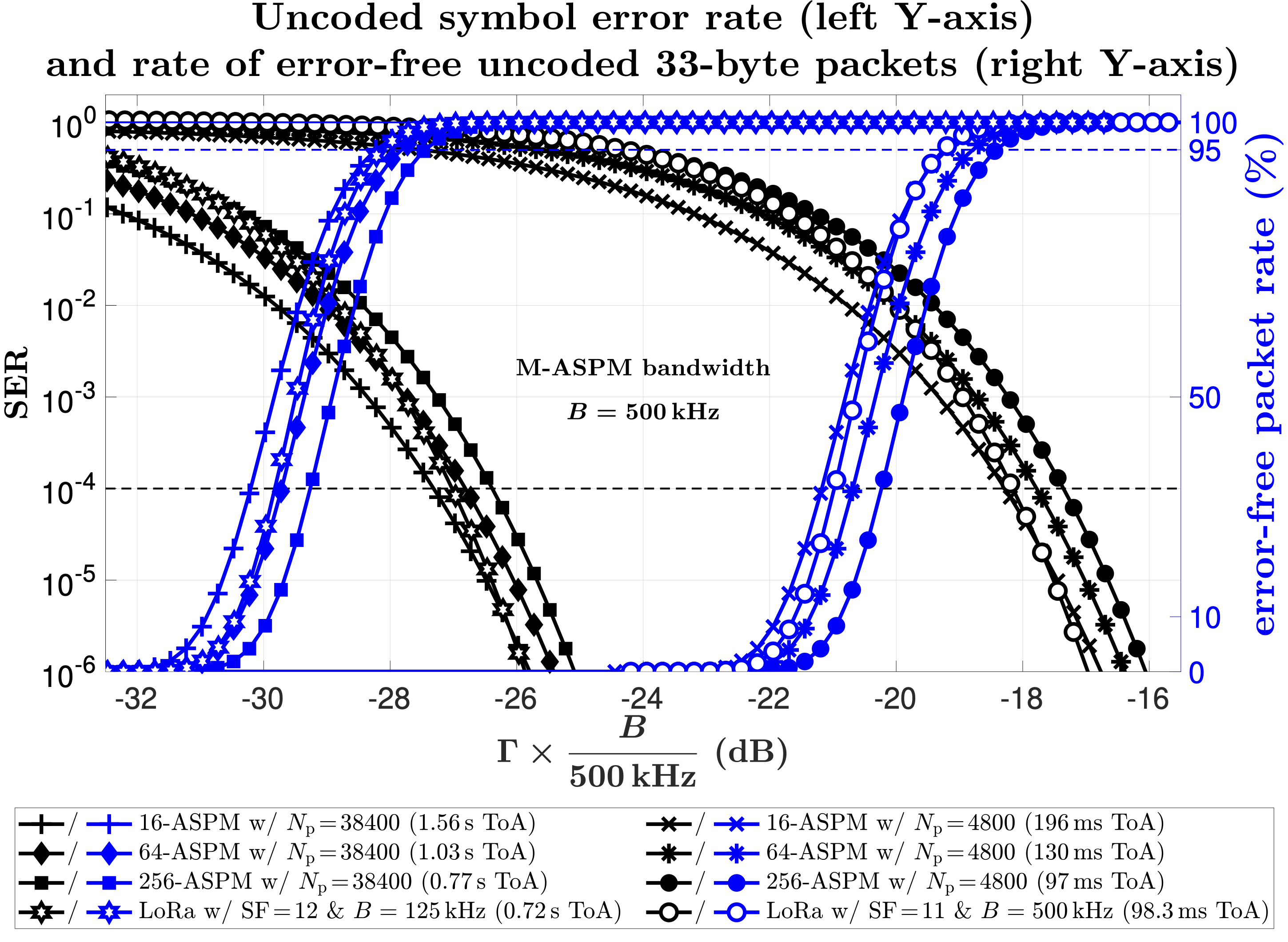}}
\caption{\boldmath Correspondence in Rx sensitivities between LoRa and M-ASPM for range of IpI values used in subsequent quantitative examples and simulations.
\label{fig:LoRa and M-ASPM}}
\end{figure}

\subsection{NUMERICAL VALUES FOR M-ASPM PACKETS} \label{subsec:numerical values}
For the prime interest of achieving large throughputs in high-gain, long-range M-ASPM configurations, in the subsequent simulations the IpI values vary in $9\,$dB range from $N_\mathrm{p}=4800$ to $N_\mathrm{p}=38400$. For $N_\mathrm{p}=4800$ and $B=500\,$kHz, the M-ASPM Rx sensitivity corresponds to that of LoRa with $\mathrm{SF}=11$ and $500\,$kHz bandwidth. On the high end, for $N_\mathrm{p}=38400$ this sensitivity corresponds to LoRa with $\mathrm{SF}=12$ and $125\,$kHz bandwidth. This is illustrated in~Fig.~\ref{fig:LoRa and M-ASPM} for 33-byte LoRa and M-ASPM packets for the ideal case of zero~CFO, perfect synchronization, and no preamble overhead. We will refer to this figure again in more detail in Section~\ref{sec:secondary tradeoffs}, when discussing the tradeoff between collision resistance and energy efficiency in M-ASPM.

As illustrated earlier in Fig.~\ref{fig:Tx packet}, the leading sequence uses the PSF of length $L^\prime=29$ and $100$\% duty cycle. Then the range of the CFO measurements is $\pm F_\mathrm{s}/(2L^\prime)$, or $\pm 30\,$ppm. The total length of the leading sequence is $N_\mathrm{LS}=220$~pulses ($4\,$ms), and the parameters of the detection algorithm are chosen according to~{\cite{Nikitin2025detection}}. The number of pulses used for obtaining the CFO values is $N_\mathrm{CFO}=64$, which provides better than $0.5\,$ppm accuracy of these measurements. This accuracy is sufficient for CFO correction of the largest PSFs used in the simulations (with $L_i=1824$), and for STO compensation in the longest payloads (e.g., $1.32\,$s or 440~pulses with $N_\mathrm{p}=4800$).

As further shown in Fig.~\ref{fig:Tx packet}, the timing sequence uses PSFs of length $L=1200$. With this choice, for the same magnitude of the pulses in the leading and timing sequences, the detection probability and the joint probability of detection with synchronization and correct channel identification are effectively equivalent.

There is limited utility (akin to ‘‘false positive" detection) in the correct detection of the packets, and obtaining the CFO, if the subsequent payload cannot be reliably synchronized and decoded. Similarly, it would be wasteful (akin to ‘‘false negative" detection) to ‘‘miss" high-quality payloads by failing to correctly detect the leading pulse sequence and obtain the CFO. Consequently, the PSF lengths~$L_i$ in the payloads are chosen so that, without collisions, the 95\% packet detection probability typically precedes the respective probability of a detected packet to be error-free by 1--2\,dB, except for the case of short 16-ASPM packets, where these probabilities are achieved at about the same SNR. Further, with a sensible technical requirement that the transmitted payload pulses have the same magnitude for all PSF channels, the difference in lengths of the payload PSFs is confined to a~1\,dB range, so that the impact of the detection probability on the total error-free throughput remains insignificant. This relation between the detection probability and the rate of error-free detected packets is appropriately shown in the presented simulation results.

\begin{figure*}[b!]
\centering{\includegraphics[width=17.2cm]{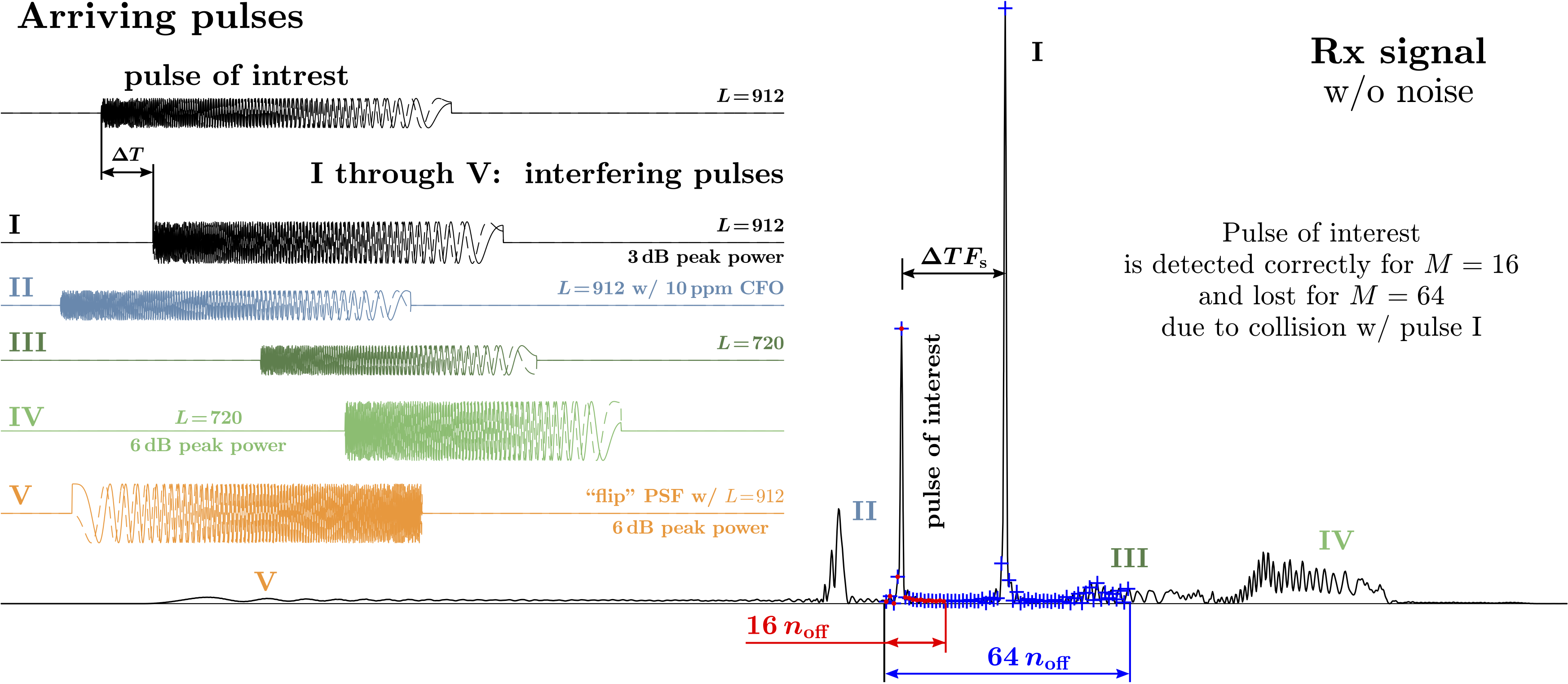}}
\caption{\boldmath Impact of payload collisions in M-ASPM decreases with increase in (i)~$M$, (ii)~difference in PSF lengths and/or temporal directions, and (iii)~magnitude of CFO.
\label{fig:payload collision impact}}
\end{figure*}

\section{IMPACT OF PACKET COLLISIONS IN M-ASPM} \label{sec:collisions}
For random transmissions, packets can partially overlap in time. The average number of packets that overlap with a given packet is the {\em average number of collisions per packet\/}~$\lambda$, or {\em collision rate\/}. If all transmitted packets have the same ToA~$T_\mathrm{packet}$, then for a constant mean transmission rate~$\mathcal{R}$ the value of~$\lambda$ is given by
\beginlabel{equation}{eq:lambda}
  \lambda = 2\mathcal{R}T_\mathrm{packet}\,.
\end{equation}

For uniform transmission time probabilities, the fraction of the packets that do not experience collisions is~$\exp(-\lambda)$. To keep this fraction above~90\%, the packet rate needs to be limited to
\beginlabel{equation}{eq:rate packets}
  \mathcal{R} <  \frac{-\ln(0.9)\, F_\mathrm{s}}{2N_\mathrm{PL}N_\mathrm{p}}\,,
\end{equation}
where $N_\mathrm{PL}$ is the number of pulses in payload. (Here we ignore the small ToA overhead due to the leading and the timing sequences.) For example, $\mathcal{R} < 0.04\,$Hz for the packet in~Fig.~\ref{fig:Tx packet}. Therefore, if all ToA packet collisions were impactful (i.e., resulting in an error), the rate of packets shown in~Fig.~\ref{fig:Tx packet} would be limited to only one packet every 25~seconds, or less than 3,500~packets/day. Even for much shorter 55-pulse payloads the restriction~(\ref{eq:rate packets}) leads to just one packet per 3~seconds, or 28,800~packets/day. For nodes transmitting 100~packets/day, this amounts to only 288~nodes, which is clearly insufficient for applications requiring high-throughput wide-area coverage. Further, extending the IpI to increase the receiver sensitivity proportionally reduces the Tx rate constraint even more.

Favorably, as demonstrated in Section~\ref{subsec:collision exposure} and further discussed below, in high-gain M-ASPM configurations the impact of collisions is no longer related to the ToA, and does not increase with the IpI. Specifically, when ${N_\mathrm{p}>2n_\mathrm{off}M}$, ${2n_\mathrm{off}M}$ replaces ${N_\mathrm{p}}$ in~(\ref{eq:rate packets}) for the packet rate constraint on co-PSF collisions in M-ASPM. Then, for example, for the packets in~Fig.~\ref{fig:Tx packet} with ${N_\mathrm{p}=38400}$, and with~$n_\mathrm{off}=5$ (the value used throughout the rest of the paper), the Tx rate constraint becomes $15$~times larger for~$M=256$ and, at the expense of doubling Tx energy per packet, $240$~times larger for $M=16$.

Indeed, with SCDS, in the Rx a given payload pulse is sampled only at~$M$ points that are spread over the time interval~$n_\mathrm{off}M/F_\mathrm{s}$. Therefore, the {\em effective\/} collision rate for M-ASPM is proportional to~$M$, and payloads with smaller values of~$M$ are less sensitive to collisions. This is illustrated in Fig.~\ref{fig:payload collision impact}, where one can see that if the pulse of interest is sampled at 16~points (for $M=16$), its position offset would be determined correctly. On the other hand, if it is sampled at 64~points (for $M=64$), then the interfering co-PSF pulse~I will cause an error.

In addition, Fig.~\ref{fig:payload collision impact} illustrates that impact of payload collisions in M-ASPM decreases when the interfering packets have the CFO different from the signal of interest, and/or employ PSFs of different lengths and/or temporal directions. For example, if the interfering pulse~I were absent, the position offset of the pulse of interest would be obtained correctly, without impact of collisions from the interfering pulses II~through~V, even if the Rx signal is sampled over the full shown time interval.

Although squaring is a nonlinear operation and the magnitudes of overlapping Rx pulses are not additive, differences in the CFO and inter-PSF lengths from the signal of interest broaden the Rx pulses and reduce their magnitude in general, as signified by the Rx pulses II~through~IV. Then the interfering signal becomes more dispersed in time, and its impact on the quality of the packets of interest becomes less due to the confusion among the distinct peaks in the Rx signal, and more contributory to the overall increase in the interference power (i.e., reducing the signal-to-interference-plus-noise ratio (SINR)). In particular, the impact of inter-PSFs of the opposite temporal direction from the pulse of interest (e.g., as the interfering pulse~V in~Fig.~\ref{fig:payload collision impact}) may be typically treated as Gaussian contribution to the noise.

Let us now assess the various factors contributing to the overall impact of collisions among M-ASPM packets that use a single detection channel with multiple payload channels. For simplicity, we will assume that all payloads have the same IpI and the number of pulses (i.e., the same ToA), and the total Tx rate is equally distributed among the packets with different channels.

\subsection{EFFECTIVE COLLISION RATE FOR CO-PSF PAYLOADS} \label{subsec:effective}
As discussed in Section~\ref{subsec:collision exposure}, a co-PSF payload collision will be impactful only when the time intervals used for decimated sampling collide. Then, when ${N_\mathrm{p}>2n_\mathrm{off}M}$, the {\em effective\/} average number of collisions for a payload with the ToA~$T_\mathrm{pl} = N_\mathrm{PL}N_\mathrm{p}/F_\mathrm{s}$ is
\beginlabel{equation}{eq:pllambda}
  \lambda_\mathrm{pl} = \lambda\, \frac{T_\mathrm{pl}}{T_\mathrm{packet}} \frac{2n_\mathrm{off}M}{N_\mathrm{p}} = 4n_\mathrm{off}M\, N_\mathrm{PL}\, \frac{\mathcal{R}}{F_\mathrm{s}}\,,
\end{equation}
and $\lambda_\mathrm{pl}$ is independent of the IpI~$N_\mathrm{p}$. Therefore, for a given~$M$, an increase in the M-ASPM's receiver sensitivity does not exacerbate collisions and does not reduce the data throughput. In other words, while the payload ToA~$N_\mathrm{p}\,N_\mathrm{PL}/F_\mathrm{s}$ is proportional to the processing gain, with SCDS the ``effective" payload ToA remains bounded by~$2n_\mathrm{off}M\,N_\mathrm{PL}/F_\mathrm{s}$.

The fraction of the packets that are not affected by collisions is~$\exp(-\lambda_\mathrm{pl})$. Then, for example, to keep the packet loss due to collisions in a single-channel M-ASPM below~10\%, for 33-byte payloads (i.e., for $N_\mathrm{PL}=33\times 8/\log_2 M$) the packet rate would be constrained as
\beginlabel{equation}{eq:rate}
  \mathcal{R} \lesssim  \frac{-\ln(0.9)\, F_\mathrm{s}}{4n_\mathrm{off}M\, N_\mathrm{PL}} \approx \frac{32\,\log_2 M}{M}\,\mathrm{Hz}\,.
\end{equation}
This amounts to about $1\,$Hz (86,400~packets/day) for 256-ASPM, or $8\,$Hz ( 691,200~packets/day) for 16-ASPM.

Let us for now disregard the interference from the detection channel. Then the rate (probability) of error-free detected packets $\widetilde{P}_\mathrm{efdp}\left(\Gamma^\prime\right)$ under collisions can be assessed as
\beginlabel{equation}{eq:co-PSF}
  \widetilde{P}_\mathrm{efdp}\left( \Gamma^\prime \right) \gtrsim  P_\mathrm{efdp}\left(\Gamma^\prime\right)\, \exp\left( -\lambda_\mathrm{pl}\right),
\end{equation}
where $P_\mathrm{efdp}\left(\Gamma^\prime\right)$ is such rate without collisions, $\Gamma^\prime$~is the peak SNR, and $\lambda_\mathrm{pl}$~is expressed by~(\ref{eq:pllambda}). The approximate equality between the left- and right-hand sides of~(\ref{eq:co-PSF}) would be expected when all interfering packets are sufficiently, but not excessively strong (e.g., about~3--9\,dB relative to the packet of interest).

The left-hand sides of~(\ref{eq:co-PSF}) will be generally larger for weaker interfering packets. Since the measured pulse-position offset is determined by the largest among $M$~samples (see~\eqref{eq:pl sampling}), a stronger pulse of interest is more likely to ``win" against a weaker interfering pulse even when their sampling intervals collide. For example, if the interfering Rx pulse~I in Fig.~\ref{fig:payload collision impact} were the pulse of interest, its position offset would be determined correctly even if the Rx signal is sampled over the full shown time interval. (Here one can draw a parallel with the ``capture effect" in LoRa modulation~\cite{Heusse2023performance}.)

From now on, to distinguish between the respective rates (probabilities) with and without collisions, we mark those with collisions by overhead tildes.

\subsection{MULTIPLE PAYLOAD CHANNELS} \label{subsec:multiple}
By design, the PSFs used in the detection channel are quasi-orthogonal to those used in the payloads. Therefore, the impact of the detection channel on the rate of error-free detected packets under collisions would be mainly due to its contribution to the payloads' SINR. To quantify this contribution, we can use the {\it complementary energy overhead of the detection channel\/}~$\kappa$, defined as the (average) ratio of the energy of the payload channels to the total energy of the packets. For example, $\kappa\approx 0.975$ for the 440-pulse packet shown in Fig.~\ref{fig:Tx packet}, and it would decrease to $\kappa\approx 0.8$ for $N_\mathrm{PL}=55$.

With this, for multiple payload channels, $\widetilde{P}_\mathrm{efdp}\left(\Gamma^\prime\right)$ can be estimated as
\beginlabel{equation}{eq:multiple PSFs}
  \widetilde{P}_\mathrm{efdp}\left( \Gamma^\prime \right) \gtrsim  P_\mathrm{efdp}\left( \bar{\Gamma}^\prime_{\!1\!-\!\varepsilon\kappa} \right)\, \exp\left( -\varepsilon\lambda_\mathrm{pl}\right),
\end{equation}
where $\Gamma^\prime$ is the peak SNR and $\bar{\Gamma}^\prime_{\!1\!-\!\varepsilon\kappa}$ is the ``effective" payload SINR that can be expressed as
\beginlabel{equation}{eq:SINR varepsilon}
  \bar{\Gamma}^\prime_{\!1\!-\!\varepsilon\kappa} =  \left( \frac{1}{\Gamma^\prime} +  \frac{1-\varepsilon\kappa}{\gamma^\prime} \right)^{-1},
\end{equation}
where $\gamma^\prime$ is the total peak signal-to-interference ratio (SIR) and $1/N_\mathrm{ch}\le \varepsilon\le 1$ is the {\em cross-channel collision factor\/} (CcCF).

The CcCF quantifies the overall orthogonality among the payload channels, and for a single channel $\varepsilon=1$. Then the decrease in the effective SINR is due only to the detection channel. For $N_\mathrm{ch}$~strongly orthogonal payload channels $\varepsilon=1/N_\mathrm{ch}$, reducing the effective collision rate by~$N_\mathrm{ch}$ times, and the remaining~$N_\mathrm{ch}-1$~channels contribute to the decrease in the effective SINR.

\subsection{COLLISIONS IN DETECTION CHANNEL} \label{subsec:leading}
In this paper we adopt the algorithm for detection of M-ASPM packets described in~{\cite{Nikitin2025detection}}. Due to the highly nonlinear nature of the signal processing in this algorithm, a thorough analysis of the impact of collisions on the detection channel is a rather complicated task that should be addressed elsewhere. In the context of the work presented in the current paper, our main interest is a rough assessment of this impact in relation to the payload collisions and, specifically, for interfering packets of high power.

In~{\cite{Nikitin2025detection}}, for asynchronous noncoherent detection the in-phase and quadrature digital signals $I[k]$ and $Q[k]$ in the receiver are filtered with a matched filter~$\zeta^\prime[k]$ of the leading sequence's PSF~$\hat{\zeta}^\prime[k]$ to obtain~$y_\mathrm{nc}[k]$:
\beginlabel{equation}{eq:Rx raw}
  y_\mathrm{nc} = \left(I+\mathrm{i}\,Q\right)\ast \zeta^\prime\,.
\end{equation}
For the Rx signal~$y^2_\mathrm{nc}[k]$ the respective MPA output~$\bar{p}[k]$ is obtained with the modulus~$N^\prime_\mathrm{p}=L^\prime$ and the smoothing factor~$K$ as
\beginlabel{equation}{eq:MPA in}
  \bar{p}[k] = \frac{K\!-\!1}{K} \bar{p}[k-N^\prime_\mathrm{p}] + \frac{1}{K} y^2_\mathrm{nc}[k].
\end{equation}
This MPA output is used as the input to QTFs for troughs to obtain the detection fence (threshold)~$\check{\alpha}[k]$ and, together with~$\check{\alpha}[k]$, as an input for the pulse counting function~$\CalphaP_3[k]$. Further, the MEA~$z[k]$ of the pulse counting function is computed as
\begin{equation} \label{eq:MEA pc}
  z[k] = \frac{K_\mathrm{pc}\!-\!1}{K_\mathrm{pc}} z[k-N^\prime_\mathrm{p}] + \frac{1}{K_\mathrm{pc}} \CalphaP_3[k],
\end{equation}
and the detection occurs when~$z[k]$ exceeds a given {\it rate threshold\/}. In particular, in~{\cite{Nikitin2025detection}} we use the rate threshold value~$2/3$ and the smoothing factor~$K_\mathrm{pc}$ for the MEA of the pulse counting function is obtained as $K_\mathrm{pc}=(N_\mathrm{LS}-2K)/\ln{3}$. Then, by measuring $y_\mathrm{nc}[k]$ at $N_\mathrm{CFO}$~consecutive indices corresponding to the peaks in~$y^2_\mathrm{nc}[k]$ preceding the detection, we determine~$\Delta{f}_\mathrm{c}$ in the $\pm F_\mathrm{s}/(2N^\prime_\mathrm{p})$ range with $F_\mathrm{s}/(2N^\prime_\mathrm{p}N_\mathrm{CFO})$ accuracy.

\subsubsection{Modification of QTF fencing in detection algorithm to improve its robustness under heavy collisions} \label{subsubsec:QTF}
However, collisions with strong interfering leading sequences add a significant DC bias to $\bar{p}[k]$~given by~(\ref{eq:MPA in}). Therefore, to improve robustness of the detection algorithm under heavy collisions, in the current implementation $\bar{p}[k]$~is used only for computing the QTF's slew rate parameter~$\mu[k]$. For obtaining~$\check{\alpha}[k]$ and~$\CalphaP_3[k]$, we subtract from~$\bar{p}[k]$ the ``MPA background"~$\bar{p}_\mathrm{bg}[k]$ expressed as
\beginlabel{equation}{eq:MPA bg}
  \bar{p}_\mathrm{bg}[k] = \frac{K\!-\!1}{K} \bar{p}_\mathrm{bg}[k-N^\prime_\mathrm{p}\pm 1] + \frac{1}{K} y^2_\mathrm{nc}[k],
\end{equation}
(i.e., the MPA computed with the modulus offset by $\pm 1$), and reduce the scaling parameter in the QTF fencing to~$\beta=3$.

The minimal $\pm 1$~modulus offset ensures approximate equality of the timescales in the MPA filtering given by~(\ref{eq:MPA in}) and~(\ref{eq:MPA bg}). This is required for accurate background calculations under highly non-stationary conditions of heavy collisions. At the same time, this offset sufficiently reduces the magnitude of peaks in~$\bar{p}_\mathrm{bg}[k]$ that correspond to the leading pulse sequences. A more detailed discussion of the MPA properties, and their relation to the detection algorithm, can be found in~\cite{Nikitin2025detection}.

\begin{figure}[!t]
\centering{\includegraphics[width=8.6cm]{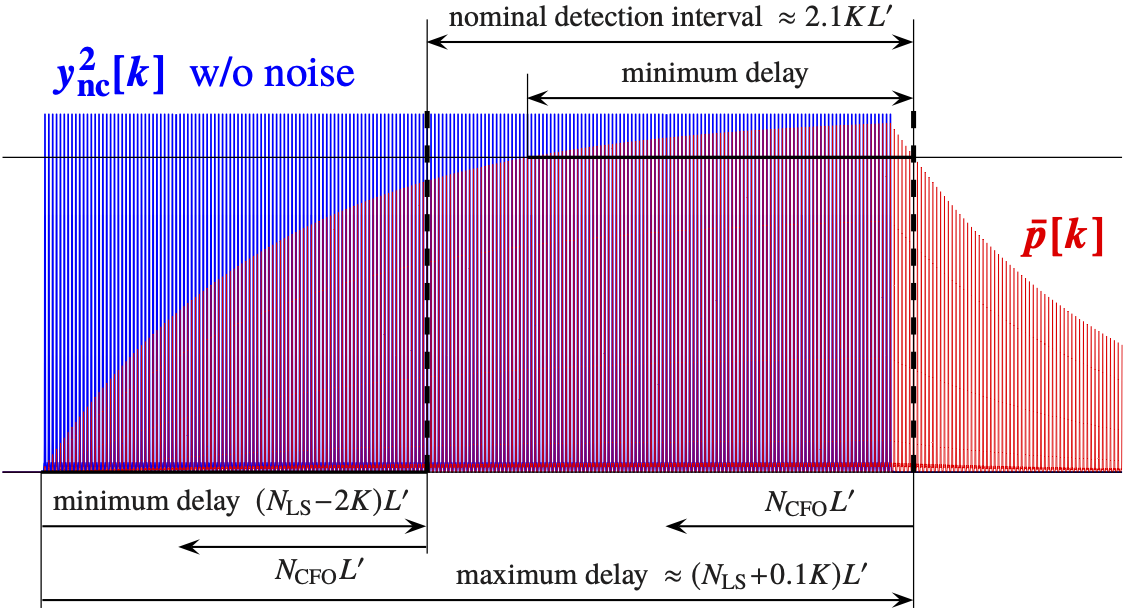}}
\caption{\boldmath Detection delays and nominal detection interval for leading sequence.
\label{fig:collision interval}}
\end{figure}

\begin{figure}[!b]
\centering{\includegraphics[width=8.6cm]{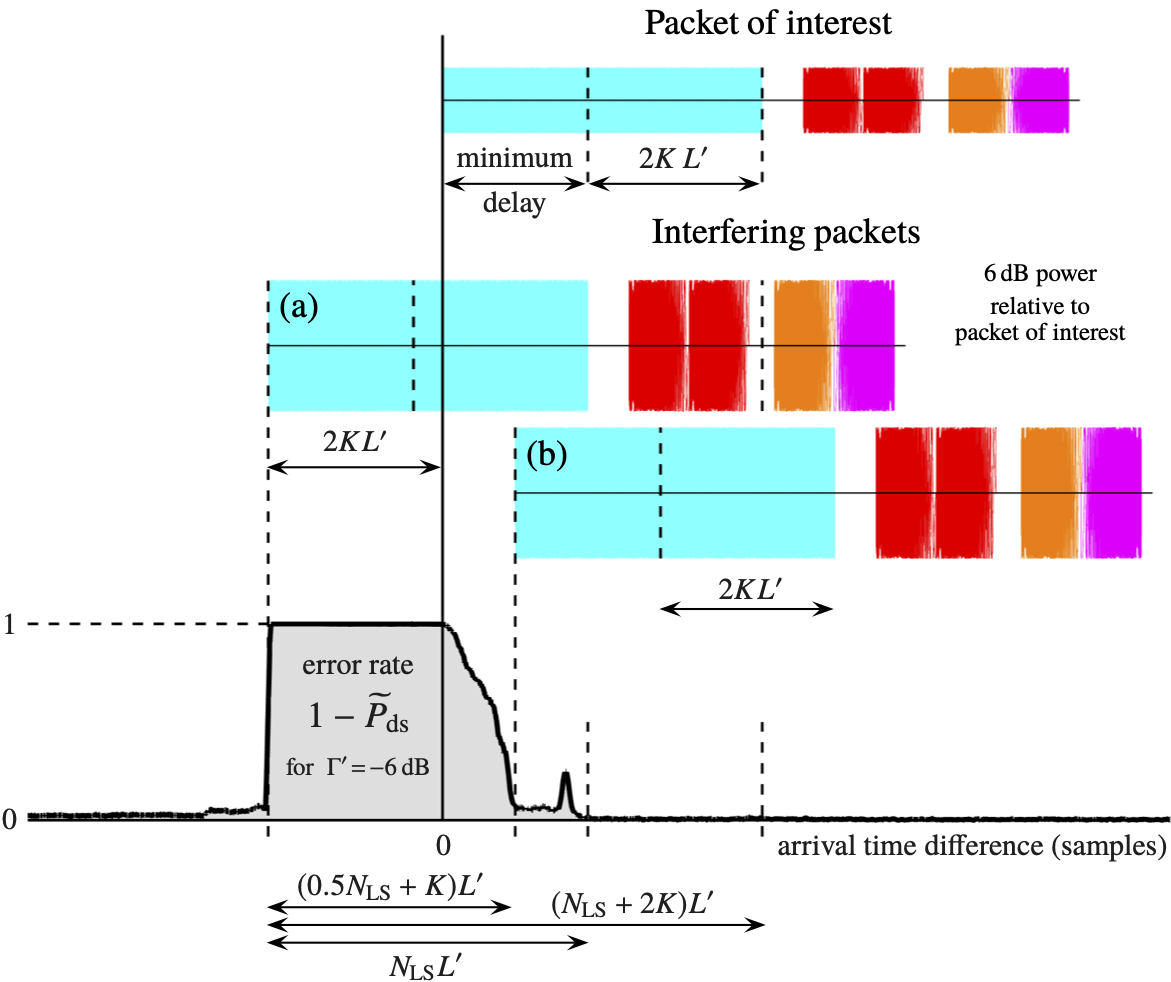}}
\caption{\boldmath Impact of trailing interfering packets (represented by packet~(b)) rapidly decays with increased arrival delay even when these packets are significantly stronger than packet of interest.
\label{fig:front collisions}}
\end{figure}

\subsubsection{Effective rate of collisions in detection channel} \label{subsubsec:detection lambda}
We would like to note that, since the MPA~$\bar{p}[k]$ consists of narrow pulses separated by~$N^\prime_\mathrm{p}$, most of the single and some of the multiple collisions among leading sequences can be resolved. In the subsequent simulations, however, to simplify distinction of the packets of interest from the interfering packets we detect only the first packet among those arriving within a designated time interval.

This time interval is indicated as {\em nominal detection interval\/} in~Fig.~\ref{fig:collision interval}. Note that the minimum delay in the detection time from the beginning of the leading sequence is $(N_\mathrm{LS}\!-\!2K)L^\prime$ samples, and it occurs for zero threshold value. The presence of noise raises the QTF threshold and increases the delay, and this increase is larger for weaker signals. For simplicity, let us assume that the additional delay is only due to the elevated QTF threshold, and not the missing peaks in the pulse counting function. Then, for the parameters of the detection algorithm used in the simulations, the maximum additional delay for which the detection can still occur is about $2.1KL^\prime$~samples. Thus, the designated time interval for detection of the packet of interest is chosen from the minimum through maximum delay.

\begin{figure}[!b]
\centering{\includegraphics[width=8.6cm]{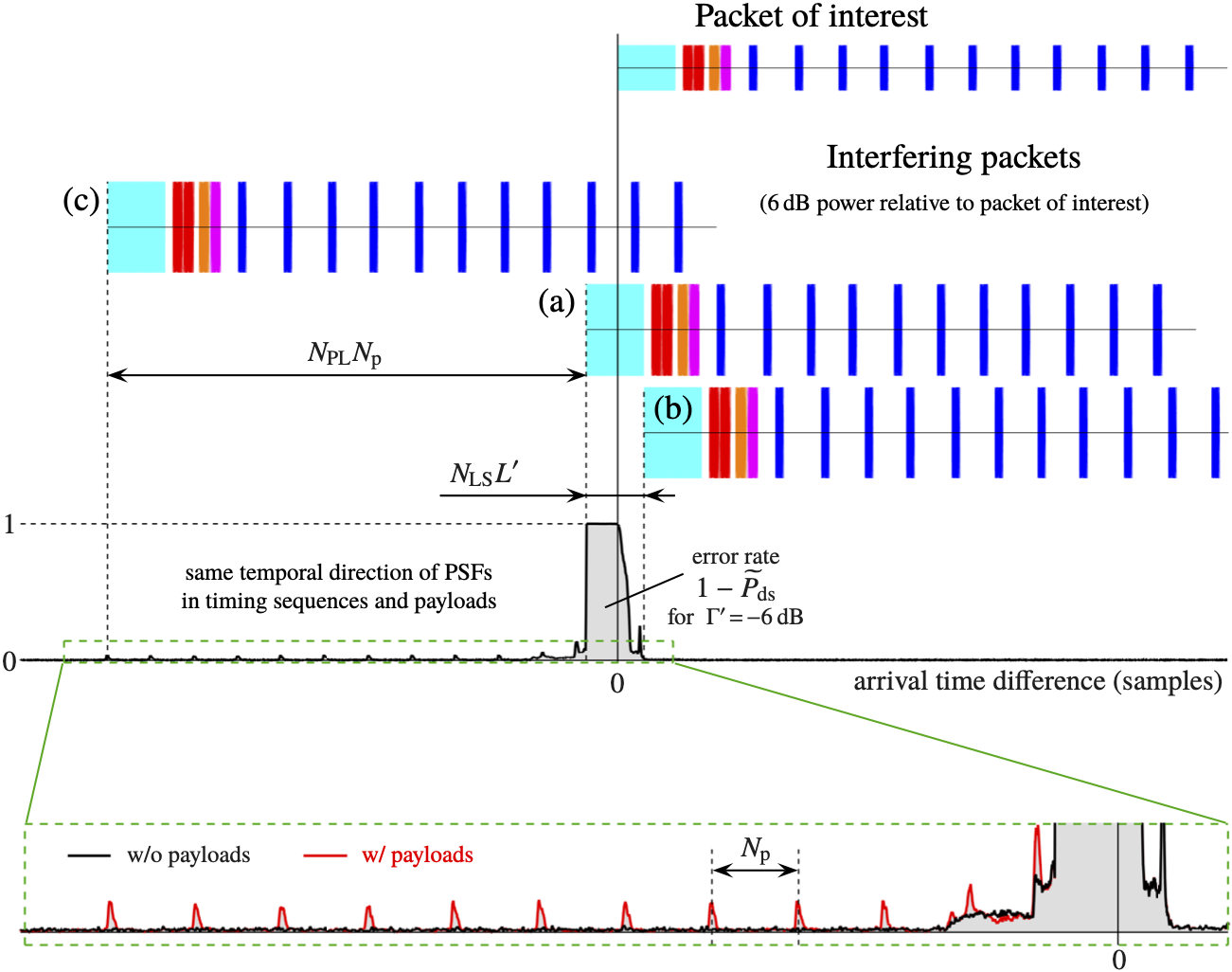}}
\caption{\boldmath Impact of collisions with payloads on detection probability. Imperfect orthogonality between payload pulses and those in timing sequence increases error rate of synchronization (and/or correct channel selection) when these pulses collide. 
\label{fig:front collisions pl}}
\end{figure}

With this, a rough upper-bound estimate for the average number~$ \lambda^\prime$ of impactful collisions among leading pulse sequences can be made as the rate of collisions of the leading pulse sequence with the nominal detection interval, e.g., as $\lambda^\prime = (N_\mathrm{LS}+2K)\, L^\prime\, \mathcal{R}/F_\mathrm{s}$. However, since we detect the first to arrive packet within the nominal detection interval, the impact of trailing interfering packets rapidly decays with increased arrival delay even when these packets are significantly stronger than the packet of interest.

This is illustrated in Fig.~\ref{fig:front collisions}. Here, for a constant peak AWGN SNR~$\Gamma^\prime=-6\,$dB, we obtain the detection error rate as a function of the time difference between arrivals of the packet of interest and a single strong (6\,dB) interfering packet without payload. We use the simulation settings and the parameter values of the detection channel as described in Section~\ref{sec:simulation settings}. For each value of the time difference, we conduct 170~trials with different combinations of position offsets of the third pulse in the timing sequence, and differently generated noise and CFO values. In~Fig.~\ref{fig:front collisions}, thus obtained error rate is plotted as the central moving average for $3L^\prime$~time samples. As one can see, for a lagging interfering packet (packet~(b)) the detection error probability rapidly decays with the increased lag. Therefore, for the base collision rate estimate in the detection channel~$\lambda^\prime$ we will use
\beginlabel{equation}{eq:LS colllisions}
  \lambda^\prime = N_\mathrm{LS}\, L^\prime\, \frac{\mathcal{R}}{F_\mathrm{s}} \approx  \frac{320}{MN_\mathrm{PL}}\,\lambda_\mathrm{pl}\,.
\end{equation}

\subsubsection{Impact on detection channel from interfering payloads} \label{subsubsec:LS PL}
Since the PSFs used in the detection channel are quasi-orthogonal to those used in the payloads, the main impact on the detection channel from the payloads is expected to be due to their contribution to the overall SINR. However, collisions with payloads may also raise the effective collision rate in the detection channel. In particular, imperfect orthogonality between the payload pulses and those in the timing sequence increases error rate of synchronization (and/or correct channel selection) when these pulses collide. 

This is illustrated in~Fig.~\ref{fig:front collisions pl}, which repeats the simulations presented in Fig.~\ref{fig:front collisions} for 16-ASPM packets with short 11-pulse payloads that use PSFs with the same temporal direction as the PSFs in the timing sequence. One can see in the lower panel of the figure that collisions with payloads produce $N_\mathrm{PL}+1$ additional small peaks in the detection error rate for interfering packets that precede the packet of interest by as much as ${N_\mathrm{PL}N_\mathrm{p}+2KL^\prime}$~samples. The net result of such collisions can be treated as an increase in the effective collision rate for the detection channel.

Therefore, the overall rate of the packet detection with correct channels selection under packet collisions can be assessed as
\beginlabel{equation}{eq:ds}
  \widetilde{P}_\mathrm{ds}\left( \Gamma^\prime \right) \gtrsim P_\mathrm{ds}\left( \bar{\Gamma}^\prime_{b} \right)\,
  \exp\left(-a\lambda^\prime \right),
\end{equation}
where $a$~is of order unity and $\bar{\Gamma}^\prime_{b}$ is the peak effective SINR of the detection channel expressed as
\beginlabel{equation}{eq:SINR b}
  \bar{\Gamma}^\prime_{b} =  \left( \frac{1}{\Gamma^\prime} +  \frac{b}{\gamma^\prime} \right)^{-1},
\end{equation}
where $\gamma^\prime$ is the total peak SIR and $0\le b\le 1$.

\subsection{OVERALL COLLISION IMPACT} \label{subsec:overall}
Consequently, the lower bound for the total error-free throughput~$\widetilde{P}_\mathrm{teft}\left(\Gamma^\prime\right)$ under collisions can be approximated as the product of such bounds for the rate of packet detection with correct channels selection~$\widetilde{P}_\mathrm{ds}\left(\Gamma^\prime\right)$ and the rate of error-free detected packets $\widetilde{P}_\mathrm{efdp}\left(\Gamma^\prime\right)$:
\beginlabel{equation}{eq:overall}
  \widetilde{P}_\mathrm{teft}\left( \Gamma^\prime \right) \gtrsim P_\mathrm{ds}\left( \bar{\Gamma}^\prime_{b} \right)\,
  P_\mathrm{efdp}\left( \bar{\Gamma}^\prime_{\!1\!-\!\varepsilon\kappa} \right)\,
  \exp\left(-a\lambda^\prime-\varepsilon\lambda_\mathrm{pl}\right),
\end{equation}
where $\Gamma^\prime$ is the peak SNR, $\kappa$~is the complementary energy overhead of the detection channel, $\varepsilon$ is the CcCF, and $\bar{\Gamma}^\prime_{b}$ and $ \bar{\Gamma}^\prime_{\!1\!-\!\varepsilon\kappa}$ are the effective SINRs of the detection and payload channels, respectively.

From~(\ref{eq:LS colllisions}) the relation between~$a\lambda^\prime$ and~$\varepsilon\lambda_\mathrm{pl}$ can be expressed as
\beginlabel{equation}{eq:lambdas ratio}
  \frac{a\lambda^\prime}{\varepsilon\lambda_\mathrm{pl}}  \approx  \frac{2\log_2 M}{M}\, \frac{a}{\varepsilon}\, \frac{20}{N_\mathrm{bytes}}\,,
\end{equation}
where~$N_\mathrm{bytes}$ is the payload size in (uncoded) bytes.
This relation is explored in more detail in the ensuing sections, when presenting the simulation results.

\section{COMMON SIMULATION SETTINGS} \label{sec:simulation settings}
In each of the simulations that follow, the packet of interest and the interfering packets have the same payload sizes (and thus the same ToA~$T_\mathrm{packet}$) and $M$~values, and the values $m_j\in \{0,1,2,\dots,M\!-\!1\}$ for the position offsets of each payload pulse are generated with uniform probability.

For analytical tractability, we assume the pure ALOHA packet time scheduling~\cite{Abramson70ALOHA}, i.e., Poisson traffic. For each trial, the number of interfering packets that arrive at rate~$\mathcal{R}$ within the time interval~$\Delta{T}$, which includes $\pm T_\mathrm{packet}$ range relative to the packet of interest, is drawn from the Poisson distribution with the rate (expectation) value~$\mathcal{R}\Delta{T}$. Then the arrival times of the interfering packets are randomly chosen with uniform probability within~$\Delta{T}$.

For each packet, the CFO values are generated with uniform probability in the $\pm 30\,$ppm range. (Except when considering, as in Fig.~\ref{fig:CFO impact}, the impact of the CFO range on collisions, where this range varies.) The initial STO values are randomly chosen in the $\pm 0.5$~interval.

At each trial, we attempt to detect and synchronize the packet of interest and decode its payload for different AWGN SNR values that vary in $0.5\,$dB increments in a range that spans, depending on a particular simulation, 13~to~21\,dB. Unless specified otherwise, for obtaining the average success rates, each trial is repeated 4,000~times for non-zero interference rates and 1,000~times in the collision-free cases.

\subsection{POWER OF INTERFERING PACKETS} \label{subsec:remote nodes}
In general, as was discussed in Section~\ref{sec:collisions}, for a given transmit rate the impact of collisions increases with power of interfering packets. To assess such increase, in our simulations we vary the power of interfering packets from 0~to~6\,dB relative to the packet of interest.

\begin{figure}[!b]
\centering{\includegraphics[width=8.6cm]{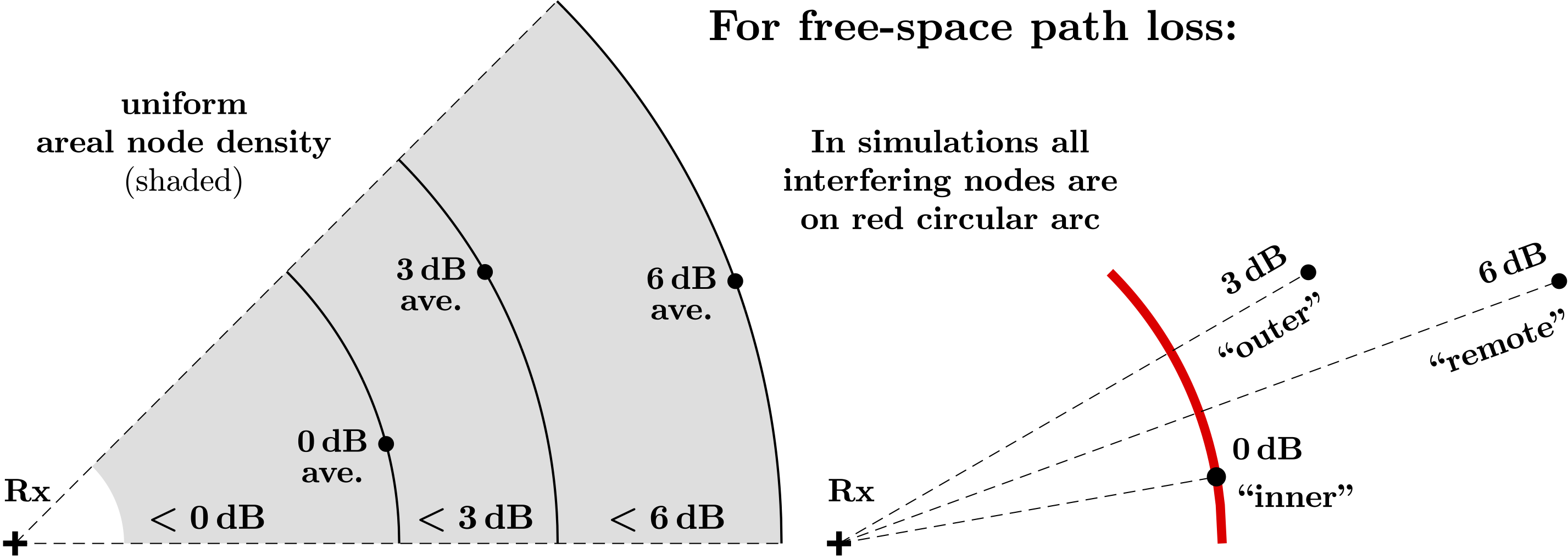}}
\caption{\boldmath Designation of inner, outer, and remote nodes in simulations.
\label{fig:density}}
\end{figure}

If all uplink nodes transmit at the same power, then the interference power is higher for nodes with larger path attenuation (e.g., more remote from the gateway). This is illustrated in the left-hand side of Fig.~\ref{fig:density} for equal-power nodes distributed with uniform density over the shaded area. For simplicity, in the simulations we maintain the same power of interfering packets relative to the packet of interest, and designate the nodes under 0\,dB, 3\,dB, and 6\,dB interference levels as ``inner," ``outer," and ``remote," respectively. This corresponds to the node placement illustrated in the right-hand side of Fig.~\ref{fig:density}.

Note that such designation implies equal transmit power of all nodes. If, for example, this power is proportional to the path attenuation of the respective nodes, then all received packets would have equal powers and should be considered ``inner."

\subsection{DETECTION CHANNEL} \label{subsec:detection channel}
For consistency and ease of comparability, we use the same leading sequence in all simulations. As illustrated earlier in Fig.~\ref{fig:Tx packet}, the leading sequence uses the PSF of length $L^\prime=29$ and $100$\% duty cycle. Then the range of the CFO measurements is $\pm F_\mathrm{s}/(2L^\prime)$, or $\pm 30\,$ppm. The total length of the leading sequence is $N_\mathrm{CFO}=220$~pulses ($4\,$ms), and the parameters of the detection algorithm are chosen according to~{\cite{Nikitin2025detection}}. Specifically, the MPA smoothing factor is~$K = 60$, the rate threshold value is~$2/3$, and the smoothing factor~$K_\mathrm{pc}$ for the MEA of the pulse counting function is $K_\mathrm{pc}=(N_\mathrm{LS}-2K)/\ln{3}=91$. Further, the QTF fencing is modified as described in~Section~\ref{subsubsec:QTF}.

The number of pulses available for obtaining the CFO values is limited by the minimum delay, $N_\mathrm{CFO}\le N_\mathrm{LS}-2K=100$. For the simulations we choose $N_\mathrm{CFO}=64$, which provides better than $0.5\,$ppm accuracy of the CFO measurements. This accuracy is sufficient for CFO correction of the largest PSFs used in the simulations (with $L_i=1824$), and for STO compensation in the longest payloads (e.g., $1.32\,$s or 440~pulses with $N_\mathrm{p}=4800$).

As further shown in Fig.~\ref{fig:Tx packet}, the timing sequence uses PSFs of length $L=1200$. With this, for the same magnitude of the pulses in the leading and timing sequences, the detection probability and the joint probability of detection with synchronization and correct channel identification are effectively equivalent.

For the position offset increment of the third pulse in the timing sequence we use~$n^\prime_\mathrm{off}=11$.

\subsection{PAYLOAD CHANNELS} \label{subsec:payload channels}
In payload channels we maintain the same peak, rather than average, power. Therefore, to keep the spread of average payload powers within a~1\,dB range, the differences in payload pulse duty cycles are also confined to this range. (Note that maintaining the average payload power allows us to significantly extend the range of pulse duty cycles, thus reducing the CcCF. This would require, however, different peak Tx powers for different payload channels.)

In most simulations we use payload PSFs of lengths $L_i\in \{720,768,816,864,912\}$, but in the simulations presented in Figs.~\ref{fig:ten channels 2D} and~\ref{fig:rates55 2D} these lengths are doubled. We refer to the respective payload channels as 1~through~5 if the temporal direction of the payload PSF coincides with that of the synchronization PSFs, and as 6~through~10 for the opposite temporal direction (``flip" PSFs).

For ease of mental calculations in comparing the results of different simulations, for payloads we use the values of $N_\mathrm{PL}$ and $N_\mathrm{p}$ that are multiples of 55~and~4,800, respectively, and vary in the $9\,$dB range. (Note that a 55-pulse payload corresponds to 27.5~and 41.25~uncoded bytes for $M=16$ and $M=64$, respectively.)

\subsection{PACKET RATES} \label{subsec:packet rates}
We vary the rates of packets arriving at the Rx in $13.8\,$dB range as multiples of $\mathcal{R}_0=2\times 10^5$~packets/day, from~$\mathcal{R}_0$ to~$24\mathcal{R}_0$ ($4.8\times 10^6$~packets/day). Then, in different simulations with different $N_\mathrm{PL}$~and~$N_\mathrm{p}$ values, the respective ToA average numbers of collisions per packet vary from~$\lambda=6.15$ to~$\lambda=49.2$.

\begin{figure*}[t!]
\centering{\includegraphics[width=17.2cm]{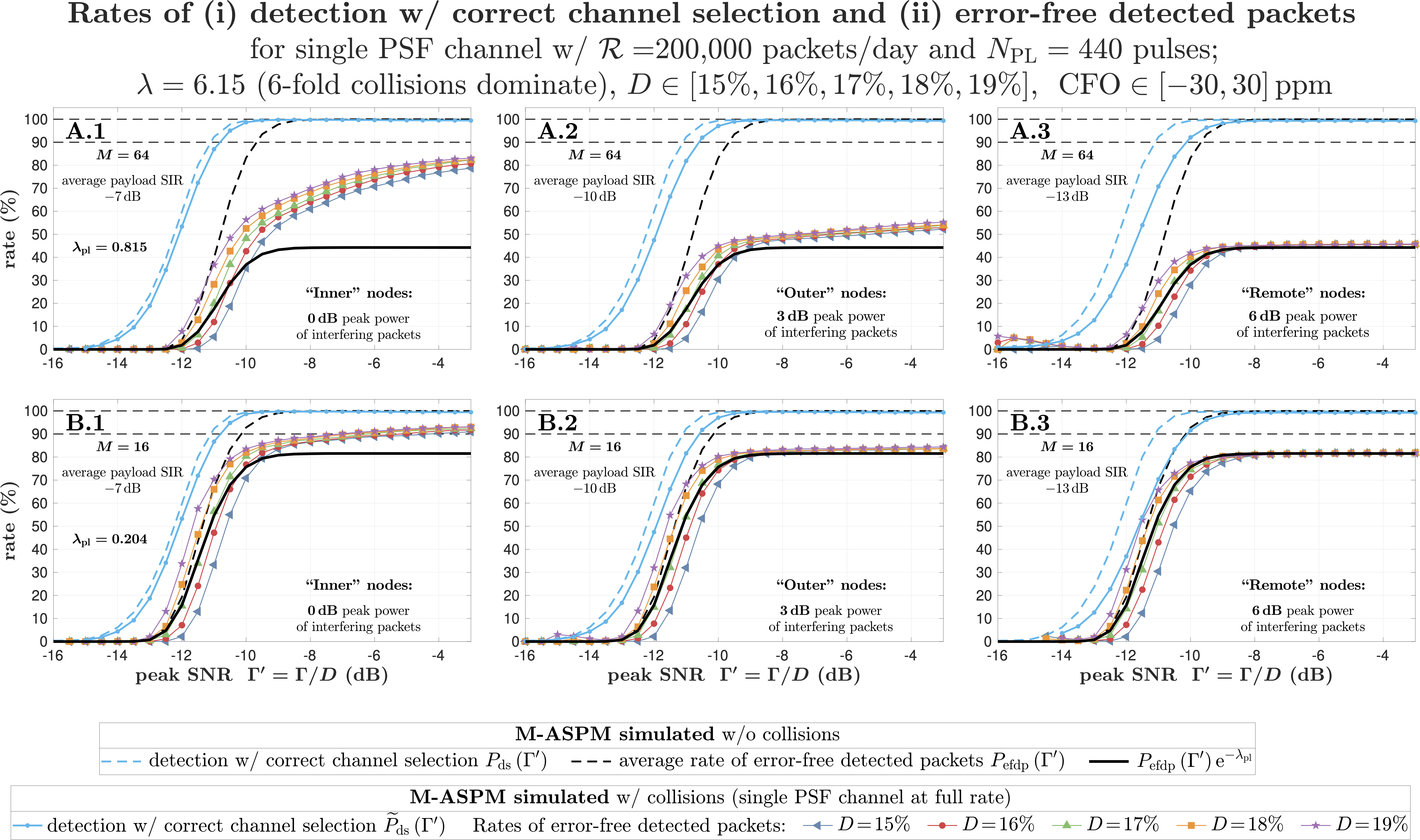}}
\caption{\boldmath Impact of co-PSF (single payload channel) collisions for 16- and 64-ASPM packets with long ($N_\mathrm{PL}=440$) payloads.
\label{fig:self-collisions}}
\end{figure*}

\begin{figure}[!b]
\centering{\includegraphics[width=8.6cm]{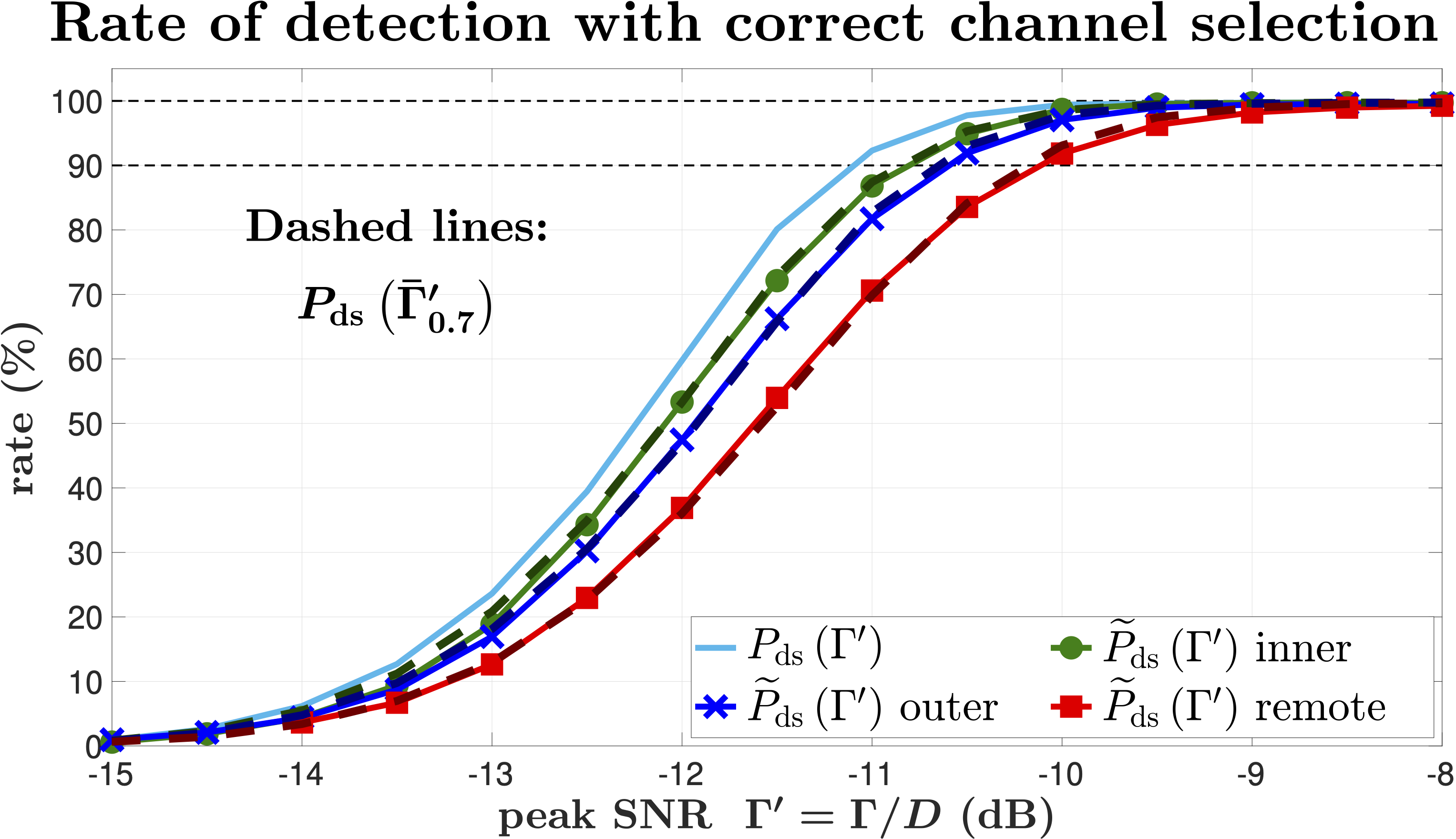}}
\caption{\boldmath Impact of collisions on detection channel for simulations presented in Fig.~\ref{fig:self-collisions}.
\label{fig:ds collisions}}
\end{figure}

\begin{figure}[!b]
\centering{\includegraphics[width=8.6cm]{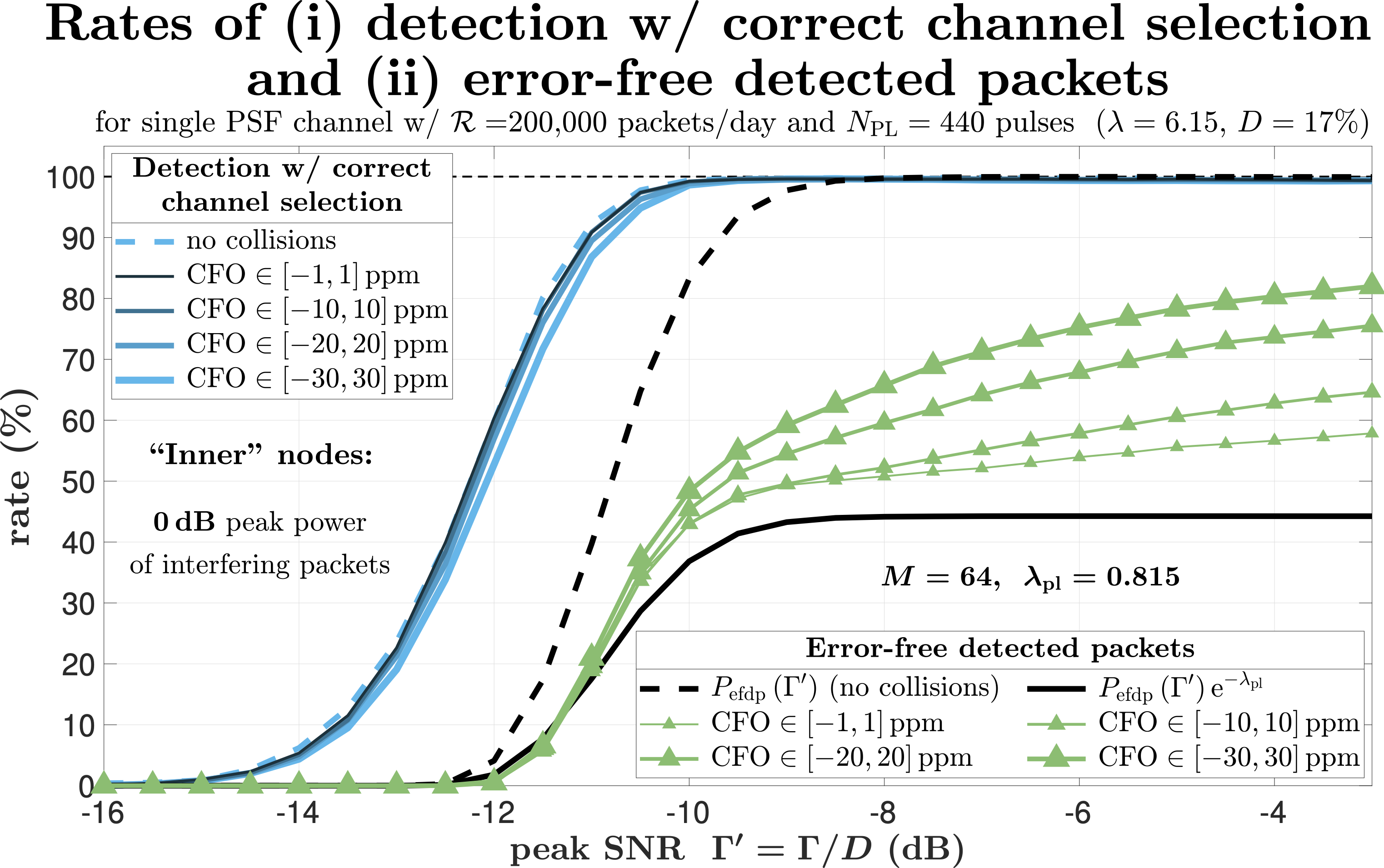}}
\caption{\boldmath Impact of partial orthogonality due to CFO on error-free packet rate under co-PSF collisions. (Compare with panel~A.1 in Fig.~\ref{fig:self-collisions}.)
\label{fig:CFO impact}}
\end{figure}

\begin{figure*}[t!]
\centering{\includegraphics[width=17.2cm]{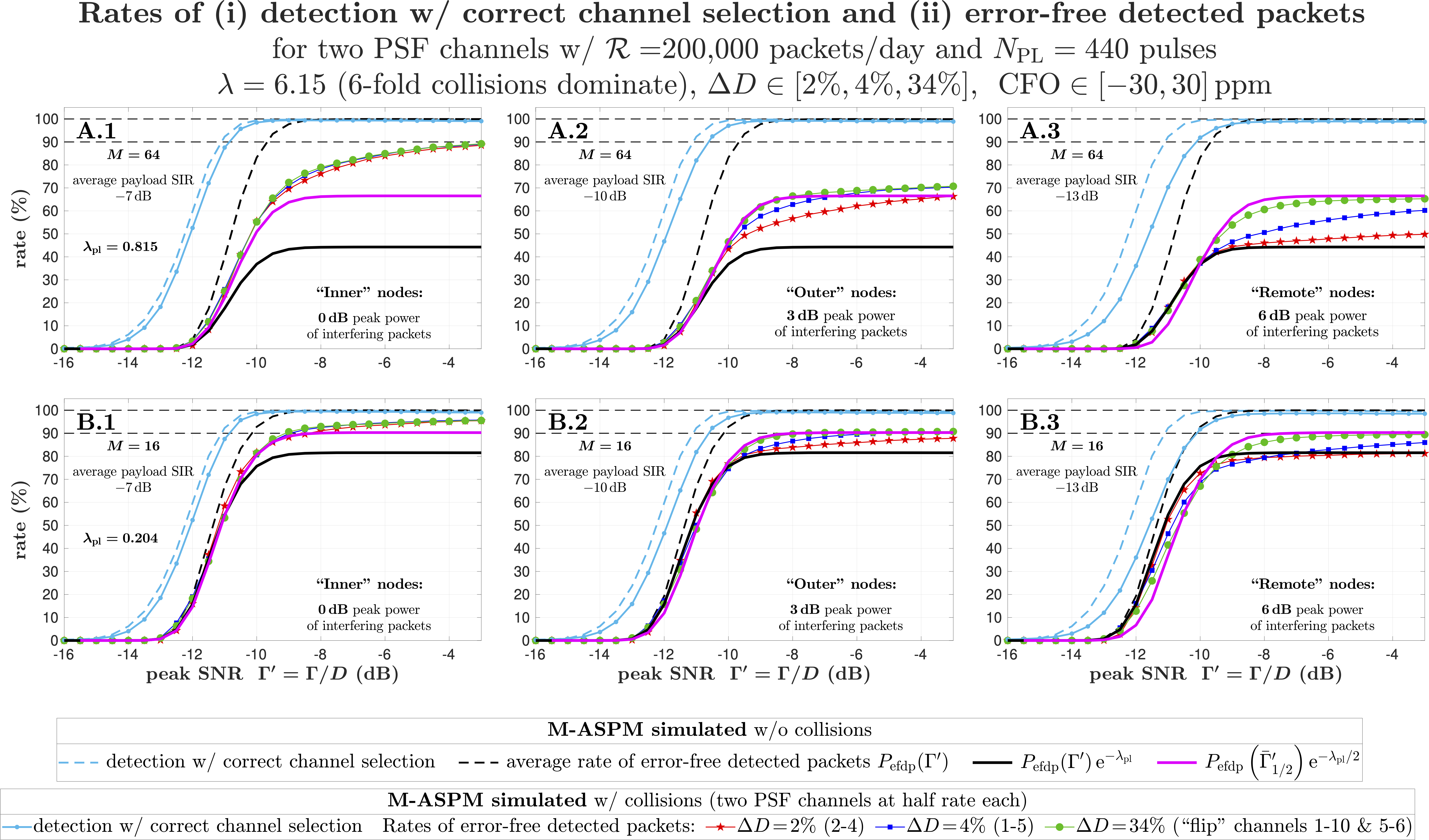}}
\caption{\boldmath Impact of partial orthogonality between PSFs in two-channel configurations. For ``flip" channels CcCF decreases to about $1/2$.
\label{fig:two channels}}
\end{figure*}

\begin{figure*}[b!]
\centering{\includegraphics[width=17.2cm]{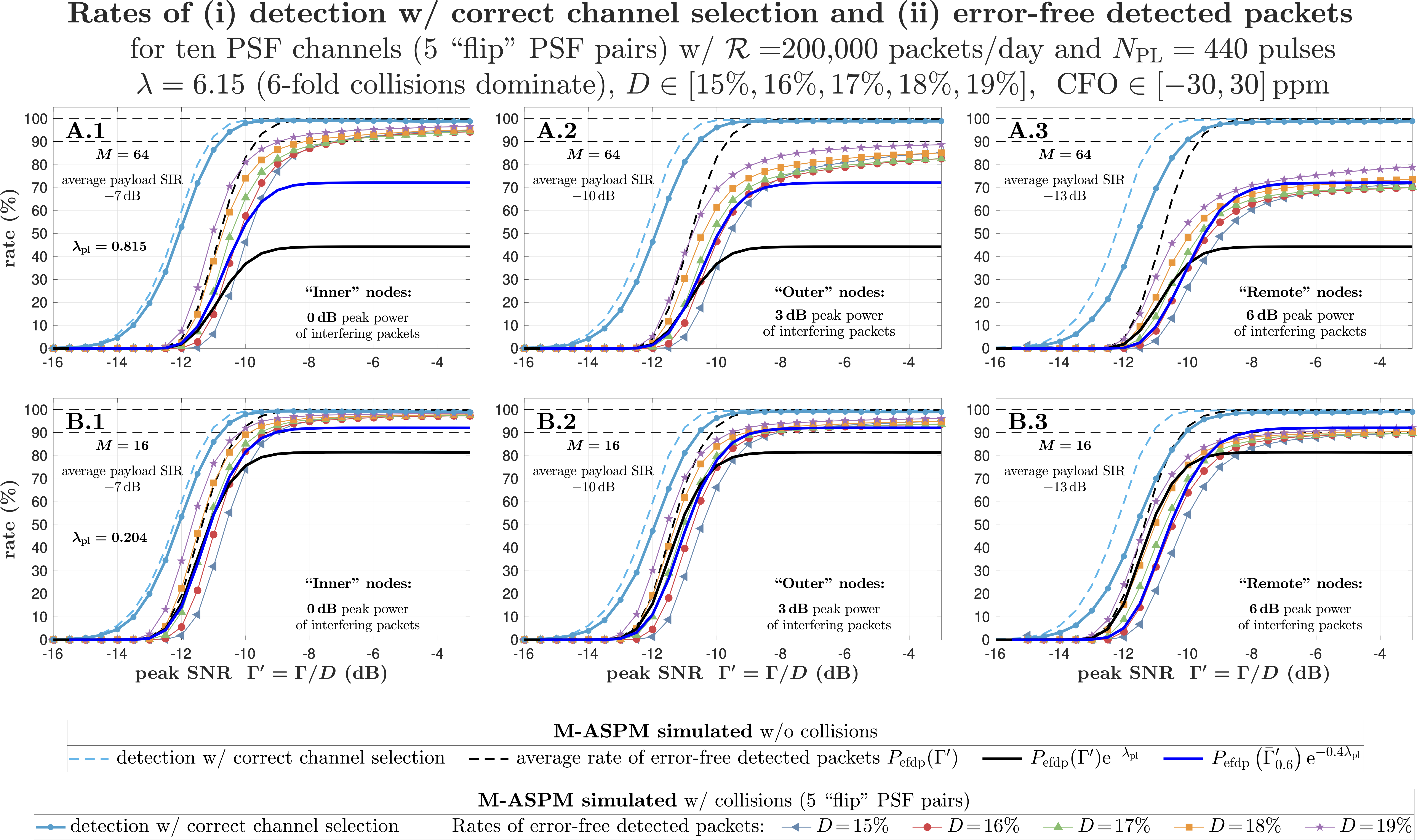}}
\caption{\boldmath Even with small spread in pulse duty cycles, ten-channel configuration further reduces CcCF and offers additional reduction in collision impact. 
\label{fig:ten channels}}
\end{figure*}

\section{BENCHMARK SIMULATION RESULTS} \label{sec:simulation results}

\subsection{SINGLE PAYLOAD CHANNEL} \label{subsec:co-PSF}
The simulations presented in Fig.~\ref{fig:self-collisions} are for co-PSF collisions among 16- and 64-ASPM packets with long ($N_\mathrm{PL}=440$) payloads. For the IpI $N_\mathrm{p}=4800$ and the packet transmission rate $\mathcal{R}=\mathcal{R}_0=2\times 10^5\,$packets/day the average ToA collision rate for such packets is $\lambda=6.15$. The effective collision rates $\lambda_\mathrm{pl}$ are calculated according to~(\ref{eq:pllambda}) as $\lambda_\mathrm{pl}=0.815$ and $\lambda_\mathrm{pl}=0.204$ for $M=64$ and $M=16$, respectively.

In all panels, the dashed cyan lines plot the probability (rate) of packet detection with correct channel selection without collisions, and the solid cyan lines plot this probability under collisions.

The dashed black lines show the average (among all ten payload channels) rate of error-free detected packets without collisions. As can be seen in the figure, the 95\% packet detection probability precedes the respective average probability of a detected packet to be error-free by about 1--1.5\,dB. Since the difference in payload pulse duty cycles is confined to a~1\,dB range, without collisions the impact of the detection probability on the total error-free throughput is insignificant.

For obtaining the average rates  under collisions, for each payload channel each trial is repeated 2,000 times. For each pulse duty cycle, the rates of error-free detected packets under collisions are plotted as averages of the respective ``flip" channels. For comparison, the solid black lines show the approximate lower bound for average rates of error-free detected packets computed according to~(\ref{eq:co-PSF}).

\subsubsection{Impact of packet collisions on detection channel} \label{subsubsec:ds collisions}
As follows from~(\ref{eq:LS colllisions}), for the long $N_\mathrm{PL}=440$ payloads used in Fig.~\ref{fig:self-collisions} the effective collision rate in the detection channel is negligibly small in comparison with that among payloads (i.e., $\lambda^\prime\ll \lambda_\mathrm{pl}$) for both $M=16$ and $M=64$. Consequently, as shown in Fig.~\ref{fig:ds collisions}, the impact of increased interference power on the detection probability is mainly due to decrease in the effective SINR and can be approximated with $b=0.7$ in~(\ref{eq:ds}) as
\beginlabel{equation}{eq:Pds self}
  \widetilde{P}_\mathrm{ds}\left(\Gamma^\prime\right) \approx P_\mathrm{ds}\left(\bar{\Gamma}^\prime_{0.7}\right).
\end{equation}

\subsubsection{Impact of CFO range} \label{subsubsec:CFO}
Differences in the CFO between the interfering packets and the signal of interest broaden the Rx pulses and reduce their magnitude in general, as signified by the Rx pulse~II in Fig.~\ref{fig:payload collision impact}. In other words, non-zero CFO introduces some degree of orthogonality between the packet of interest and the interfering packets. As can be seen in Fig.~\ref{fig:self-collisions}, the impact of such (rather small) partial orthogonality is significant only for the inner nodes. Further, as illustrated in Fig.~\ref{fig:CFO impact}, this impact becomes smaller with decrease in the CFO range.

\begin{figure*}[t!]
\centering{\includegraphics[width=17.2cm]{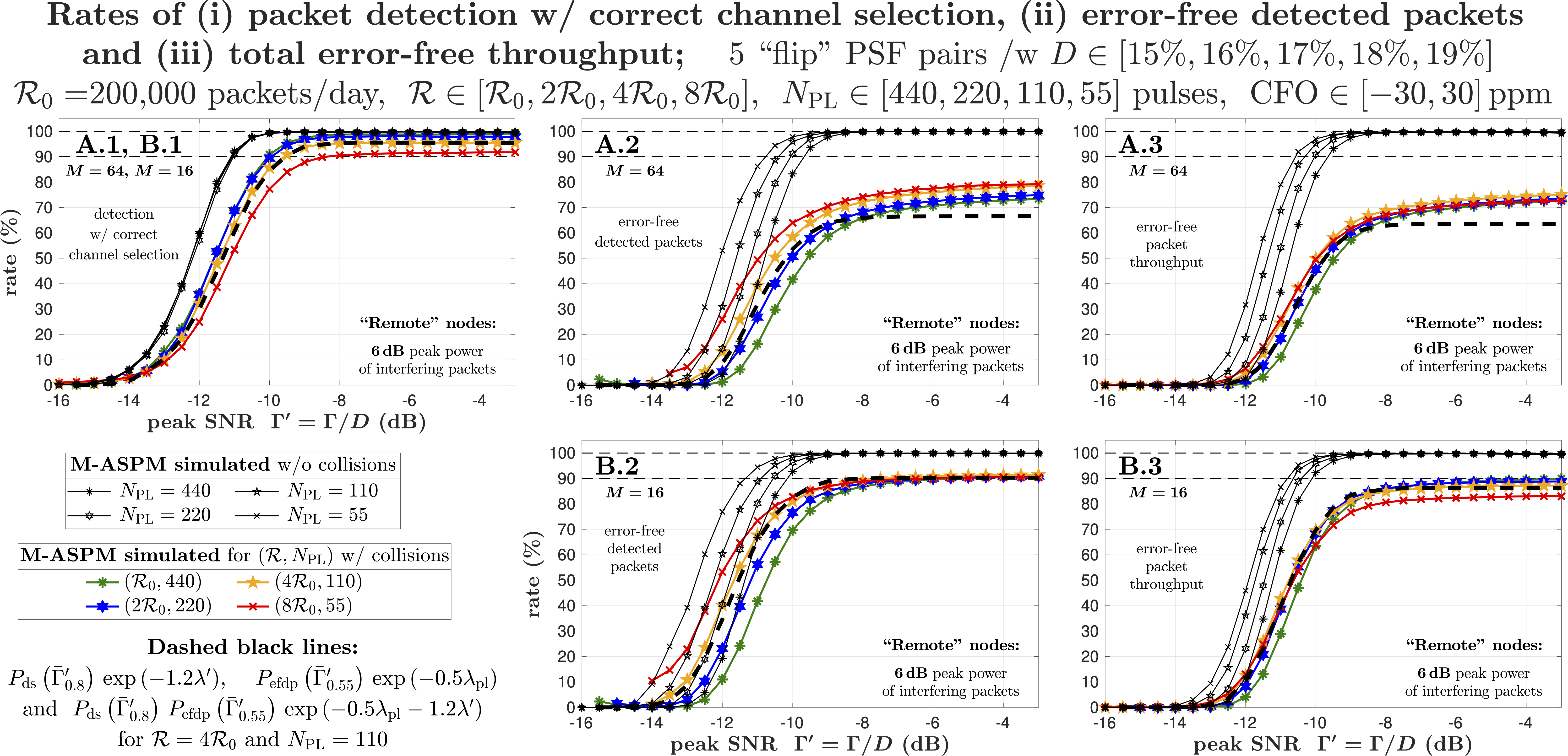}}
\caption{\boldmath As long as effective collision rate~$\lambda^\prime$ in detection channel stays relatively small, effective packet collision rate remains approximately proportional to~$\lambda_\mathrm{pl}$ and thus to product $\mathcal{R}N_\mathrm{PL}$. Then total error-free throughput remains almost unchanged for different packet rates if $\mathcal{R}N_\mathrm{PL}\approx\const$.
\label{fig:tradeoffs}}
\end{figure*}

\subsection{TWO PAYLOAD CHANNELS} \label{subsec:inter-PSF}
The simulations presented in Fig.~\ref{fig:two channels} illustrate the impact of partial orthogonality between PSFs in two-channel configurations. 

As in Fig.~\ref{fig:self-collisions}, the solid black lines show the approximate lower bound for average rates of error-free detected packets computed according to~(\ref{eq:co-PSF}), while the solid magenta lines show this bound computed according to~(\ref{eq:multiple PSFs}) with $\kappa=1$ and $\varepsilon=1/2$. As can be seen in Fig.~\ref{fig:two channels}, larger difference in PSF lengths results in stronger orthogonality and smaller CcCF. In particular, for two channels with ``flip" PSFs the CcCF decreases to about $1/2$.

\subsection{TEN PAYLOAD CHANNELS USING FIVE ``FLIP" PSF PAIRS} \label{subsec:5D2}
The simulations presented in Fig.~\ref{fig:ten channels} demonstrate that further reduction in the collision impact can be achieved when the packet rates are equally shared among ten payload channels that use five ``flip" PSF pairs of different lengths. As can be seen, even with a rather small spread in pulse duty cycles (from 15\% to 19\%) this ten-channel configuration further reduces the CcCF and offers overall improvement over two channels.

The solid blue lines show the approximate lower bound for average rates of error-free detected packets computed according to~(\ref{eq:co-PSF}) with $\kappa=1$ and $\varepsilon=0.4$.

\begin{figure}[!b]
\centering{\includegraphics[width=8.6cm]{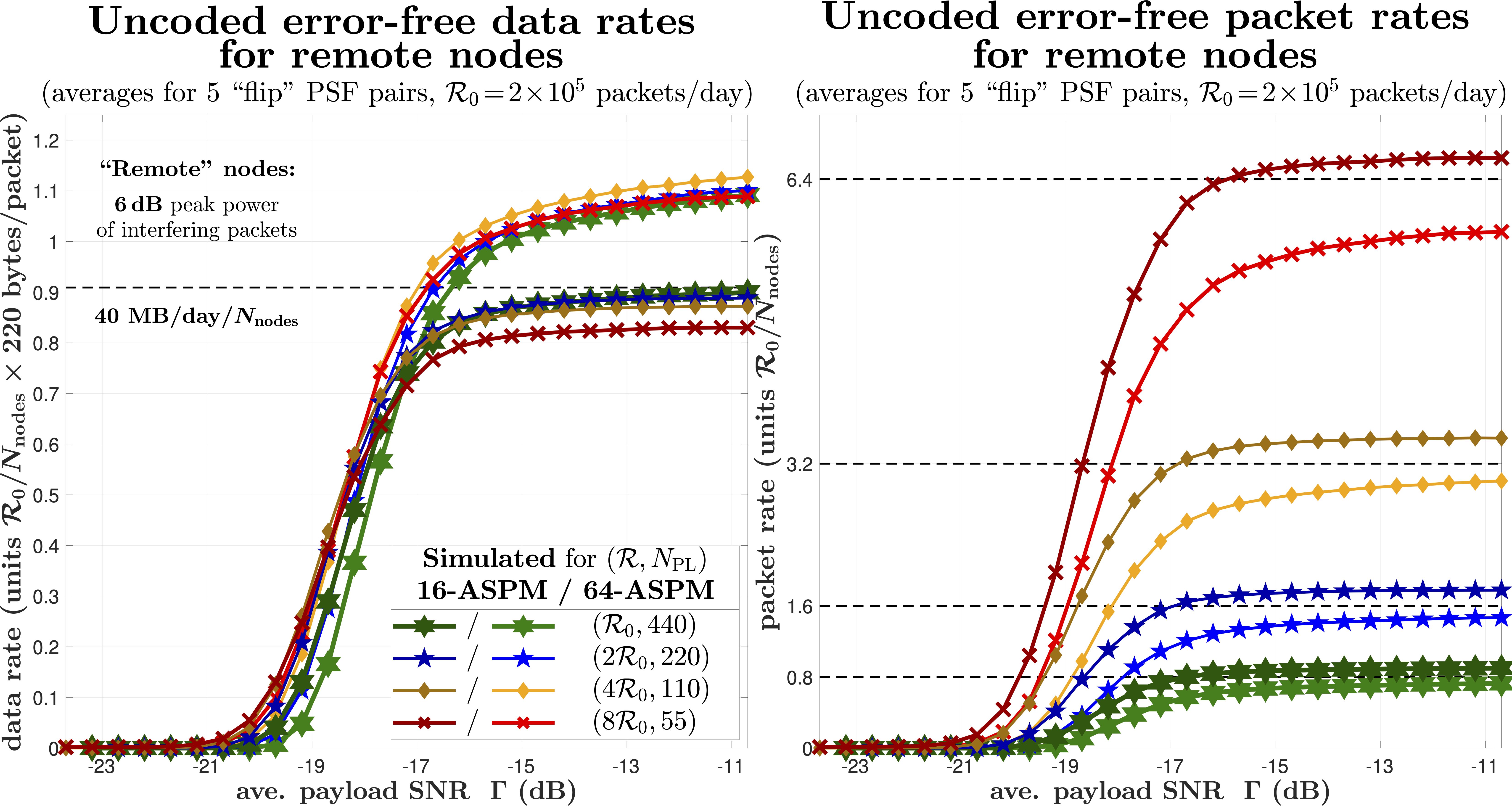}}
\caption{\boldmath For simulations presented in Fig.~\ref{fig:tradeoffs} total error-free data throughput is proportional to product $\mathcal{R}N_\mathrm{PL}$ (left) and error-free packet rate is inversely proportional to~$N_\mathrm{PL}$ (right).
\label{fig:tradeoffs 2}}
\end{figure}

\section{ADDITIONAL ILLUSTRATIONS FOR ASSESSMENTS MADE IN SECTION~\ref{sec:collisions}} \label{sec:main tradeoffs}
To illustrate validity of assessments for the lower bounds on the rates under collisions provided in Section~\ref{sec:collisions}, in this section we use simulated results for remote nodes only. As was discussed in Section~\ref{subsec:effective}, for inner and outer nodes the collision impact will be smaller.

\subsection{TOTAL ERROR-FREE THROUGHPUT} \label{subsec:constant data}
As long as the effective collision rate~$\lambda^\prime$ in the detection channel stays relatively small, it follows from (\ref{eq:overall})~and~(\ref{eq:pllambda}) that, for a given~$M$, the effective packet collision rate remains approximately proportional to~$\lambda_\mathrm{pl}$ and thus the product $\mathcal{R}N_\mathrm{PL}$. At the same time, the probability of error-free detected packets without collisions
\beginlabel{equation}{eq:Pefdp}
  P_\mathrm{efdp}=\left(1\!-\!P_\mathrm{s}\right)^{N_\mathrm{PL}}  \!\approx  1\!-\!N_\mathrm{PL}P_\mathrm{s} \approx 1 \quad \mathrm{for} \quad N_\mathrm{PL}P_\mathrm{s}\ll 1\,.
\end{equation}
Then the total error-free throughput remains almost unchanged for a wide range of different packet rates and payload sizes if $\mathcal{R}N_\mathrm{PL}\approx\const$.

This is confirmed and illustrated by the simulations presented in Fig.~\ref{fig:tradeoffs}. Here the plotted rates under collisions are obtained as averages for all 10~channels, with each trial for each channel repeated 2,000 times. For reference, the dashed black lines show the approximate lower bounds on the rates computed according to~(\ref{eq:ds}), (\ref{eq:multiple PSFs})~and~(\ref{eq:overall}) for $\mathcal{R}=4\mathcal{R}_0$ and $N_\mathrm{PL}=110$. (With the parameter values $a=1.2$, $b=0.8$, $\varepsilon=0.5$ and $\kappa=0.9$.) Further, as shown in Fig.~\ref{fig:tradeoffs 2}, for these simulations the total error-free data throughput is proportional to the product $\mathcal{R}N_\mathrm{PL}$ (left), and the error-free packet rate is inversely proportional to~$N_\mathrm{PL}$ (right).

\subsection{MANYFOLD INCREASE IN IPI} \label{subsec:manyfold}
It follows from~(\ref{eq:ASPM SER binom EbN0}) that, for a given~$M$, in the ideal case of zero CFO and STO the symbol error probability~$P_\mathrm{s}$ of noncoherent M-ASPM in AWGN channel remains unchanged if
\beginlabel{equation}{eq:sensitivity}
\Gamma N_\mathrm{p} = \Gamma^\prime D N_\mathrm{p} = \Gamma^\prime L = \const\,.
\end{equation}
Therefore, if we are to maintain the same signal quality for a given path attenuation (e.g., the range), any increase in the payload IpI and/or PSF length proportionally decreases the average and/or peak, respectively, payload power. (Note that, when~(\ref{eq:sensitivity}) holds, the total packet energy and the energy overhead of the detection channel remain unchanged for any IpI and PSF length values.)

\addtocounter{figure}{1}
\begin{figure*}[!b]
\centering{\includegraphics[width=17.2cm]{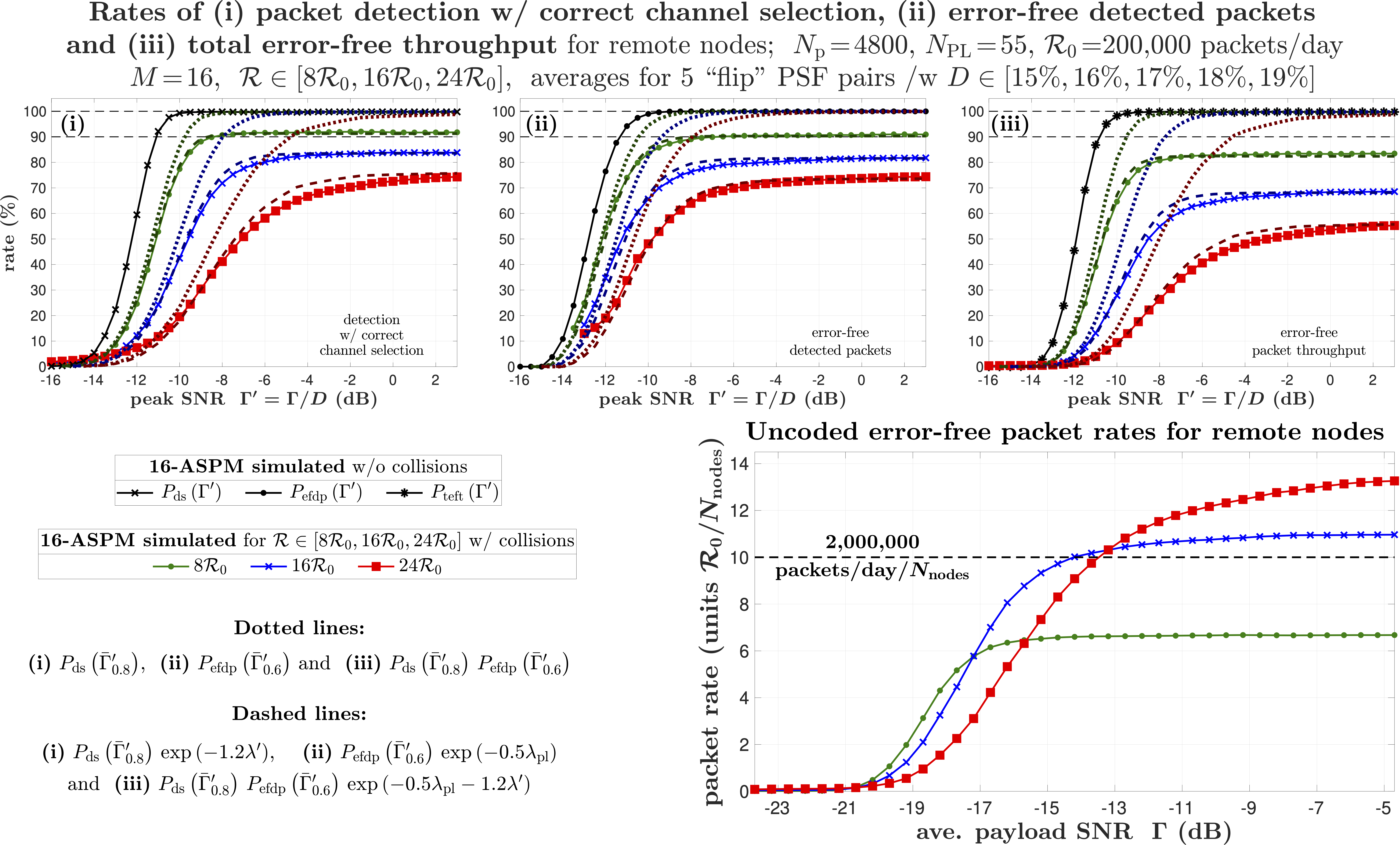}}
\caption{\boldmath For $M=16$ and $N_\mathrm{PL}=55$ collisions in detection channel and among payloads make similar contributions into overall collision impact.
\label{fig:rates55}}
\end{figure*}

\addtocounter{figure}{-2}
\begin{figure}[!t]
\centering{\includegraphics[width=8.6cm]{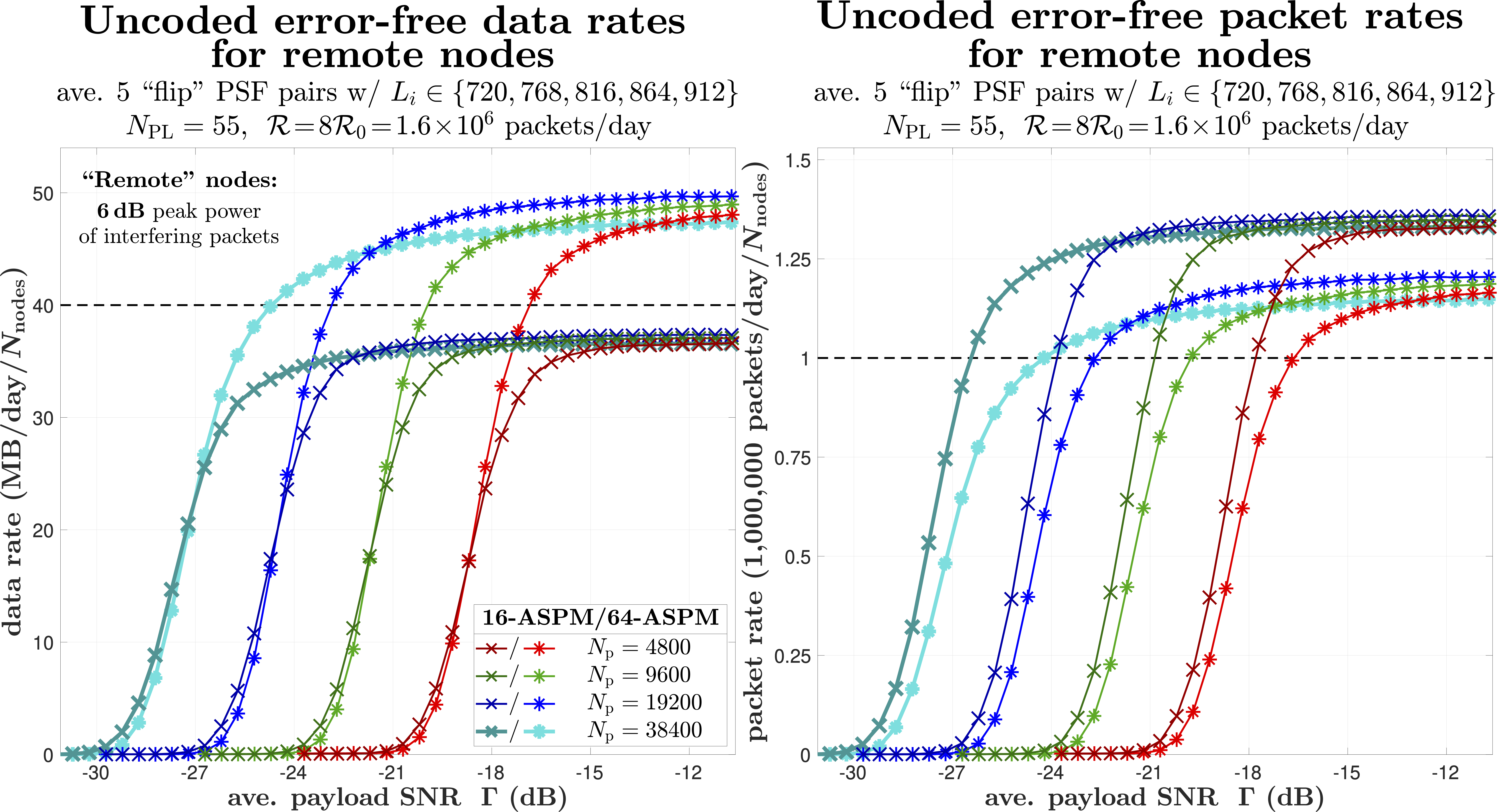}}
\caption{\boldmath Extending IpI in M-ASPM, while proportionally increasing receiver sensitivity and ToA, does not exacerbate impact of collisions. Apparent slight reduction in throughput for $N_\mathrm{p}=38400$ is caused by cumulative error in STO compensation for longer payloads due to finite precision of CFO measurements in detection channel.
\label{fig:tradeoffs 55Np}}
\end{figure}
\addtocounter{figure}{1}

In the simulations presented in Fig.~\ref{fig:tradeoffs 55Np} the PSF lengths are the same as in the previous simulations ($L_i\in\{720,768,816,864,912\}$), and thus the CcCF is expected to remain unchanged. Therefore, while the ToA collisions increase from 6-fold (for  $N_\mathrm{p}=4800$) to almost 50-fold (for  $N_\mathrm{p}=38400$), the total error-free throughput is expected to remain effectively the same for all IpI values.

However, since the accuracy of the CFO measurements in the detection channel is only about~$0.5\,$ppm, the cumulative error in the STO compensation causes the apparent slight reduction in throughput for $N_\mathrm{p}=38400$, when the payload ToA becomes~$1.32\,$s.

\subsection{HIGH TX PACKET RATES WITH LARGE IMPACT ON DETECTION CHANNEL} \label{subsec:detection rates}
As follows from~(\ref{eq:lambdas ratio}), for large~$M$ (e.g., $M\ge 64$) and/or long payloads the contribution of the detection channel collisions into the overall collision impact remains comparatively small. However, for short payloads with small~$M$ this contribution becomes significant. For example, from~(\ref{eq:lambdas ratio}) with $M=16$
\beginlabel{equation}{eq:lambdas ratio M16}
  \frac{a\lambda^\prime}{\varepsilon\lambda_\mathrm{pl}}  \approx  \frac{a}{2\varepsilon}\, \frac{20}{N_\mathrm{bytes}}\,,
\end{equation}
and for payloads with $N_\mathrm{bytes}\lesssim 10a/\varepsilon$ the contribution of detection channel collisions into the overall collision impact exceeds that of payload channels.

This is illustrated by the simulations presented in Fig.~\ref{fig:rates55}, where $a\lambda^\prime/(\varepsilon\lambda_\mathrm{pl})  \approx 0.9$ for $a=1.2$ and $\varepsilon=1/2$, and the detection and payload channels make similar contributions into the total collision impact. As further illustrated in Fig.~\ref{fig:largeSNRrates}, the maximum achievable error-free packet rate is thus limited by the detection channel, as it cannot exceed the rate of packet detection with correct channel selection.

Consequently, M-ASPM configurations with high packet rates may require considering a tradeoff between the Tx energy efficiency and the maximum achievable error-free packet rate. For example, for 33-byte packets with~$\varepsilon=1/2$ in Fig.~\ref{fig:largeSNRrates}, at the expense of about~70\% increase in the energy per packet (taking into account the overhead of the detection channel), for the same packet loss due to collisions 16-ASPM offers a 5-fold increase in the error-free throughput over 256-ASPM. (At the same time the computational cost of payload processing according to~(\ref{eq:payloadsampling}) is reduced by 8~times. Then, unless the computational cost of synchronization exceeds about~3/4 of 16-ASPM payload processing in the Rx, such 5-fold packet rate increase would not raise the overall computational burden on the gateway.)

\begin{figure}[!t]
\centering{\includegraphics[width=8.6cm]{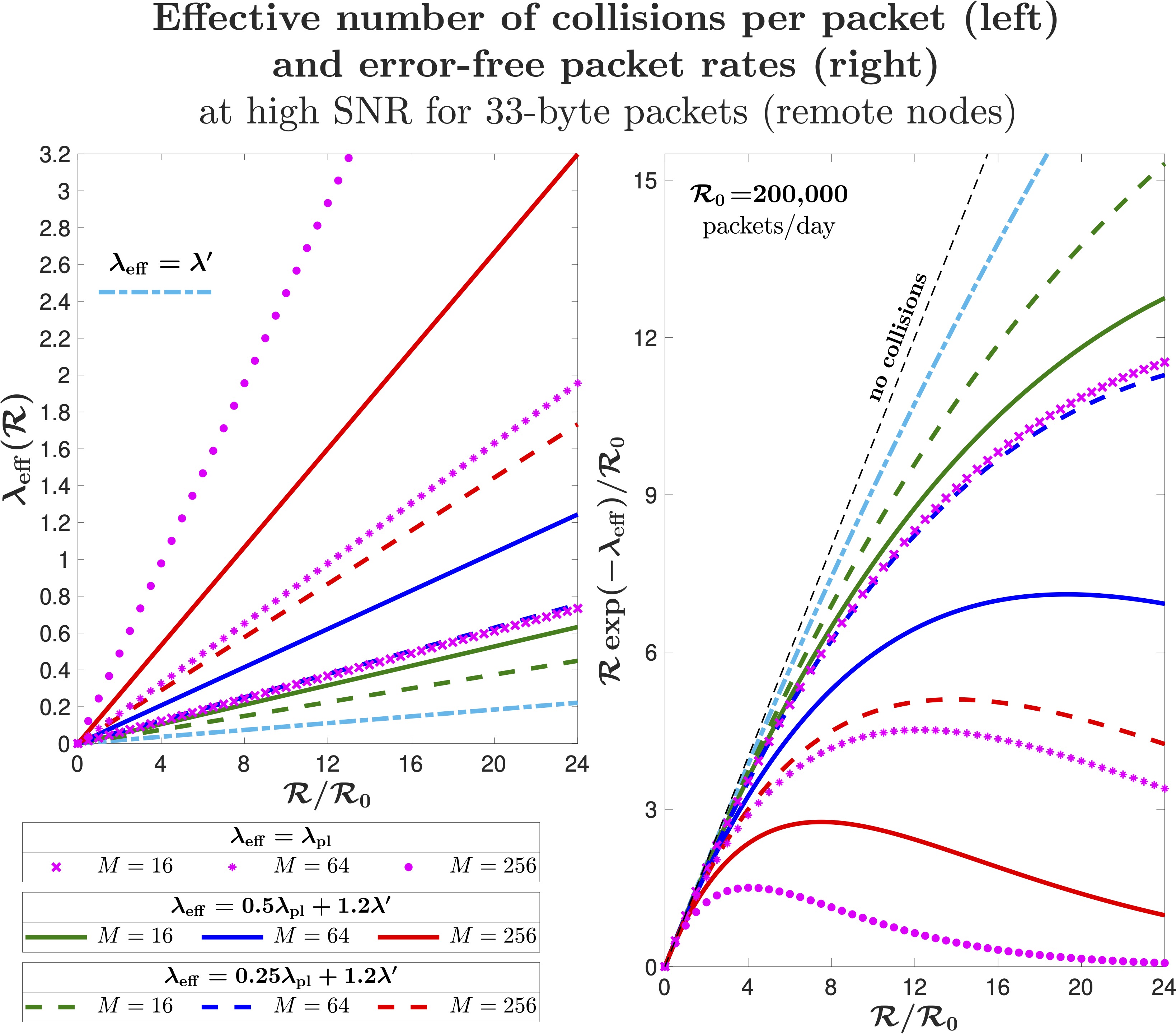}}
\caption{\boldmath Upper bound on error-free packet rates is determined by impact of collisions in detection channel.
\label{fig:largeSNRrates}}
\end{figure}

\section{TRADEOFFS AND OPTIONS TO CONSIDER} \label{sec:secondary tradeoffs}

\subsection{PARAMETERS OF DETECTION CHANNEL} \label{subsec:detection parameters}
When a single detection channel is shared among multiple quasi-orthogonal channels used for payloads, the maximum achievable error-free packet rate is limited by the detection channel. Thus, the parameters of the detection channel should be chosen based on the requirements and practical constraints of particular network configurations, including the choices of the signal bandwidth and/or the carrier frequency. The appropriate modifications to the detection channel configuration can be made based on the descriptions given in~\cite{Nikitin2025detection} and in Sections~\ref{subsec:synchronization}, \ref{subsec:flip correction}, and~\ref{subsubsec:QTF} of the current paper.

For instance, while shortening the leading sequence reduces the effective rate of collisions (see~(\ref{eq:LS colllisions})), it also decreases the detection sensitivity and raises the impact of reduction in the SINR due to interfering packets. Indeed, for the equality of the detection probabilities $P_\mathrm{ds}\left( \bar{\Gamma}^\prime_{b} \right)=P_\mathrm{ds}\left( x\Gamma^\prime \right)$, from the condition
\beginlabel{equation}{eq:SINR equality}
 \left( \frac{x}{\Gamma^\prime} +  \frac{b}{\gamma^\prime} \right)^{-1} \!= \Gamma^\prime
\end{equation}
it follows that
\beginlabel{equation}{eq:SIR limit}
  x =  1 - b \frac{\Gamma^\prime}{\gamma^\prime}\,.
\end{equation}
Then $x$~expresses the loss in the detection sensitivity that is due to the effective SINR of the detection channel. Unless $\Gamma^\prime/\gamma^\prime\ll 1$, this loss can be quite significant.

For example, it can be noticed in the simulations presented in this paper that $P_\mathrm{ds} \left( -11\,\mathrm{dB} \right)\approx 90$\% in the absence of interference. Then for $b=0.7$ the peak SIR $\gamma^\prime=-5.3$ (that corresponds to the average payload SIR $-13\,$dB in Figs.~\ref{fig:self-collisions}, \ref{fig:two channels}, \ref{fig:ten channels} and \ref{fig:ten channels 2D}) results in a rather small $x=0.9\,$dB decrease in sensitivity at~90\% detection probability. This can be seen in the right-hand panels of Figs.~\ref{fig:self-collisions}, \ref{fig:two channels}, \ref{fig:ten channels}, \ref{fig:ten channels 2D} and, in particular, in Fig.~\ref{fig:ds collisions}. On the other hand, in Fig.~\ref{fig:rates55} the average values of $\gamma^\prime$ are $-6.1\,$dB, $-9.1\,$dB. and $-10.8\,$dB for the packet rates $8\mathcal{R}_0$, $16\mathcal{R}_0$, and $24\mathcal{R}_0$, respectively. Then for $b=0.8$ the respective sensitivity losses at~90\% detection probability are $1.3\,$dB, $3.1\,$dB, and $6.3\,$dB. This can be observed for the dotted curves in the upper left panel of Fig.~\ref{fig:rates55}.

At the same time, for the benchmark detection algorithm in~\cite{Nikitin2025detection} the detection sensitivity is approximately proportional to~$N_\mathrm{LS}^{0.7}$. Then, at the expense of doubling the ToA of the leading sequence (and thus $\lambda^\prime$), for $N_\mathrm{LS}=440$ the 90\% detection sensitivity would improve to about~$-13.1\,$dB. Consequently, for the $\gamma^\prime$ values of $-6.1\,$dB, $-9.1\,$dB, and $-10.8\,$dB in Fig.~\ref{fig:rates55}, the respective sensitivity losses would decrease to about $0.7\,$dB, $1.6\,$dB, and $2.8\,$dB. This can be a favorable tradeoff for longer payloads.

\begin{figure}[!b]
\centering{\includegraphics[width=8.6cm]{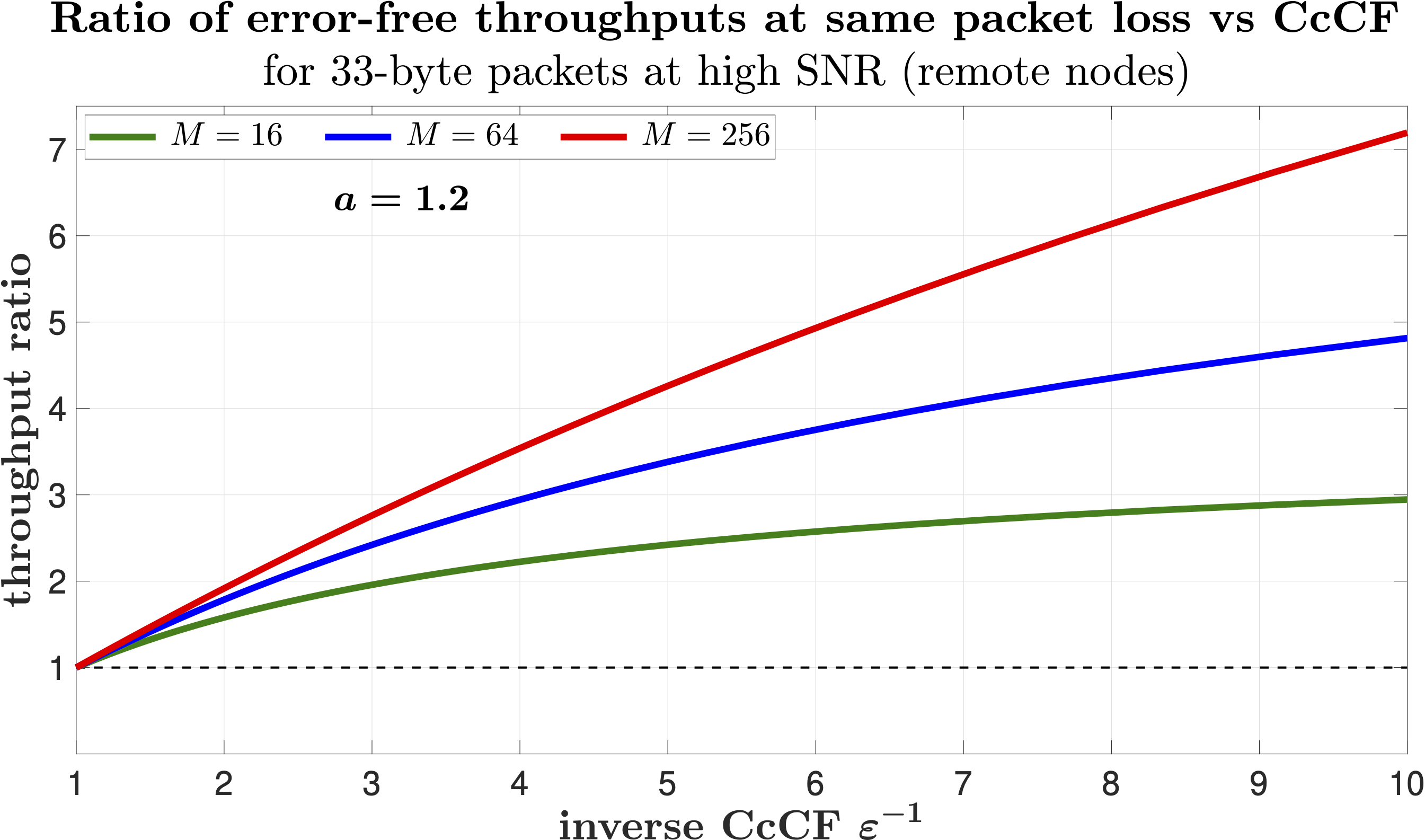}}
\caption{\boldmath CcCF reduction for short packets leads to greater increase in error-free throughput at high SNR for larger $M$ values.
\label{fig:largeSNRratios}}
\end{figure}

\addtocounter{figure}{1}
\begin{figure*}[!b]
\centering{\includegraphics[width=17.2cm]{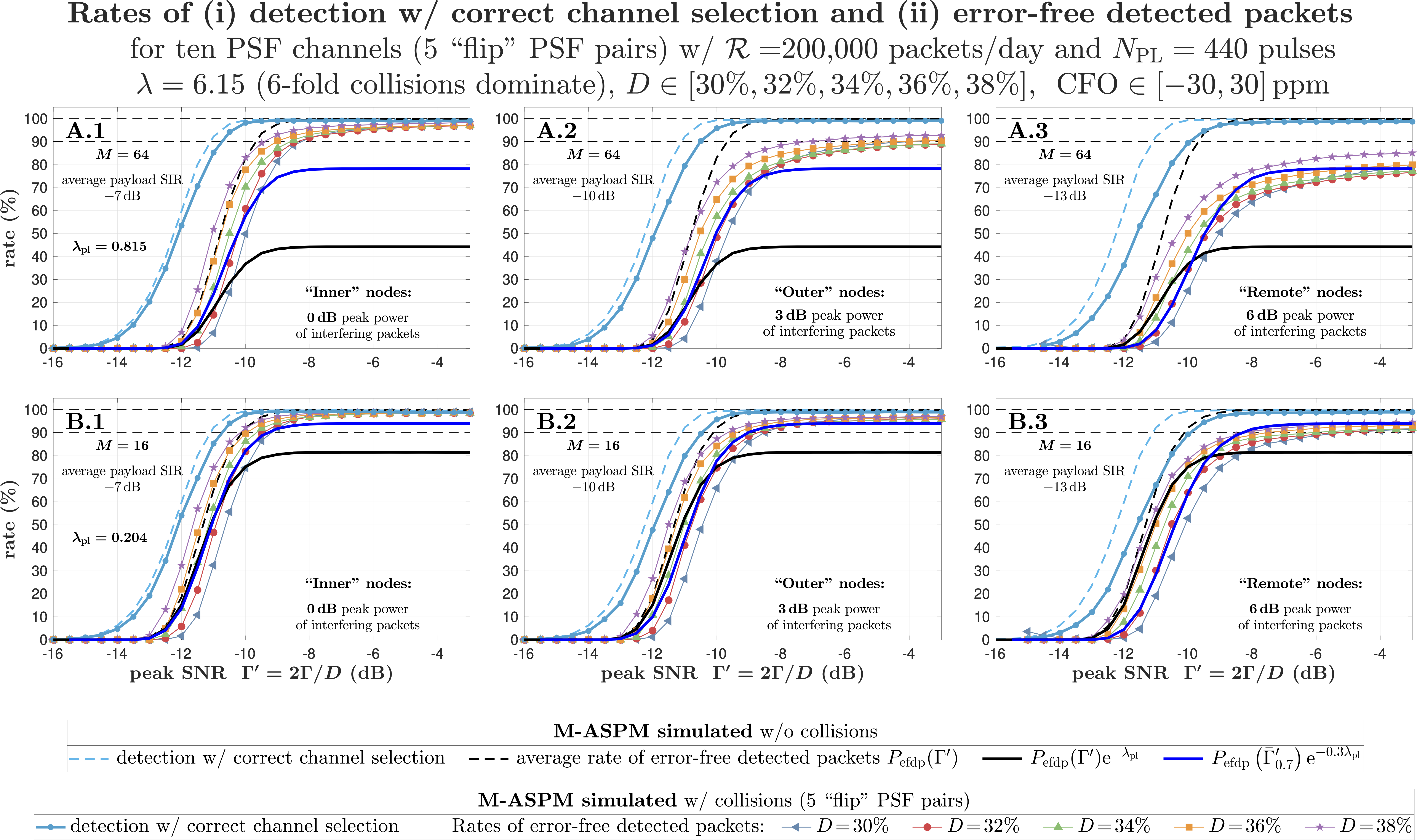}}
\caption{\boldmath Proportional duty cycle increase in all payload channels offers moderate reduction in CcCF. (Compare with Fig.~\ref{fig:ten channels}.)
\label{fig:ten channels 2D}}
\end{figure*}

\subsection{CCCF REDUCTION} \label{subsec:CcCF reduction}
When $\lambda^\prime\ll \lambda_\mathrm{pl}$ (e.g., long payloads), a smaller CcCF results in proportionally smaller effective packet collision rate. However, a smaller value of~$\varepsilon$ increases the value of $1\!-\!\varepsilon\kappa$ and, therefore, decreases the effective payload SINR~$\bar{\Gamma}^\prime_{\!1\!-\!\varepsilon\kappa}$. This tradeoff may need to be considered for some particular configurations. Specifically, for the equality of the probabilities $P_\mathrm{efdp}\left( \bar{\Gamma}^\prime_{b} \right)=P_\mathrm{efdp}\left( y\Gamma^\prime \right)$ the sensitivity loss~$y$ can be expressed as
\beginlabel{equation}{eq:SIR reduction payload}
  1 - \frac{\Gamma^\prime}{\gamma^\prime} < y =  1 - (1\!-\!\varepsilon\kappa) \frac{\Gamma^\prime}{\gamma^\prime} <1\,.
\end{equation}
Therefore, for a large number of strongly orthogonal payload channels $\gamma^\prime$ needs to remain sufficiently larger than~$\Gamma^\prime$.

For shorter payloads the contribution of the detection channel into the overall collision impact becomes comparatively larger. Consequently, reduction in the CcCF would be typically more beneficial for payloads with greater~$M$ values. For example, as illustrated in Fig.~\ref{fig:largeSNRratios}, for 33-byte packets and the same packet loss due to collisions, a 10-fold CcCF reduction only triples the ratio of error-free throughputs at high SNR for 16-ASPM, while increases this ratio by 7~times for 256-ASPM.

\subsection{PSF LENGTH INCREASE IN PAYLOAD CHANNELS} \label{subsec:duty cycle}
The effective collision rate of M-ASPM payloads does not depend on the PSF length. However, if it is desired to use the same magnitude of Tx pulses throughout the packet, increasing the length of payload PSFs can degrade the overall collision performance.

For example, to maintain the sensitivity matching between detection and payload channels, in~\cite{Nikitin2025detection} the duration of the leading sequence $N_\mathrm{LS}\propto L_i^{1.4}$, where~$L_i$ is the length of the payload PSF. Then doubling the value of~$L_i$ increases $N_\mathrm{LS}$ by approximately $2.6$~times. Further, to maintain reliable synchronization and payload channel selection, a larger~$L_i$ requires a proportionally greater length~$L$ of the pulses in the timing sequence. This combined increase raises both the effective collision rate~$\lambda^\prime$ of the detection channel and its energy overhead.

\addtocounter{figure}{-2}
\begin{figure}[!t]
\centering{\includegraphics[width=8.6cm]{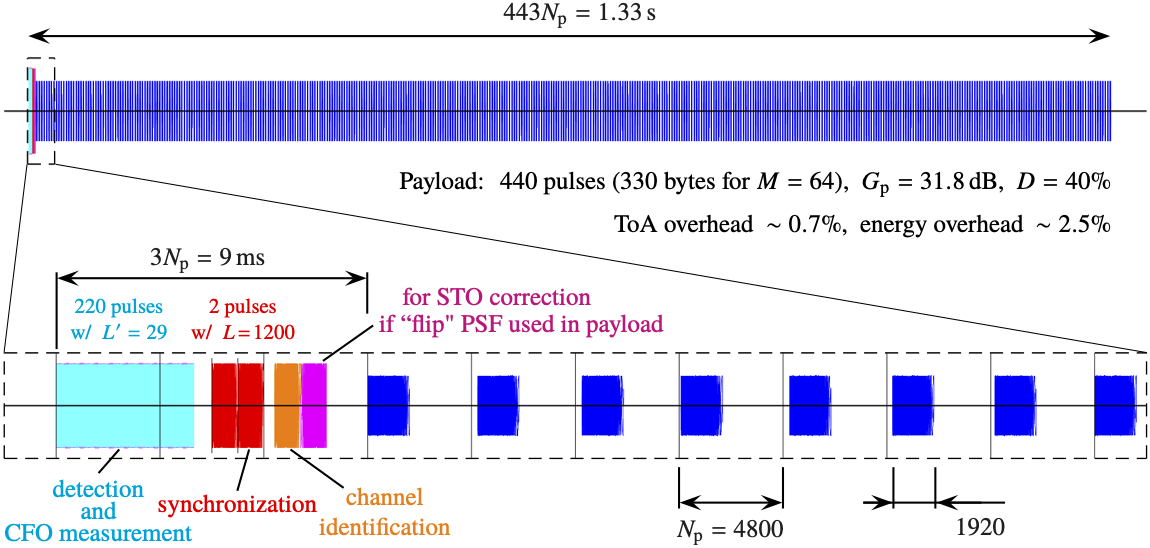}}
\caption{\boldmath Tx M-ASPM packet with doubled payload pulse duty cycle. (Compare with Fig.~\ref{fig:Tx packet}.)
\label{fig:Tx packet 2D}}
\end{figure}
\addtocounter{figure}{1}

If, however, we allow for different peak powers in the detection and payload channels, we can increase the lengths of the payload PSFs while proportionally decreasing the peak power of the payload pulses. This preserves the ToA and energy overheads of the detection channel, as well as the average payload power, and does not reduce the lower bound for the total error-free throughput. For example, in Fig.~\ref{fig:Tx packet 2D} we illustrate the M-ASPM packet obtained by doubling the payload PSF length of the packet shown in Fig.~\ref{fig:Tx packet}.

For a single payload channel (i.e., $\varepsilon=1$), such increase in the PSF length has very little impact on the collision performance. For multiple payload channels, however, using ``flip" PSF pairs with large differences in length among the pairs allows us to significantly reduce the CcCF.

In Fig.~\ref{fig:ten channels 2D} we repeat the simulations presented in Fig.~\ref{fig:ten channels} where, for simplicity, we simply double the PSF lengths of all payload channels. For obtaining the average rates under collisions, for each payload channel each trial is repeated 2,000 times, and the rates of error-free detected packets are plotted as averages of the respective ``flip" channels. Since the differences in lengths of the PSFs with the same temporal direction are still rather small (with $\Delta{D}$ range 2\%~to~8\%, instead of 1\%~to~4\%), such proportional duty cycle increase in all payload channels offers only moderate reduction in the CcCF. This can be seen in the results presented in Fig.~\ref{fig:ten channels 2D} .

Further, as was discussed in Section~\ref{subsec:CcCF reduction} and illustrated in Fig.~\ref{fig:largeSNRratios}, such rather small reduction in~$\varepsilon$ provides only marginal increase in error-free throughput at high SNR for short 16-ASPM packets. This is confirmed and illustrated in Fig.~\ref{fig:rates55 2D}.

\subsection{TRADEOFF BETWEEN COLLISION RESISTANCE AND TX ENERGY EFFICIENCY} \label{subsec:energy for rate}
For given IpI $N_\mathrm{p}$ and payload size~$N_\mathrm{bytes}$, the spread in the receiver sensitivity for~$M$ in the range $16\le M\le 256$ 
is rather small. For example, as can be seen in Fig.~\ref{fig:LoRa and M-ASPM}, the differences in the receiver sensitivities for different $M$~values are confined to less than~$1\,$dB range. Further, this sensitivity spread is largely unaffected by the IpI and payload size values. Therefore, the payload energy is roughly proportional to~$1/\log_2M$.

On the other hand, the effective payload collision rate~$\lambda_\mathrm{pl}$ is proportional to $M/\log_2 M$. Then its reduction by decreasing the~$M$ value comes at the expense of the Tx energy efficiency.

At the same time, the energy overhead of the detection channel is proportional to~$\log_2M$. Consequently, for short payloads (i.e., small $N_\mathrm{bytes}$) the benefit of improving the collision resistance may outweigh the reduction in the Tx energy efficiency.

For example, for the numerical values used in Fig.~\ref{fig:LoRa and M-ASPM} both the packet energy and the ToA of 256-ASPM packets are on par with those of LoRa with $\mathrm{SF}=12$ and $125\,$kHz bandwidth. (In fact, with an 8-frame preamble~\cite{Xhonneux2022maximum, Maleki24CSStutorial} the ToA of a 33-byte LoRa packet exceeds that of 256-ASPM by about~36\%.)

However, $\lambda_\mathrm{eff}\approx 3.7\times 10^{-2}\lambda$ for 33-byte 256-ASPM packets with CcCF $\varepsilon=1/2$. Thus, even for remote nodes the effective collision rate of such packets is about 27~times smaller than the rate of ToA packet collisions. Note that this rate is less than the ToA collision rate of LoRa packets consisting of a single frame (instead of 22~frames required for 33-byte $\mathrm{SF}=12$ LoRa packets, plus preamble).

At the same time, as was mentioned in Section~\ref{subsec:detection rates} (see Fig.~\ref{fig:largeSNRrates}), for $\varepsilon=1/2$ 33-byte 16-ASPM packets offer a 5-fold increase in the error-free throughput over 256-ASPM, at the expense of doubling the ToA and about 70\% increase in the energy per packet (taking into account the energy overhead of the detection channel).

\begin{figure*}[!t]
\centering{\includegraphics[width=17.2cm]{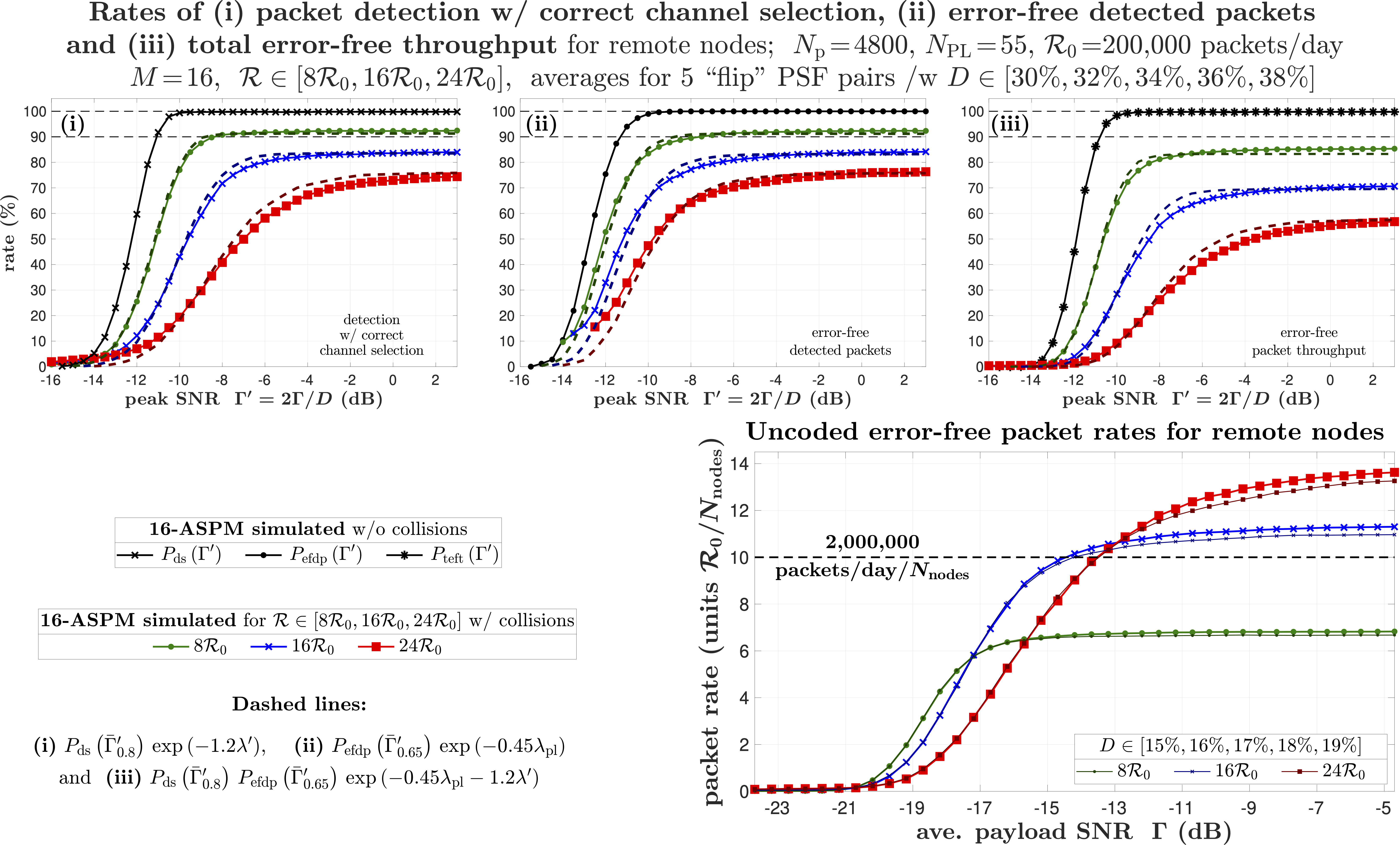}}
\caption{\boldmath For high 16-ASPM packet rates and short payloads proportional duty cycle increase in all payload channels offers only marginal improvement in total throughput. (Compare with Fig.~\ref{fig:rates55}.)
\label{fig:rates55 2D}}
\end{figure*}

\begin{figure*}[!b]
\centering{\includegraphics[width=17.2cm]{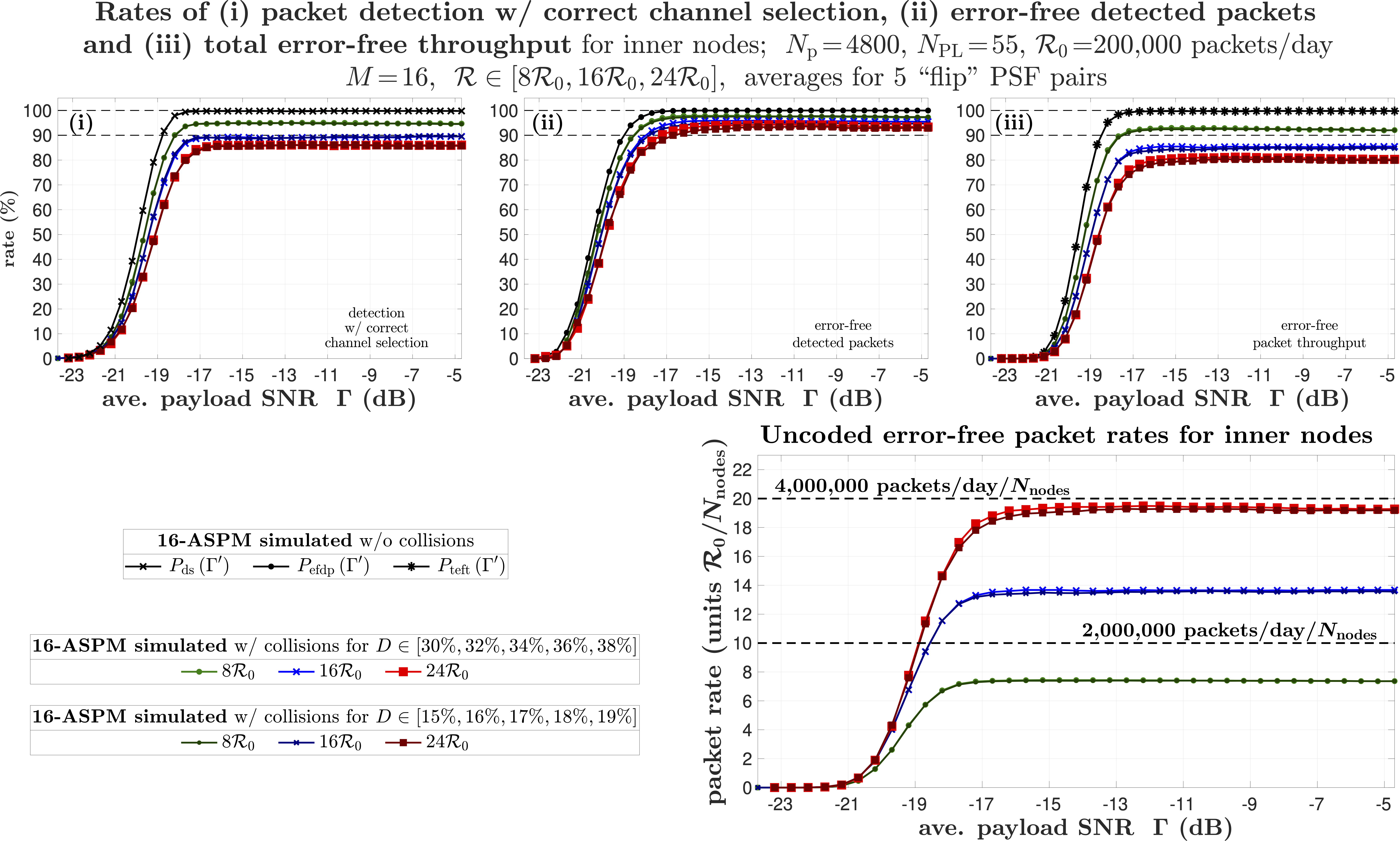}}
\caption{\boldmath Tx power control significantly increases total error-free throughput for short high-rate packets. (Compare with Fig.~\ref{fig:rates55 2D}.)
\label{fig:rates55 1Dvs2D}}
\end{figure*}

\subsection{TX POWER CONTROL} \label{subsec:power control}
During synchronization we determine the power of the synchronization peaks, i.e., the values of $\bar{z}^\prime_k$ and $z^\prime_{k+L}$ in~(\ref{eq:two largest peaks}). Therefore, as discussed in~\cite{Nikitin2024implementation}, during packet detection we also obtain a measure of the Rx signal quality. By adjusting transmit powers of the nodes we can maintain this measure approximately equal for all packets received by the gateway, ensuring that the Tx power is proportional to the path attenuation for the respective nodes. Consequently, all received packets would have similar powers and can be considered ``inner."

As can be seen from the simulations presented in Figs.~\ref{fig:self-collisions}, \ref{fig:two channels},~\ref{fig:ten channels} and~\ref{fig:ten channels 2D}, the impact of collisions for inner nodes is significantly smaller than for outer and/or remote nodes. Smaller power of interfering packets increases the SIR, and even partial orthogonality (e.g., as shown in Fig.~\ref{fig:CFO impact}, due to CFO differences) reduces the CcCF.

If there is a constraint on the peak power of payload pulses, Tx power control can be implemented within this constraint by changing pulse duty cycles in payloads. However, while for a given~$N_\mathrm{PL}$ the value of~$M$ and/or payload pulse duty cycle do not impact the Tx energy efficiency, one needs to be cognitive that the computational cost of Rx payload processing is $\mathcal{O}(N_\mathrm{bytes}LM/\log_2 M)$.

To further illustrate the benefits of Tx power control, in Fig.~\ref{fig:rates55 1Dvs2D} the simulations presented in Figs.~\ref{fig:rates55} and~\ref{fig:rates55 2D} are repeated for the inner nodes. As one can see, for example, for $\mathcal{R}=24\mathcal{R}_0$ the power control increases the total error-free throughput at high SNR from less than~60\% to about~80\%, while the 90\%~sensitivity loss for $P_\mathrm{efdp}$ is reduced from over~$6\,$dB to less than~$1\,$dB.

Also note that with power control doubling the payload pulse duty cycles does not provide noticeable benefits for high-rate packets with small payloads, and shorter PSFs can be used to reduce the computational burden on the gateway.

\section{SUMMARY AND IMPLICATIONS FOR LARGE-SCALE LPWAN DEPLOYMENTS} \label{sec:implications}
In this work we demonstrate that impact of collisions on error-free M-ASPM packet rates can remain effectively invariant to a wide range (e.g., over two orders of magnitude) changes in payload processing gain and spectral efficiency. In particular, with SCDS, when ${N_\mathrm{p}>2n_\mathrm{off}M}$ and M-ASPM operates in the spread-spectrum region, ${2n_\mathrm{off}M}$ replaces ${N_\mathrm{p}}$ in~(\ref{eq:rate packets}) for the packet rate constraint on co-PSF collisions in M-ASPM. Then any further increase in the processing gain beyond $G_\mathrm{p}=5n_\mathrm{off}M/8$, while proportionally extending the ToA and reducing the spectral efficiency, does not exacerbate collisions. For example, with $M=16$ and the highest processing gain $G_\mathrm{p}=40.8\,$dB used in the simulations, the effective payload collision rate for co-PSF collisions is 240~times smaller than the rate of ToA collisions. Therefore, under constrained transmit power, we can maintain the capacity of an uplink-focused M-ASPM network while extending its range.

In dense uplink-dominant networks such as smart utility metering, agriculture monitoring, and municipal sensing infrastructures, network capacity is often constrained not by the raw link budget but by collision-limited throughput and gateway congestion. Under such conditions, increasing sensitivity typically reduces effective capacity due to longer packet durations. The invariance of the effective collision rate to the M-ASPM processing gain (demonstrated, in particular, in Sections~\ref{sec:ASPM collision scaling} and \ref{sec:collisions}, and in Fig.~\ref{fig:tradeoffs 55Np}) suggests that M-ASPM can extend communication range without proportionally reducing network throughput under high node density. This structural property may reduce gateway density requirements and/or increase supported node count per gateway in collision-limited deployments.

As an illustration, consider the following simplified example. Imagine a single-gateway network with 10\%~packet loss due to collisions, where we are to raise the processing gain by~$9\,$dB. This may be needed to compensate for increased shadowing, indoor placement of the nodes, physical range extension, higher ambient noise, required reduction in the average Tx power, and/or other factors. If this 8-fold gain increase is accompanied by the respective rise in the collision impact, the packet loss becomes~57\%, and the throughput is reduced by more than half, to~43\%. In M-ASPM, however, the increased processing gain does not cause throughput degradation (see Fig.~\ref{fig:tradeoffs 55Np}). At the same time, if the 57\%~packet loss is still acceptable, the number of uplink M-ASPM nodes can be further raised 8-fold, almost quadrupling the total gateway capacity.

In this paper, we further establish how a single collision-resistant detection channel can be shared by multiple quasi-orthogonal PSF channels used for payloads.  Such sharing is especially useful for scaling and economizing long-range, high-throughput M-ASPM network configurations. The detection channel is insensitive to relatively large CFO and, in addition to detecting arrivals of packets, provides corrections for the impact of the CFO and the sampling time offsets on the received signal, and performs packet synchronization and payload channel identification. Further, except for very short payloads with small $M$~values, the contribution of the detection channel into the overall collision impact remains relatively small. For example, in the presented simulations we demonstrate that millions of packets a day, randomly transmitted from a wide area, can be received by a  sole gateway with a single $500\,$kHz frequency channel.

When interfering packets are not excessively strong (e.g., below 9\,dB relative to the packet of interest), the approximate lower bounds on the rates (probabilities) of (i)~packet detection with correct channel selection, (ii)~error-free detected packets, and (iii)~total error-free throughput in such M-ASPM configurations with a shared single detection channel can be assessed according to~(\ref{eq:ds}), (\ref{eq:multiple PSFs})~and~(\ref{eq:overall}), respectively. This analysis of diverse collision contributions enables quantification of various tradeoffs available for designing M-ASPM networks that satisfy particular technical requirements and/or constraints.

The presented analytical results are illustrated and verified by extensive simulations for high packet collision rates in wide ranges of payload sizes, payload processing gains, and powers of the noise and the interfering packets. In particular, in various simulations for a single gateway with a $500\,$kHz frequency channel, the transmit packet rates vary from $2\times 10^5$ to $4.8\times 10^6$ packets per day, with the average number of ToA collisions per packet varying from $6.15$~to~$49.2$. The payload sizes are in the $27.5$~to~$330$ uncoded bytes range, payload processing gains vary from $31.8\,$dB~to~$40.8\,$dB, payload SNR spans from $-31.1\,$dB to $-4.7\,$dB, and the average payload SIR values are in the $-18.5\,$dB~to~$-7\,$dB range.

For analysis and simulations presented in this paper, a pure ALOHA protocol is implied. It is important to point out, however, that M-ASPM collision resistance arises from its PHY-level properties, and not from any adaptation of the signal processing to a particular traffic model. This enables a general inference that M-ASPM provides a structurally distinct scaling behavior compared to conventional LPWAN modulations, decoupling range extension from collision-induced throughput degradation. This property provides a number of additional options and tradeoffs for economizing various LPWAN/LoRaWAN network configurations. Figure~\ref{fig:implications} illustrates one of such options. For simplicity, this illustration assumes a single gateway with a constrained spectral band, and the range achievable for a particular receiver sensitivity under a given Tx power constraint. It also assumes effectively the same other physical conditions, e.g., the same antenna gains and various system attenuations such as insertion, path, and matching losses, etc.

\begin{figure}[!b]
\centering{\includegraphics[width=8.6cm]{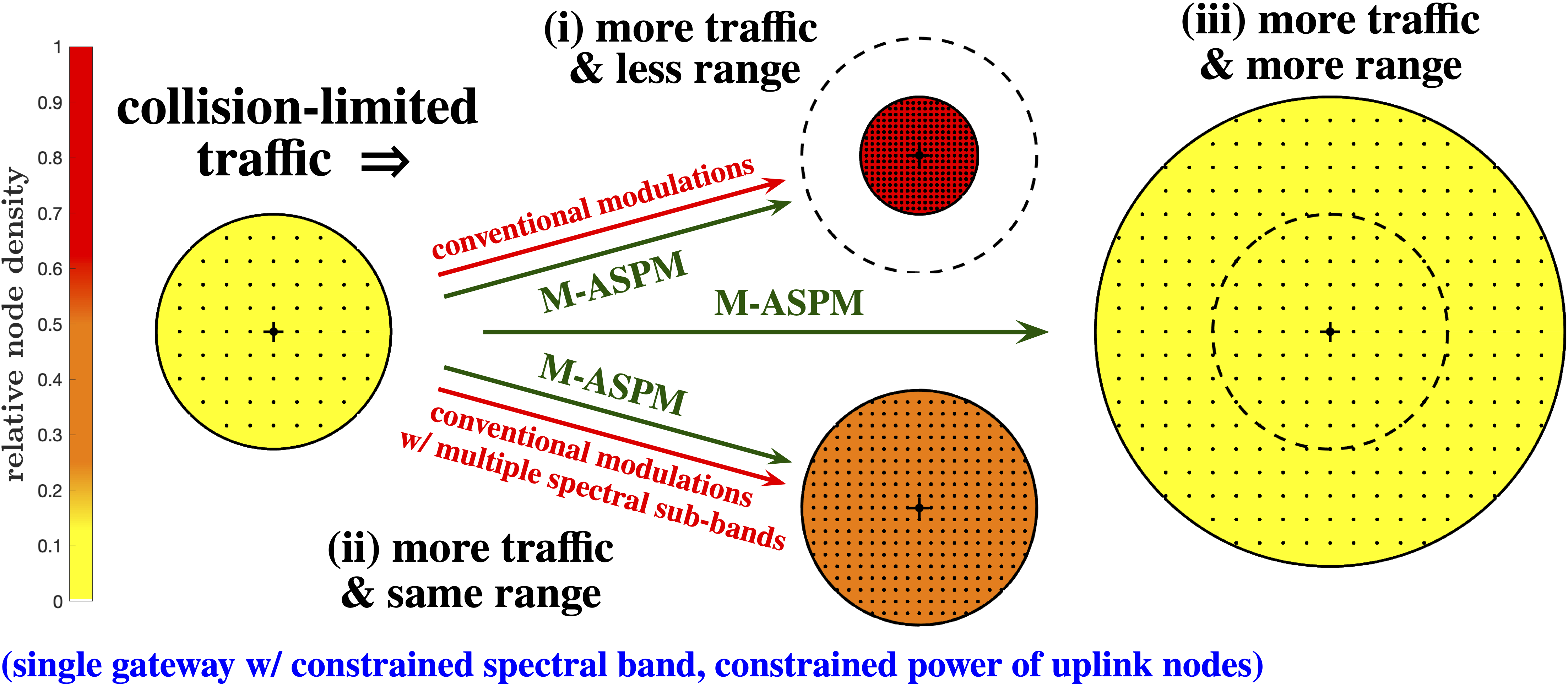}}
\caption{\boldmath By decoupling range extension from collision-induced throughput degradation, M-ASPM extends options for scaling and economizing LPWAN/LoRaWAN networks.
\label{fig:implications}}
\end{figure}

Under collision-limited operation, for a given bandwidth, larger data traffic requires reduction in collision exposure per bit. In conventional modulations, when collision exposure increases with the ToA, for a given bandwidth such reduction inevitably leads to smaller receiver sensitivity and, therefore, smaller range (case~(i)).

Alternatively, with conventional modulations we can reduce the bandwidth while preserving both the receiver sensitivity and the collision exposure. For example, as can be seen in Fig.~\ref{fig:contrast}, the collision exposures per bit and the receiver sensitivities for LoRa with the values $(\mathrm{SF},B)$ equal to $(12,500\,\mathrm{kHz})$, $(11,250\,\mathrm{kHz})$, and $(10,125\,\mathrm{kHz})$ are similar. Then, by using multiple sub-bands within the given spectral band, we can preserve the range while increasing the collision-limited traffic (case~(ii)).

However, this imposes additional implementation and management costs on the network. For example, it requires multi-band gateways and uplink transmitters operating in different frequency bands. Further, conventional LPWAN modulations do not allow additional range extension without exacerbating collisions (case~(iii)).

In contrast, in M-ASPM the collision exposure per bit is proportional to~$M/\log_2 M$. For example, as discussed in Section~\ref{sec:ASPM collision scaling} and illustrated in Fig.~\ref{fig:contrast} therein, changing~$M$ from~$256$ to~$16$ provides 8-fold reduction in the M-ASPM collision exposure per bit, without changes in the Tx spectral band.

Then, since the receiver sensitivity can be changed without impacting the collision susceptibility, the respectively increased M-ASPM's collision-limited traffic can be supported for all ranges shown in Fig.~\ref{fig:implications}, including case~(iii). While, due to increased path attenuation, extending the range unavoidably increases the energy consumption of remote uplink nodes, it reduces the number of gateways needed for wide areal coverage, and can significantly decrease the overall implementation and management costs of the network.

While a full cross-layer system evaluation is beyond the scope of this paper, the analytical and simulation results presented herein lay the groundwork for further network-level M-ASPM studies incorporating gateway density scaling, diverse traffic models, and regulatory constraints.

\section{CONCLUSION} \label{sec:conclusion}
Unlike conventional LPWAN modulations, where increased sensitivity exacerbates collision exposure due to extended ToA, M-ASPM allows decoupling of the receiver sensitivity from effective collision scaling. This structural distinction enables range extension without respective degradation of throughput under collision-limited operation.

In this work, we demonstrate analytically and verify by simulation that, with SCDS, in M-ASPM the effective payload collision rate remains nearly invariant to wide-range changes in processing gain and spectral efficiency. The proposed shared detection channel architecture further enables scalable multi-channel payload configurations with reduced implementation overhead.

These results provide a PHY-level foundation for future system-level studies and potential deployment of collision-resilient LPWAN architectures based on M-ASPM.

\appendices
\section{ACRONYMS} \label{app:acronyms}
ACF:~autocorrelation function;
A/D:~Analog-to-Digital;
ASPM: Aggregate Spread Pulse Modulation;
AWGN:~Additive White Gaussian Noise;
BER:~Bit Error Rate;
CcCF:~Cross-channel Collision Factor;
CFO:~Carrier Frequency Offset;
D/A:~Digital-to-Analog;
IpI:~Interpulse Interval;
LO:~Local Oscillator;
LoRa:~Long Range (modulation technique for LPWANs based on chirp spread spectrum);
LoRaWAN:~Long Range Wide Area Network;
LPWAN:~Low-Power Wide Area Network;
M-ASPM:~M-ary ASPM;
MEA:~Modulo Exponential Averaging;
MPA:~Modulo Power Averaging;
PHY:~physical layer;
ppm:~parts per million;
PSD:~Power Spectral Density;
PSF:~Pulse Shaping Filter;
QTF:~Quantile Tracking Filter;
RC:~Raised-Cosine;
Rx:~receiver;
SCDS:~Synchronized Corrected Decimated Sampling;
SER:~Symbol Error Rate;
SF:~Spreading Factor (for LoRa);
SFO:~Sampling Frequency Offset;
SIR:~Signal-to-Interference Ratio;
SINR:~Signal-to-Interference-plus-Noise Ratio;
SNR:~Signal-to-Noise Ratio;
STO:~Sampling Time Offset;
TBP:~Time-Bandwidth Product;
ToA:~Time-on-Air;
Tx:~transmitter.

\section{COMMENTS ON NOTATIONS} \label{app:notations}
Whenever a particular notation is introduced in the paper, it is immediately defined. Some notations are confined to the specific sections. The notations that may appear multiple times throughout the paper include:

\begin{description}
\setlength{\labelwidth}{8mm}
\item[{$\alpha$}] {upper fence/threshold}
\item[{$B$}] {bandwidth}
\item[{$\beta$}] {RC roll-off factor or scaling parameter in QTF fencing}
\item[{$D$}] {pulse duty cycle}
\item[{$\Delta{f}_\mathrm{c}$}] {CFO value}
\item[{$\varepsilon$}] {CcCF}
\item[{$\eta$}] {spectral efficiency}
\item[{$f_\mathrm{b}$}] {bit rate}
\item[{$f_\mathrm{c}$}] {carrier/LO frequency}
\item[{$F_\mathrm{s}$}] {sample rate}
\item[{$f_\mathrm{p}$}] {pulse rate}
\item[{$\gamma$}] {SIR}
\item[{$\bar{\Gamma}$}] {SINR}
\item[{$\Gamma$}] {SNR}
\item[{$k$}] {sample index (in digital signal representations)}
\item[{$K$}] {MEA/MPA averaging factor}
\item[{$\kappa$}] {detection channel complementary energy overhead}
\item[{$L$}] {PSF length}
\item[{$\lambda$}] {average number of ToA collisions per packet}
\item[{$\lambda_\mathrm{pl}$}] {effective rate of collisions for payloads}
\item[{$\lambda^\prime$}] {effective rate of collisions for detection channel}
\item[{$M$}] {number of states in M-ary encoding}
\item[{$\mu$}] {QTF rate parameter}
\item[{$N_\mathrm{p}$}] {average interpulse interval}
\item[{$N_\mathrm{CFO}$}] {number of pulses for CFO measurement}
\item[{$N_\mathrm{LS}$}] {total number of pulses in leading sequence}
\item[{$N_\mathrm{PL}$}] {number of pulses in payload}
\item[{$\mathcal{N}_\mathrm{s}$}] {oversampling factor}
\item[{$\bar{p}$}] {MPA output}
\item[{$P_\mathrm{d}$}] {probability of packet detection}
\item[{$P_\mathrm{ds}$}] {joint probability of detection and synchronization}
\item[{$P_\mathrm{efdp}$}] {probability of error-free detected packet}
\item[{$P_\mathrm{s}$}] {symbol error probability}
\item[{$P_\mathrm{teft}$}] {total error-free throughput}
\item[{$\mathcal{R}$}] {Tx packet rate}
\item[{$y_\mathrm{nc}$}] {received pulse train (noncoherent detection)}
\end{description}

In the mathematical notations we reserve the letters ``$\zeta$", ``$g$", and ``$h$" for pulse shaping filters, with $g$~and~$h$ being the real and the imaginary parts, respectively, of~$\zeta$. For example, we denote the finite impulse response of a PSF applied to a designed pulse train as $\hat{\zeta}[k]$, where $k$~is the sample index. As in this paper we assume the single-sideband M-ASPM, the PSF components $g$ and $h$ are related to each other through the Hilbert transform, e.g., $h(t)=\pm H(g)(t)$ (in analog domain) or $h[k]=\pm H\{g[k]\}$ (in digital representation).

We find it convenient to use the ``hat" operator for $\hat{\zeta}[k]$, $\hat{g}[k]$, and $\hat{h}[k]$ to distinguish them from their respective matched filters $\zeta[k]=\hat{\zeta}^\ast[-k]$, $g[k]=\hat{g}[-k]$, and $h[k]=\hat{h}[-k]$. (As is common, the superscript asterisk denotes the complex conjugate.) We also use the hat symbol in Section~{\ref{sec:ASPM collision scaling}} to denote the designed pulse train~$\hat{x}[k]$. Further, we use the overhead ``check" symbol for quantities related to the signal's troughs (such as, e.g., the detection threshold~$\check{\alpha}[k]$ obtained from the outputs of QTFs for troughs).

To distinguish between the respective rates (probabilities) with and without collisions, we mark those with collisions by overhead tildes.

We reserve the letters $i$, $j$, $k$, $m$, $n$, $L$, $M$ and $N$ for variables (including subscripts) that take integer values.

\section*{ACKNOWLEDGMENT}
This research used the ALICE High Performance Computing Facility at the University of Leicester.


\end{document}